\documentclass[12pt]{article}
\usepackage{epsfig,amsmath,euscript}

\renewcommand{\arraystretch}{1.2}

\def\pslash{\rlap{\hspace{0.02cm}/}{p}}
\def\nslash{\rlap{\hspace{0.02cm}/}{n}}
\def\vslash{\rlap{\hspace{0.02cm}/}{v}}

\def\F{{\EuScript F}}
\def\H{{\EuScript H}}
\def\X{{\EuScript X}}

\begin{document}

\begin{titlepage}

\begin{flushright}
CLNS~04/1858\\
{\tt hep-ph/0402094}\\[0.2cm]
February 9, 2004
\end{flushright}

\vspace{0.7cm}
\begin{center}
\Large\bf 
Factorization and Shape-Function Effects in\\
Inclusive B-Meson Decays
\end{center}

\vspace{0.8cm}
\begin{center}
{\sc S. W. Bosch, B. O. Lange, M. Neubert, and G. Paz}\\
\vspace{0.7cm}
{\sl Institute for High-Energy Phenomenology\\
Newman Laboratory for Elementary-Particle Physics, Cornell University\\
Ithaca, NY 14853, U.S.A.}
\end{center}

\vspace{1.0cm}
\begin{abstract}
\vspace{0.2cm}\noindent
Using methods of effective field theory, factorized expressions for 
arbitrary $\bar B\to X_u\,l^-\bar\nu$ decay distributions in the 
shape-function region of large hadronic energy and moderate hadronic 
invariant mass are derived. Large logarithms are resummed at 
next-to-leading order in renormalization-group improved perturbation 
theory. The operator product expansion is employed to relate moments of 
the renormalized shape function with HQET parameters such as $m_b$, 
$\bar\Lambda$ and $\lambda_1$ defined in a new physical subtraction 
scheme. An analytic expression for the asymptotic behavior of the shape 
function is obtained, which reveals that it is not positive definite. 
Explicit expressions are presented for the charged-lepton energy 
spectrum, the hadronic invariant mass distribution, and the spectrum in 
the hadronic light-cone momentum $P_+=E_H-|\vec{P}_H|$. A new method for 
a precision measurement of $|V_{ub}|$ is proposed, which combines good 
theoretical control with high efficiency and a powerful discrimination 
against charm background.
\end{abstract}
\vfil

\end{titlepage}

\section{Introduction}

Inclusive $B$-meson decays into light final-state particles, such as the 
semileptonic process $\bar B\to X_u\,l^-\bar\nu$ and the radiative 
process $\bar B\to X_s\gamma$, are of great importance to the 
determination of the Cabbibo--Kobayashi--Maskawa (CKM) matrix element 
$|V_{ub}|$ and play a prominent role in the search for New Physics at the 
$B$ factories. The ever increasing accuracy of experimental data on the 
inclusive decay rates and spectra provides a strong incentive to 
continuously improving the theoretical description of these processes. To 
keep up with the reduction of experimental errors, it will soon be 
necessary to have theoretical predictions for inclusive decay 
distributions with uncertainties below the 10\% level.

There are several challenges facing theorists pursuing this ambitious 
goal. Due to experimental cuts one is generally faced with a situation 
where the hadronic final states are constrained to have large energy 
$E_H\sim M_B$ but only moderate invariant mass 
$s_H\sim M_B\Lambda_{\rm QCD}$. In this kinematic region (called the 
``shape-function region''), the decay rates and spectra are obtained 
using a twist expansion \cite{Neubert:1993ch,Bigi:1993ex,Mannel:1994pm}, 
which is considerably more complicated than the conventional heavy-quark 
expansion employed in the calculation of inclusive rates for 
$\bar B\to X_c\,l^-\bar\nu$ decays 
\cite{Chay:1990da,Bigi:1992su,Manohar:1993qn}. In the twist expansion, 
infinite sets of power corrections are resummed into non-perturbative 
``shape functions'' describing the internal structure of the $B$ meson. 
At the same time, the short-distance corrections are also more 
complicated than in most other applications of heavy-quark expansions, 
because the kinematics implies the relevance of three separated mass 
scales: $M_B$ (``hard''), $\sqrt{M_B\Lambda_{\rm QCD}}$ 
(``hard-collinear''), and $\Lambda_{\rm QCD}$ (``soft''). Large 
logarithms of ratios of these scales arise at every order in perturbation 
theory and must be resummed 
\cite{Falk:1993vb,Korchemsky:1994jb,Akhoury:1995fp}. To properly 
disentangle the physics associated with these scales requires a 
sophisticated effective field-theory machinery \cite{Bauer:2000ew}, which 
has only been fully developed recently. A systematic treatment consists 
of matching QCD onto soft-collinear effective theory (SCET) in a first 
step, in which hard quantum fluctuations are integrated out. In a second 
step, SCET is matched onto heavy-quark effective theory (HQET), and 
hard-collinear modes are integrated out. The resulting expressions for 
inclusive differential decay rates have the factorized form 
$d\Gamma\sim H\,J\otimes S$ \cite{Korchemsky:1994jb,Akhoury:1995fp}. The 
function $H$ contains the hard corrections, the jet function $J$, which 
describes the properties of the final-state hadronic jet, contains the 
hard-collinear effects, and the shape function $S$ accounts for the 
internal soft dynamics in the $B$ meson 
\cite{Neubert:1993ch,Bigi:1993ex}. The $\otimes$ symbol implies a 
convolution over a light-cone momentum variable $\omega$ associated with 
the residual momentum of the $b$ quark inside the $B$ meson.

Up until now, no complete next-to-leading order predictions for inclusive
decay rates and spectra in the shape-function region have been presented 
in the literature. In the present paper we close this gap. We go beyond 
previous work in several important ways. First, we complete the matching 
calculations for the two-step matching QCD\,$\to$\,SCET\,$\to$\,HQET at
next-to-leading order in perturbation theory. We then derive 
renormalization-group equations governing the dependence of the functions 
$H$, $J$, and $S$ on the renormalization scale, and solve these equations
analytically in momentum space. Next, we derive several model-independent 
properties of the shape function $S$, which were so far unknown. In 
particular, we present the precise form of the relations between 
renormalized shape-function moments and HQET parameters such as 
$\bar\Lambda$ and $\lambda_1$, and we give an analytical formula for the 
asymptotic behavior of the shape function. An unexpected outcome of this 
analysis is the finding that the shape function is not positive definite, 
but acquires a negative radiative tail at large values of $|\omega|$. We 
also discuss how to derive shape-function independent relations between 
different decay spectra, which are free of spurious Landau-pole 
singularities. This improves on similar relations that can be found in 
the literature \cite{Leibovich:1999xfx,Neubert:2001sk,Leibovich:2001ra}. 
Finally, we propose a new method for a high-precision determination of 
$|V_{ub}|$, which offers several advantages over previous approaches for
extracting this important CKM parameter.

The remainder of this paper is organized as follows: In 
Section~\ref{sec:fact}, we present a derivation of the factorization 
formula for inclusive rates in the shape-function region using the 
position-space formulation of SCET. The one-loop matching calculations 
needed to derive perturbative expressions for the functions $H$, $J$, 
and $S$ are presented in Section~\ref{sec:match}. In 
Section~\ref{sec:RGevol}, we derive evolution equations for these 
functions and present their exact analytic solutions in momentum space. A 
comprehensive discussion of the properties of the shape function is given 
in Section~\ref{sec:properties}. While much of the discussion in this 
paper is necessarily rather technical, our results have important 
implications for the phenomenology of inclusive $B$-meson decays, 
including in particular the determination of $|V_{ub}|$. We have 
therefore organized the paper in such a way that the reader interested 
mainly in the applications of our formalism can skip the conceptual parts 
of the discussion in Sections~\ref{sec:fact}--\ref{sec:properties} and 
proceed directly with Section~\ref{sec:recap}, where we summarize our 
findings and collect all formulae needed for phenomenology. The following 
Sections~\ref{sec:rates}, \ref{sec:relations}, and \ref{sec:appls} 
contain explicit predictions for various decay rates and spectra, which 
are of relevance to experimenters.

During the final stages of this work, a paper by Bauer and Manohar 
appeared \cite{Bauer:2003pi}, which overlaps with parts of our analysis. 
These authors derive expressions for the hard-scattering kernels, the jet 
function, and the renormalized shape function which agree with our 
results reported in (\ref{Hres}), (\ref{Jrescale}), and (\ref{Sonshell}) 
below. They also obtain an expression for the anomalous dimension of the 
shape function, which coincides with our result in (\ref{ourgamma}). In 
many other aspects our analysis goes beyond that of \cite{Bauer:2003pi}. 
In particular, using a technique published some time ago by two of us 
\cite{Lange:2003ff}, we succeed to obtain an exact analytic solution of 
the evolution equation for the shape function in momentum space. At 
leading-logarithmic order we confirm an expression found earlier by 
Balzereit, Mannel, and Kilian (apart from a small mistake) 
\cite{Balzereit:1998yf}. While in \cite{Bauer:2003pi} resummed 
expressions are only given for large-$N$ moments of decay spectra, we 
present results for arbitrary decay distributions in physical phase 
space. We also disagree with Bauer and Manohar on the conclusion that 
moments of the renormalized shape function cannot be related to HQET 
parameters. In fact, we derive the explicit form of such relations.

\section{Factorization theorem for inclusive decay rates}
\label{sec:fact}

Using the optical theorem, the hadronic physics relevant to the inclusive 
semileptonic decay $\bar B\to X_u\,l^-\bar\nu$ can be related to a 
hadronic tensor $W^{\mu\nu}$ defined via the discontinuity of the forward 
$B$-meson matrix element of a correlator of two flavor-changing weak 
currents $J^\mu=\bar u\gamma^\mu(1-\gamma_5) b$ 
\cite{Chay:1990da,Bigi:1992su,Manohar:1993qn}. We define
\begin{equation}\label{WandTdef}
   W^{\mu\nu} = \frac{1}{\pi}\,\mbox{Im}\,
    \frac{\langle\bar B(v)|\,T^{\mu\nu}\,|\bar B(v)\rangle}{2 M_B} \,,
    \qquad
   T^{\mu\nu} = i \int d^4x\,e^{iq\cdot x}\,\,
    \mbox{T}\,\{ J^{\dagger\mu}(0), J^\nu(x) \} \,.
\end{equation}
Here $v$ is the $B$-meson velocity and $q$ the momentum carried by the 
lepton pair. The current correlator receives contributions from widely 
separated energy and distance scales. To extract the dependence on the 
large $b$-quark mass, it is convenient to rescale the heavy-quark field 
according to $b(x)=e^{-im_b v\cdot x}\,b'(x)$. The field $b'(x)$ carries 
the residual momentum $k=p_b-m_b v=O(\Lambda_{\rm QCD})$. Then the phase 
factor in (\ref{WandTdef}) becomes 
$e^{i(q-m_b v)\cdot x}\equiv e^{-ip\cdot x}$. In the parton model, 
$p=m_b v-q$ corresponds to the momentum of the jet of light partons into 
which the $b$-quark decays. 

As long as at least some components of the jet momentum are large 
compared with $\Lambda_{\rm QCD}$, the current correlator can be 
evaluated using a short-distance expansion. The simplest case arises 
if all components of $p^\mu\sim m_b\gg\Lambda_{\rm QCD}$. Integrating 
out the light-quark propagator then produces bilocal operators of the 
form $\bar b'(x)\dots b'(0)$ with $x\sim 1/m_b$, which can be expanded in 
terms of local operators $O_n$ multiplied by coefficient functions 
$\widetilde D_n(m_b,v\cdot x,x^2)$. Performing the integration over $x$ 
gives functions $D_n(m_b,v\cdot p,p^2)=\int d^4x\,e^{-ip\cdot x}\,%
\widetilde D_n(m_b,v\cdot x,x^2)$ depending on the different 
short-distance scales \cite{Bigi:1992su,Manohar:1993qn}. The $B$-meson 
matrix elements of the local operators $O_n$ can be evaluated using 
techniques of HQET \cite{Neubert:1993mb}. Thus, the operator product 
expansion produces an expansion in inverse powers of the heavy-quark 
mass. This expansion is relevant to the computation of inclusive decay 
rates without restrictive cuts on kinematic variables. Such rates are 
governed by parton kinematics as imprinted through the dependence on the 
jet momentum $p$. The unique operator of lowest dimension is $\bar h h$ 
(with $h$ the effective heavy-quark field in HQET), whose forward matrix 
element between $B$-meson states is unity. The Wilson coefficient of this 
operator has been computed to $O(\alpha_s)$ in \cite{DeFazio:1999sv}. At 
next-to-leading order in the heavy-quark expansion two new operators 
arise, which correspond to the kinetic energy and chromo-magnetic 
interaction of a heavy quark. Their Wilson coefficients have been 
computed at tree level in \cite{Bigi:1992su,Manohar:1993qn}.

A more complicated situation arises in the region of phase space in which 
the energy of the hadronic final states is much larger than its invariant 
mass, meaning that some components of the vector $p^\mu$ are much larger 
than others. In this case, the current correlator can be expanded in 
non-local light-cone operators 
\cite{Neubert:1993ch,Bigi:1993ex,Mannel:1994pm}. The formalism is most
transparent when presented using the language of soft-collinear effective 
theory (SCET) \cite{Bauer:2000ew}. To simplify the kinematics it is 
convenient to work in the $B$-meson rest frame, where $v^\mu=(1,0,0,0)$, 
and to choose the lepton momentum $\vec{q}$ along the negative $z$-axis, 
so that $\vec{p}$ points in the $z$ direction. Next, we define two 
light-like vectors $n^\mu=(1,0,0,1)$ and $\bar n^\mu=(1,0,0,-1)$ 
satisfying $n\cdot\bar n=2$, $n\cdot v=\bar n\cdot v=1$. Any 4-vector can 
be expanded in the light-cone basis as
\begin{equation}
   p^\mu = (n\cdot p)\,\frac{\bar n^\mu}{2}
    + (\bar n\cdot p)\,\frac{n^\mu}{2} + p_\perp^\mu
   \equiv p_+^\mu + p_-^\mu + p_\perp^\mu \,,
\end{equation}
where $p_\perp\cdot n=p_\perp\cdot\bar n=0$. In the shape-function 
region, the jet momentum (and likewise the momentum of the hadronic final
state) scales like $p^\mu\sim E(\lambda,1,\sqrt{\lambda})$, where 
$\lambda\sim\Lambda_{\rm QCD}/E$ is the SCET expansion parameter and 
$E\sim m_b\gg\Lambda_{\rm QCD}$. (For the jet momentum, $p_\perp=0$ by 
choice of the coordinate system.) It follows that $p_-^\mu\sim E$ is a 
hard scale whereas $p_+^\mu\sim\Lambda_{\rm QCD}$ is a long-distance 
hadronic scale. The jet invariant mass, $p^2\sim E\Lambda_{\rm QCD}$, 
defines a hybrid, intermediate short-distance scale. We refer to a 
momentum with these scaling properties as ``hard-collinear''. The 
appropriate effective field theory for integrating out the short-distance 
fluctuations associated with the hard scale $p_-$ is called 
SCET$_{\rm I}$ and has been discussed in detail in the literature 
\cite{Bauer:2000yr,Bauer:2001yt,Chay:2002vy,Beneke:2002ph}. Here we use 
the coordinate-space formulation of SCET developed in 
\cite{Beneke:2002ph,Hill:2002vw}.

Below a matching scale $\mu_h\sim m_b$, the semileptonic current can be 
expanded as
\begin{equation}\label{current}
   \bar u(x)\gamma^\mu(1-\gamma_5) b'(x)
   = \sum_{i=1}^3 \int ds\,\widetilde C_i(s)\,
   \bar\X(x+s\bar n)\,\Gamma_i^\mu\,\H(x_-) + \dots \,,
\end{equation}
where the dots denote higher-order terms in the SCET expansion, which can 
be neglected at leading power in $\Lambda_{\rm QCD}/m_b$. The soft 
heavy-quark field $\H(x_-)=S_s^\dagger(x_-)\,h(x_-)$ and the 
hard-collinear light-quark field 
$\X(x)=S_s^\dagger(x_-)\,W_{hc}^\dagger(x)\,\xi(x)$ are SCET building 
blocks that are invariant under a set of homogeneous soft and 
hard-collinear gauge transformations 
\cite{Hill:2002vw,Beneke:2002ni,Becher:2003qh}. The objects $S_s$ and 
$W_{hc}$ are soft and hard-collinear Wilson lines \cite{Bauer:2001yt}. 
Soft fields in SCET are multi-pole expanded and only depend on the minus 
component of the position vector $x$. Finally, the position-space Wilson 
coefficient functions $\widetilde C_i(s)$ depend on the variable $s$ 
defining the position of the hard-collinear field. Here and below we 
denote functions in position space with a tilde, which is omitted from 
the corresponding Fourier-transformed functions in momentum space. A 
convenient basis of Dirac structures in (\ref{current}) is 
\begin{equation}
   \Gamma_1^\mu = \gamma^\mu(1-\gamma_5) \,, \qquad
   \Gamma_2^\mu = v^\mu(1+\gamma_5) \,, \qquad
   \Gamma_3^\mu = \frac{n^\mu}{n\cdot v}\,(1+\gamma_5) \,.
\end{equation}
The current correlator in (\ref{WandTdef}) then becomes
\begin{equation}\label{T0}
   T^{\mu\nu} = i \int d^4x\,e^{-ip\cdot x}
   \sum_{i,j=1}^3 \int ds\,dt\,\widetilde C_j^*(t)\,\widetilde C_i(s)\,\,
    \mbox{T} \left\{ \bar\H(0)\,\bar\Gamma_j^\mu\,\X(t\bar n),
    \bar\X(x+s\bar n)\,\Gamma_i^\nu\,\H(x_-) \right\} + \dots \,.
\end{equation}
In a second step, the hard-collinear fluctuations associated with the
light-quark jet can be integrated out by matching SCET onto HQET at an 
intermediate scale $\mu_i\sim\sqrt{m_b\Lambda_{\rm QCD}}$. At leading 
order the SCET Lagrangian (when written in terms of the gauge-invariant 
fields such as $\X$) does not contain interactions between hard-collinear 
and soft fields. Since the external $B$-meson states only contain soft 
constituents, we can take the vacuum matrix element over the 
hard-collinear fields, defining a jet function
\begin{equation}\label{Jdef}
   \langle\,\Omega|\,\mbox{T} \left\{ \X_k(t\bar n),\bar\X_l(x+s\bar n)
   \right\} |\Omega\rangle\equiv \delta_{kl}\,
   \widetilde{\cal J}(x+(s-t)\bar n) + \dots \,,
\end{equation}
where $k,l$ are color indices, and we have used translational invariance 
to determine the dependence on the coordinate vectors. At higher orders
in SCET power counting, additional jet functions would arise, but their 
contributions can be neglected at leading power. Shifting the integration 
variable from $x$ to $z=x+(s-t)\bar n$, with $z_-=x_-$, and introducing 
the Fourier-transformed Wilson coefficient functions
\begin{equation}
   C_i(\bar n\cdot p) = \int ds\,e^{is\bar n\cdot p}\,\widetilde C_i(s)
   \,,
\end{equation}
we then obtain
\begin{equation}\label{T1}
   T^{\mu\nu} = i \sum_{i,j=1}^3 C_j^*(\bar n\cdot p)\,C_i(\bar n\cdot p)
   \int d^4z\,e^{-ip\cdot z}\,\bar\H(0)\,\bar\Gamma_j^\mu\,
   \widetilde{\cal J}(z)\,\Gamma_i^\nu\,\H(z_-) + \dots \,.
\end{equation}
In the next step, we rewrite the bilocal heavy-quark operator as 
\cite{Neubert:1993ch}
\begin{eqnarray}\label{SFops}
   \bar\H(0)\,\Gamma\,\H(z_-)
   &=& (\bar h\,S_s)(0)\,\Gamma\,e^{z_-\cdot\partial_+}\,
    (S_s^\dagger\,h)(0) 
    = \bar h(0)\,\Gamma\,e^{z_-\cdot D_+}\,h(0) \nonumber\\
   &=& \int d\omega\,e^{-\frac{i}{2}\omega\bar n\cdot z}\,
    \bar h(0)\,\Gamma\,\delta(\omega-in\cdot D)\,h(0) \,,
\end{eqnarray}
where $\Gamma$ may be an arbitrary (even $z$-dependent) Dirac structure,
and we have used the property $in\cdot D\,S_s=S_s\,in\cdot\partial$ of 
the soft Wilson line $S_s$, where $iD^\mu=i\partial^\mu+g_s A_s^\mu$ is 
the covariant derivative with respect to soft gauge transformations. When 
this expression is used in (\ref{T1}), the resulting formula for the 
correlator involves the Fourier transform of the jet function,
\begin{equation}
   \int d^4z\,e^{-ip\cdot z}\,\widetilde{\cal J}(z)
   = \pslash_-\,{\cal J}(p^2) \,,
\end{equation}
however with $p^\mu$ replaced by the combination 
$p_\omega^\mu\equiv p^\mu+\frac{1}{2}\omega\bar n^\mu$. In defining the 
momentum-space jet function ${\cal J}(p^2)$ we have taken into account 
that the matrix element in (\ref{Jdef}) vanishes when multiplied by 
$\nslash$ from either side, since $\nslash\,\X=0$. Using that 
$p_{\omega-}=p_-$, we now obtain
\begin{equation}\label{Tres}
   T^{\mu\nu} = i\sum_{i,j=1}^3 H_{ij}(\bar n\cdot p)
   \int d\omega\,{\cal J}(p_\omega^2)\,
   \bar h\,\bar\Gamma_j^\mu\,\pslash_-\Gamma_i^\nu\,
   \delta(\omega-in\cdot D)\,h + \dots \,,
\end{equation}
where $H_{ij}(\bar n\cdot p)=C_j^*(\bar n\cdot p)\,C_i(\bar n\cdot p)$ 
are called the hard functions.

In order to compute the hadronic tensor we take the discontinuity of the 
jet function,
\begin{equation}\label{jetdef}
   J(p^2) = \frac{1}{\pi}\,\mbox{Im}\,[i{\cal J}(p^2)] \,,
\end{equation}
and evaluate the $B$-meson matrix element of the soft operator using the 
HQET trace formalism, which allows us to write \cite{Neubert:1993mb}
\begin{equation}\label{Sdef}
   \frac{\langle\bar B(v)|\,\bar h\,\Gamma\,\delta(\omega-in\cdot D)\,
   h\,|\bar B(v)\rangle}{2M_B}
   = S(\omega)\,\frac12\,\mbox{tr}\left( \Gamma\,\frac{1+\vslash}{2}
   \right) + \dots 
\end{equation}
at leading power in the heavy-quark expansion. The soft function 
$S(\omega)$ coincides with the shape function $f(k_+)$ introduced in
\cite{Neubert:1993ch}. This gives the factorization formula
\begin{equation}\label{Wres}
   W^{\mu\nu} = \sum_{i,j=1}^3 H_{ij}(\bar n\cdot p)\,
   \mbox{tr}\left( \bar\Gamma_j^\mu\,\frac{\pslash_-}{2}\,\Gamma_i^\nu\,
   \frac{1+\vslash}{2} \right)
   \int d\omega\,J(p_\omega^2)\,S(\omega) + \dots \,.
\end{equation}
At leading power one could replace $\pslash_-\to\pslash$ and 
$\bar n\cdot p=2v\cdot p_-\to 2v\cdot p$ in this result.

In the final expressions (\ref{Tres}) and (\ref{Wres}) the dependence on 
the three scales $\bar n\cdot p\sim m_b$, 
$p_\omega^2\sim m_b\Lambda_{\rm QCD}$ and $\omega\sim\Lambda_{\rm QCD}$ 
has been factorized into the hard, jet, and shape functions, 
respectively. Large logarithms associated with ratios of these scales can 
be resummed by solving renormalization-group equations for the scale 
dependence of these component functions. The factorization formula 
(\ref{Wres}) was derived at tree level in 
\cite{Neubert:1993ch,Bigi:1993ex}, and was generalized to all orders in 
perturbation theory in \cite{Korchemsky:1994jb,Akhoury:1995fp}. The 
derivation presented above is equivalent to a proof of this formula 
presented in \cite{Bauer:2001yt} (see also \cite{Bauer:2000ew}). The 
limits of integration in the convolution integral are determined by the 
facts that the jet function defined in (\ref{jetdef}) has support for 
$p_\omega^2\ge 0$, and the shape function defined in (\ref{Sdef}) has 
support for $-\infty<\omega\le\bar\Lambda$ with $\bar\Lambda=M_B-m_b$, 
where $m_b$ is the heavy-quark pole mass. The argument $p_\omega^2$ of 
the jet function can be rewritten as
\begin{equation}\label{pw2}
   p_\omega^2 = p^2 + \bar n\cdot p\,\omega
   = \bar n\cdot p\,(n\cdot P_H - (\bar\Lambda-\omega))
   \equiv \bar n\cdot p\,(n\cdot P_H - \hat\omega) \,,
\end{equation}
where $P_H=M_B\,v-q=p+\bar\Lambda v$ is the 4-momentum of the hadronic 
final state, and the variable $\hat\omega=\bar\Lambda-\omega\ge 0$. 
Finally, $n\cdot P_H=E_H-|\vec{P}_H|=s_H/2E_H+O(\Lambda_{\rm QCD}^2/m_b)$ 
is a kinematic variable of order $\Lambda_{\rm QCD}$ related to the 
hadronic invariant mass and energy of the final state. The usefulness of
this variable has also been emphasized in 
\cite{Balzereit:1998yf,Mannel:1999gs,Aglietti:2002md}. We shall see below 
that expressing the convolution integral in terms of the new variable 
$\hat\omega$ eliminates any spurious dependence of the decay spectra on 
the $b$-quark pole mass. It follows that in the shape-function region the 
hadronic tensor is most naturally written as a function of the ``parton 
variable'' $\bar n\cdot p$ and the ``hadron variable'' $n\cdot P_H$, and
the inclusive spectra are governed by a combination of parton and 
hadronic kinematics.

Using the fact that the Wilson coefficients $C_i$ are real and hence 
$H_{ij}$ is symmetric in its indices, we find
\begin{eqnarray}
   \sum_{i,j=1}^3 H_{ij}\,\mbox{tr}\left( \bar\Gamma_j^\mu\,
   \frac{\pslash_-}{2}\,\Gamma_i^\nu\,\frac{1+\vslash}{2} \right)
   &=& 2 H_{11} \left( p_-^\mu v^\nu + p_-^\nu v^\mu
    - g^{\mu\nu}\,v\cdot p_-
    - i\epsilon^{\mu\nu\alpha\beta} p_{-\alpha} v_\beta \right) \\
   &&\hspace{-4.0cm}\mbox{}+ 2 H_{22}\,v\cdot p_-\,v^\mu v^\nu
    + 2 (H_{12}+H_{23})\,(p_-^\mu v^\nu + p_-^\nu v^\mu)
    + 2 (2H_{13}+H_{33})\,\frac{p_-^\mu p_-^\nu}{v\cdot p_-} \,.
    \nonumber
\end{eqnarray}
This result may be compared with the general Lorentz decomposition of the 
hadronic tensor given in \cite{DeFazio:1999sv}:
\begin{eqnarray}\label{Wdecomp}
   W^{\mu\nu} &=& W_1 \left( p^\mu v^\nu + p^\nu v^\mu
    - g^{\mu\nu}\,v\cdot p
    - i\epsilon^{\mu\nu\alpha\beta} p_\alpha v_\beta \right)
    - W_2\,g^{\mu\nu}
    \nonumber\\
   &&\mbox{}+ W_3\,v^\mu v^\nu + W_4\,(p^\mu v^\nu + p^\nu v^\mu)
    + W_5\,p^\mu p^\nu
\end{eqnarray}
We see that the structure function $W_2$ is not generated at leading 
order in the SCET expansion. Since only the Wilson coefficient $C_1$ 
is non-zero at tree-level, the structure function $W_1$ receives 
leading-power contributions at tree level, whereas $W_4$ and $W_5$ 
receive leading-power contributions at $O(\alpha_s(m_b))$. The function 
$W_3$ receives leading-power contributions only at $O(\alpha_s^2(m_b))$, 
which is beyond the accuracy of a next-to-leading order calculation.

\section{Matching calculations}
\label{sec:match}

In this section we derive perturbative expressions for the hard functions 
$H_{ij}(\bar n\cdot p)$ and the jet function $J(p_\omega^2)$ in 
(\ref{Wres}) at next-to-leading order in $\alpha_s$. To this end, we 
match expressions for the hadronic tensor obtained in full QCD, SCET, and 
HQET, using for simplicity on-shell external $b$-quark states. We also 
present results for the renormalized shape function in the parton model, 
which are needed in the matching calculation. Throughout this paper we 
use the $\overline{\rm MS}$ subtraction scheme and work in 
$d=4-2\epsilon$ space-time dimensions.

\subsection{Hard functions}

Perturbative expressions at $O(\alpha_s)$ for the structure functions 
$W_i$ in the decomposition (\ref{Wdecomp}) have been obtained in 
\cite{DeFazio:1999sv} by evaluating one-loop Feynman graphs for the 
current correlator $T^{\mu\nu}$ using on-shell external quark states with 
residual momentum $k$ (satisfying $v\cdot k=0$) in full QCD. The leading 
terms in the region of hard-collinear jet momenta are
\begin{equation}\label{Fulvia}
\begin{aligned}
   \frac12\,W_1 &= \delta(p_k^2) \left[ 1 - \frac{C_F\alpha_s}{4\pi}
    \left( 8\ln^2 y - 10\ln y + \frac{2\ln y}{1-y} + 4 L_2(1-y)
    + \frac{4\pi^2}{3} + 5 \right) \right] \\
   &\quad\mbox{}+ \frac{C_F\alpha_s}{4\pi} \left[ -4 
    \left( \frac{\ln(p_k^2/m_b^2)}{p_k^2} \right)_{\!*}^{\![m_b^2]}
    + (8\ln y-7) \left( \frac{1}{p_k^2} \right)_{\!*}^{\![m_b^2]}
    \right] + \dots \,, \\
   \frac12\,W_4 &= \delta(p_k^2)\,\frac{C_F\alpha_s}{4\pi}\,
    \frac{2}{1-y} \left( \frac{y\ln y}{1-y} + 1 \right) + \dots \,, \\
   \frac{m_b}{4}\,W_5 &= \delta(p_k^2)\,\frac{C_F\alpha_s}{4\pi}\,
    \frac{2}{1-y} \left( \frac{1-2y}{1-y}\,\ln y - 1 \right) + \dots \,, 
\end{aligned}
\end{equation}
whereas $W_2$ and $W_3$ do not receive leading-power contributions at 
this order, in accordance with our general observations made above. Here
$\alpha_s\equiv\alpha_s(\mu)$, $y=\bar n\cdot p/m_b$, and 
$p_k^2=p^2+\bar n\cdot p\,n\cdot k$. We have used that 
$2v\cdot p/m_b=\bar n\cdot p/m_b+O(\lambda)$ in the hard-collinear 
region. The star distributions are generalized plus distributions defined 
as \cite{DeFazio:1999sv}
\begin{equation}\label{star}
\begin{aligned}
   \int_{\le 0}^z\!dx\,F(x) 
   \left( \frac{1}{x} \right)_{\!*}^{\![u]}
   &= \int_0^z\!dx\,\frac{F(x)-F(0)}{x} + F(0)\,\ln\frac{z}{u} \,, \\
   \int_{\le 0}^z\!dx\,F(x)
    \left( \frac{\ln(x/u)}{x} \right)_{\!*}^{\![u]}
   &= \int_0^z\!dx\,\frac{F(x)-F(0)}{x}\,\ln\frac{x}{u} 
    + \frac{F(0)}{2}\,\ln^2\frac{z}{u} \,,
\end{aligned}
\end{equation}
where $F(x)$ is a smooth test function. For later purposes, we note the 
useful identities
\begin{equation}\label{ids}
\begin{aligned}
   \lambda \left( \frac{1}{\lambda x} \right)_{\!*}^{\![u]}
   &= \left( \frac{1}{x} \right)_{\!*}^{\![u/\lambda]}
    = \left( \frac{1}{x} \right)_{\!*}^{\![u]} + \delta(x)\,\ln\lambda 
    \,, \\
   \lambda \left( \frac{\ln(\lambda x/u)}{\lambda x}
   \right)_{\!*}^{\![u]}
   &= \left( \frac{\ln(\lambda x/u)}{x} \right)_{\!*}^{\![u/\lambda]}
    = \left( \frac{\ln(x/u)}{x} \right)_{\!*}^{\![u]}
    + \left( \frac{1}{x} \right)_{\!*}^{\![u]} \ln\lambda 
    + \frac{\delta(x)}{2}\,\ln^2\lambda \,.
\end{aligned}
\end{equation}

In order to find the hard functions $H_{ij}$, we calculate the 
discontinuity of the current correlator (\ref{T0}) between on-shell 
heavy-quark states in momentum space. We work to one-loop order in SCET, 
keeping $i,j$ fixed and omitting the Wilson coefficient functions. The 
corresponding tree diagram yields
\begin{equation}
   D^{(0)} = K\,\delta(p_k^2) \,, \qquad \mbox{with} \quad
   K = \bar u_b(v)\,\bar\Gamma_j^\mu\,\pslash_-\,\Gamma_i^\nu\,u_b(v) \,,
\end{equation}
where $u_b(v)$ are on-shell HQET spinors normalized to unity, and the 
quantity $K$ corresponds to the Dirac trace in (\ref{Wres}). The 
interpretation of this result in terms of hard, jet, and soft functions 
is that, at tree level, $J^{(0)}(p_\omega^2)=\delta(p_\omega^2)$ and 
$S_{\rm parton}^{(0)}(\omega)=\delta(\omega-n\cdot k)$. (The second 
result is specific to the free-quark decay picture.) Then the convolution 
integral $\int d\omega\,J(p_\omega^2)\,S(\omega)$ in (\ref{Wres}) 
produces $\delta(p_k^2)$, and comparison with (\ref{Fulvia}) shows that 
$H_{11}^{(0)}=1$, while all other hard functions vanish at tree level.

\begin{figure}
\begin{center}
\epsfig{file=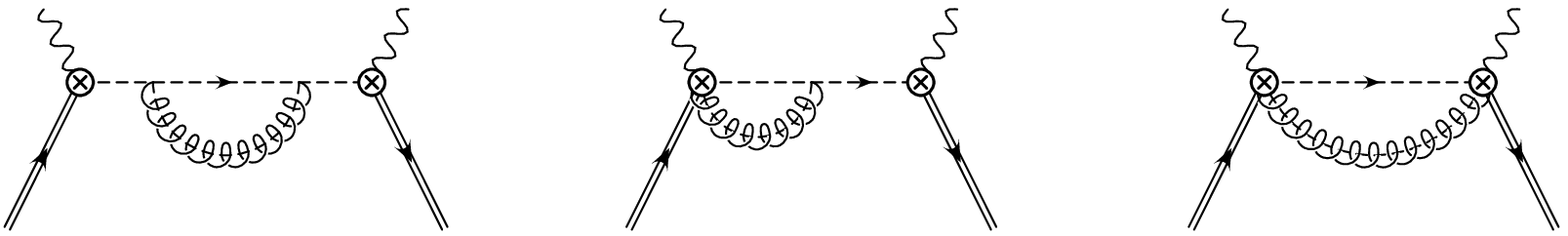,width=15cm}
\epsfig{file=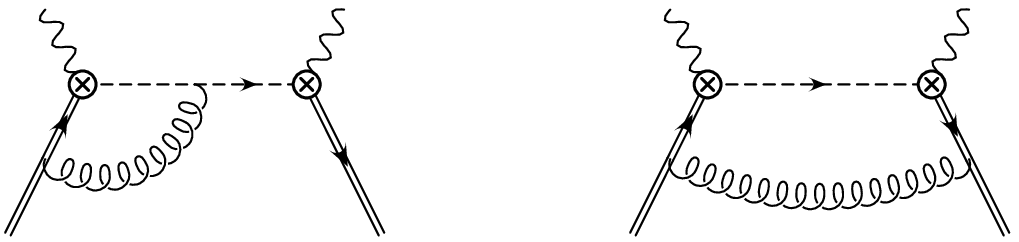,width=10cm}
\end{center}
\centerline{\parbox{14cm}{\caption{\label{fig:scet}
One-loop diagrams contributing to the current correlator in SCET. The
effective current operators are denoted by crossed circles, and 
hard-collinear propagators are drawn as dashed lines. Mirror graphs 
obtained by exchanging the two currents are not shown.}}}
\end{figure}

The diagrams contributing at one-loop order are shown in 
Figure~\ref{fig:scet}. They are evaluated using the Feynman rules of 
SCET. The first three graphs contain hard-collinear gluon exchanges, 
while the last two diagrams contain soft exchanges. The wave-function 
renormalization factors of the external heavy quarks equal 1 on-shell. 
For the sum of all hard-collinear exchange graphs, we find
\begin{equation}\label{hcloops}
   D_{hc}^{(1)} = K\,\frac{C_F\alpha_s}{4\pi} \left[
   \left( \frac{4}{\epsilon^2} + \frac{3}{\epsilon} + 7 - \pi^2 \right)
   \delta(p_k^2)
   + 4 \left( \frac{\ln(p_k^2/\mu^2)}{p_k^2} \right)_{\!*}^{\![\mu^2]}
   - \left( \frac{4}{\epsilon} + 3 \right)
   \left( \frac{1}{p_k^2} \right)_{\!*}^{\![\mu^2]} \right] .
\end{equation}
The sum of the soft contributions is given by
\begin{eqnarray}\label{sloops}
   D_{s}^{(1)} 
   &=& K\,\frac{C_F\alpha_s}{4\pi} \Bigg[
    \left( - \frac{2}{\epsilon^2} - \frac{4}{\epsilon}\,L
    + \frac{2}{\epsilon} - 4L^2 + 4L - \frac{\pi^2}{6} \right)
    \delta(p_k^2) \nonumber\\
   &&\quad\mbox{}- 8 \left( \frac{\ln(p_k^2/\mu^2)}{p_k^2} 
    \right)_{\!*}^{\![\mu^2]}
    + \left( \frac{4}{\epsilon} + 8L - 4 \right)
   \left( \frac{1}{p_k^2} \right)_{\!*}^{\![\mu^2]} \Bigg] \,,
\end{eqnarray}
where $L=\ln(\bar n\cdot p/\mu)$. The $1/\epsilon$ poles in the sum of 
the hard-collinear and soft contributions are subtracted by a 
multiplicative renormalization factor $Z_J^2$ applied to the bare current 
correlator in (\ref{T0}), where
\begin{equation}\label{ZJres}
   Z_J = 1 + \frac{C_F\alpha_s}{4\pi} \left( - \frac{1}{\epsilon^2}
    + \frac{2}{\epsilon}\,L - \frac{5}{2\epsilon} \right)
\end{equation}
is the (momentum-space) current renormalization constant in SCET 
\cite{Bauer:2000yr}. Taking the sum of (\ref{hcloops}) and (\ref{sloops}) 
after subtraction of the pole terms, and matching it with the results in 
(\ref{Fulvia}), we find that at one-loop order
\begin{equation}\label{Hres}
\begin{aligned}
   H_{11}(\bar n\cdot p) &= 1 + \frac{C_F\alpha_s}{4\pi}
    \left( -4 L^2 + 10 L - 4\ln y - \frac{2\ln y}{1-y} - 4 L_2(1-y)
    - \frac{\pi^2}{6} -12 \right) , \\
   H_{12}(\bar n\cdot p) &= \frac{C_F\alpha_s}{4\pi}\,
    \frac{2}{1-y} \left( \frac{y\ln y}{1-y} + 1 \right) , \\
   H_{13}(\bar n\cdot p) &= \frac{C_F\alpha_s}{4\pi}\,
    \frac{y}{1-y} \left( \frac{1-2y}{1-y}\,\ln y - 1 \right) . 
\end{aligned}
\end{equation}
In deriving these results we have used the identities (\ref{ids}) to 
rearrange the various star distributions. The remaining hard functions 
start at $O(\alpha_s^2)$. Using the relation $H_{ij}=C_i\,C_j$, one can 
derive from these results expressions for the Wilson coefficients in the 
expansion of the semileptonic current in (\ref{current}). We confirm the 
expressions for these coefficients given in \cite{Bauer:2000yr}.

\subsection{Jet function}

After the hadronic tensor is matched onto HQET as shown in (\ref{Wres}), 
the SCET loop graphs in Figure~\ref{fig:scet} determine the one-loop 
contributions to the product of the jet function and the shape function 
in (\ref{Wres}). We may write this product symbolically as 
$J^{(1)}\otimes S^{(0)}+J^{(0)}\otimes S^{(1)}$, where the $\otimes$ 
symbol means a convolution in $\omega$. Whereas the jet function is a 
short-distance object that can be calculated in perturbation theory, the 
shape function is defined in terms of a hadronic matrix element and 
cannot be properly described using Feynman diagrams with on-shell 
external quark states. In a second step, we must therefore extract from 
the results (\ref{hcloops}) and (\ref{sloops}) the one-loop contribution 
$J^{(1)}$ to the jet function. To this end, we must compute the 
renormalized shape function at one-loop order in the parton model. This 
will be done in the next subsection. We may, however, already anticipate 
the result for the jet function at this point, because at one-loop order 
the graphs in Figure~\ref{fig:scet} can be separated into diagrams with 
hard-collinear (first line) or soft (second line) gluon exchange. (This 
separation would be non-trivial beyond one-loop order.) The 
hard-collinear contribution in (\ref{hcloops}) thus determines the 
convolution $J^{(1)}\otimes S^{(0)}=J^{(1)}(p_k^2)$, whereas the soft 
contribution in (\ref{sloops}) corresponds to $J^{(0)}\otimes S^{(1)}$. 
It follows that the renormalized jet function is given by the 
distribution
\begin{equation}\label{Jres}
   J(p_\omega^2) = \delta(p_\omega^2)
   + \frac{C_F\alpha_s}{4\pi} \left[ (7-\pi^2)\,\delta(p_\omega^2)
   + 4 \left( \frac{\ln(p_\omega^2/\mu^2)}{p_\omega^2} 
   \right)_{\!*}^{\![\mu^2]}
   - 3 \left( \frac{1}{p_\omega^2} \right)_{\!*}^{\![\mu^2]} \right] .
\end{equation}
This result disagrees with a corresponding expression obtained in 
\cite{Mannel:2000aj}. It will often be useful to separate the dependence 
on $\bar n\cdot p$ and $n\cdot P_H$ in this result by means of the 
substitution $p_\omega^2=y\,\hat p_\omega^2$, where 
$\hat p_\omega^2=m_b(n\cdot P_H-\hat\omega)$ according to (\ref{pw2}). 
Using the identities (\ref{ids}), we find
\begin{eqnarray}\label{Jrescale}
   y\,J(p_\omega^2)\equiv\hat J(\hat p_\omega^2,y)
   &=& \delta(\hat p_\omega^2) + \frac{C_F\alpha_s}{4\pi} \Bigg[
    \big( 2\ln^2 y - 3\ln y + 7 - \pi^2 \big)\,\delta(\hat p_\omega^2)
    \nonumber\\
   &&\quad\mbox{}+ 4 \left(
    \frac{\ln(\hat p_\omega^2/\mu^2)}{\hat p_\omega^2} 
    \right)_{\!*}^{\![\mu^2]}
    + (4\ln y - 3) 
    \left( \frac{1}{\hat p_\omega^2} \right)_{\!*}^{\![\mu^2]} \Bigg] \,.
\end{eqnarray}
The jet function is non-zero only if $y\ge 0$ and 
$n\cdot P_H\ge\hat\omega$, which ensures that $p_\omega^2\ge 0$.

\subsection{Renormalized shape function}
\label{sec:SFrenorm}

Having calculated the short-distance objects $H_{ij}$ and $J$ in the 
factorization formula (\ref{Wres}) at one-loop order, we now turn to a 
study of radiative corrections to the shape function $S(\omega)$. There 
is considerable confusion in the literature about the renormalization 
properties of the shape function, and several incorrect results for its 
anomalous dimension have been published. We therefore present our 
calculation in some detail in this subsection and the following section.

\begin{figure}
\begin{center}
\epsfig{file=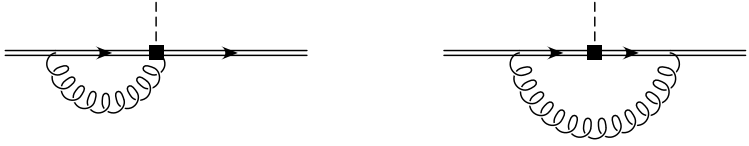,width=10cm}
\end{center}
\centerline{\parbox{14cm}{\caption{\label{fig:hqet}
Radiative corrections to the shape function. The bilocal HQET operator is
denoted by the black square. A mirror copy of the first graph is not 
shown.}}}
\end{figure}

According to (\ref{Sdef}), the shape function is defined in terms of a 
hadronic matrix element in HQET and thus cannot be calculated 
perturbatively. However, the renormalization properties of this function 
can be studied using perturbation theory. To this end, we evaluate the 
matrix element (\ref{Sdef}) in HQET using external heavy-quark states 
with residual momentum $k$. For the time being, we keep $v\cdot k$ 
non-zero to regularize infra-red singularities. The relevant one-loop 
graphs are depicted in Figure~\ref{fig:hqet}. Adding the tree 
contribution, we obtain for the matrix element of the bare shape-function 
operator $O(\omega)=\bar h\,\Gamma\,\delta(\omega-in\cdot D)\,h$ 
(expressed in terms of renormalized fields)
\begin{eqnarray}\label{Sbare}
   S_{\rm bare}(\omega)
   &=& Z_h\,\delta(\omega-n\cdot k)
    - \frac{4C_F g_s^2}{(4\pi)^{2-\epsilon}}\,\Gamma(1+\epsilon)
    \nonumber\\
   &\times& \Bigg\{ \frac{1}{\epsilon}
    \int_0^\infty\!dl\,l^{-1-2\epsilon}\,\Big[ \delta(\omega-n\cdot k)
    - \delta(\omega-n\cdot k+l) \Big] \left( 1 + \frac{\delta}{l} 
    \right)^{-\epsilon} \nonumber\\
   &&\quad\mbox{}+ \theta(n\cdot k-\omega)\,(n\cdot k-\omega)^{-\epsilon}
    (n\cdot k-\omega+\delta)^{-1-\epsilon} \Bigg\} \,,
\end{eqnarray}
where $\delta=-2v\cdot k$, and
\begin{equation}
   Z_h = 1 + \frac{4C_F g_s^2}{(4\pi)^{2-\epsilon}}\,\Gamma(2\epsilon)\,
   \Gamma(1-\epsilon)\,\delta^{-2\epsilon}
\end{equation}
is the off-shell wave-function renormalization constant of a heavy quark 
in HQET. The next step is to extract the ultra-violet poles from this 
result, which determine the anomalous dimension of the shape function. We 
define a renormalization factor through
\begin{equation}\label{Sren}
\begin{aligned}
   S_{\rm ren}(\omega) &= \int_{-\infty}^{\bar\Lambda}\!d\omega'\,
    Z_S(\omega,\omega')\,S_{\rm bare}(\omega') \,, \\
   Z_S(\omega,\omega') &= \delta(\omega-\omega')
    + \frac{C_F\alpha_s}{4\pi}\,z_S(\omega,\omega') + \dots \,.
\end{aligned}
\end{equation}
The result for $Z_S$ following from (\ref{Sbare}) must be interpreted as 
a distribution on test functions $F(\omega')$ with support on the 
interval $-\infty<\omega'\le\bar\Lambda$. We obtain
\begin{eqnarray}\label{zres}
   z_S(\omega,\omega')
   &=& \left( \frac{2}{\epsilon^2}
    + \frac{4}{\epsilon}\,\ln\frac{\mu}{\bar\Lambda-\omega}
    - \frac{2}{\epsilon} \right) \delta(\omega-\omega')
    - \frac{4}{\epsilon}\,\left( 
    \frac{\theta(\omega'-\omega)}{\omega'-\omega} \right)_{\!+}
    \nonumber\\
   &=& \left( \frac{2}{\epsilon^2} - \frac{2}{\epsilon} \right)
    \delta(\omega-\omega')
    - \frac{4}{\epsilon}\,\left( \frac{1}{\omega'-\omega}
    \right)_{\!*}^{\![\mu]} .
\end{eqnarray}
Note the peculiar dependence on the parameter $\bar\Lambda$ setting the 
upper limit on the integration over $\omega'$ in (\ref{Sren}), which 
combines with the plus distribution to form a star distribution in the 
variable $(\omega'-\omega)$. 

We can now determine the renormalized shape function from (\ref{Sren}). 
The result must once again be interpreted as a distribution, this time on 
test functions $F(\omega)$ integrated over a {\em finite\/} interval 
$-\Lambda_{\rm had}\le\omega\le\bar\Lambda$. In practice, the value of
$\Lambda_{\rm had}$ is set by kinematics or by virtue of some 
experimental cut (see Sections~\ref{sec:rates} and \ref{sec:appls} 
below). The result is
\begin{eqnarray}
   S_{\rm parton}(\omega) &=& \delta(\omega-n\cdot k)\,\left\{ 1
    - \frac{C_F\alpha_s}{\pi} \left[ \frac{\pi^2}{24}
    + L_2\!\left( \frac{-\delta}{\Lambda_{\rm had}+n\cdot k} \right)
    \right] \right\} \nonumber\\
   &&\mbox{}- \frac{C_F\alpha_s}{\pi}\,\Bigg\{
    \left[ \frac{\theta(n\cdot k-\omega)}{n\cdot k-\omega} \left(
    \ln\frac{n\cdot k-\omega}{\mu} + \ln\frac{n\cdot k-\omega+\delta}{\mu}
    \right) \right]_+ \nonumber\\
   &&\quad\mbox{}+ \delta(n\cdot k-\omega)\,
    \ln^2\frac{\Lambda_{\rm had}+n\cdot k}{\mu}
    + \frac{\theta(n\cdot k-\omega)}{n\cdot k-\omega+\delta}
    + \delta(n\cdot k-\omega)\,\ln\frac{\delta}{\mu} \Bigg\} \,. \quad
\end{eqnarray}
While it was useful to keep the heavy quark off-shell in the calculation 
of the ultra-violet renormalization factor, the limit 
$\delta=-2v\cdot k\to 0$ can be taken in the result for the renormalized 
shape functions without leading to infra-red singularities. This gives
\begin{eqnarray}\label{Sonshell}
   S_{\rm parton}(\omega) &=& \delta(\omega-n\cdot k)\,\left( 1
    - \frac{C_F\alpha_s}{\pi}\,\frac{\pi^2}{24} \right) \nonumber\\
   &&\mbox{}- \frac{C_F\alpha_s}{\pi} \left[
    2 \left( \frac{1}{n\cdot k-\omega} \ln\frac{n\cdot k-\omega}{\mu}
    \right)_{\!*}^{\![\mu]} 
    + \left( \frac{1}{n\cdot k-\omega} \right)_{\!*}^{\![\mu]}
    \right] ,
\end{eqnarray}
where the star distributions must now be understood as distributions in 
the variable $(n\cdot k-\omega)$.

We stress that these results for the renormalized shape function are 
obtained in the parton model and can in no way provide a realistic 
prediction for the functional form of $S(\omega)$. This should be 
obvious from the fact that our results depend on a single ``hadronic 
parameter'' $n\cdot k$, corresponding to a fixed residual momentum of the 
heavy quark. Only the dependence on the ultra-violet renormalization 
scale $\mu$ can be trusted. However, the one-loop result in 
(\ref{Sonshell}) is needed to complete the matching calculation of the 
jet function described in the previous subsection, which can legitimately 
be performed with on-shell external $b$-quark states. Given the 
expression for the renormalized shape function, we obtain
\begin{eqnarray}
   \int d\omega J^{(0)}(p_\omega^2)\,S^{(1)}(\omega)
   &=& \frac{1}{\bar n\cdot p}\,S^{(1)}(-p^2/\bar n\cdot p) \\
   &=& - \frac{C_F\alpha_s}{\pi} \left[
    \frac{\pi^2}{24}\,\delta(p_k^2)
    + \frac{2}{\bar n\cdot p} \left( \frac{\bar n\cdot p}{p_k^2}
    \ln\frac{p_k^2}{\bar n\cdot p\,\mu} \right)_{\!*}^{\![\mu]} 
    + \frac{1}{\bar n\cdot p} \left( \frac{\bar n\cdot p}{p_k^2}
    \right)_{\!*}^{\![\mu]} \right] . \nonumber
\end{eqnarray}
With the help of the identities (\ref{ids}) this can be shown to be equal 
to the finite part of (\ref{sloops}), as we claimed above.

\section{Renormalization-group resummation}
\label{sec:RGevol}

Equations (\ref{Hres}) and (\ref{Jres}) determine the short-distance 
objects $H_{ij}$ and $J$ in the factorization formula (\ref{Wres}) at
one-loop order in perturbation theory. However, there is no common choice 
of the renormalization scale $\mu$ that would eliminate all large 
logarithms from these results. Likewise, the shape function, being a 
hadronic matrix element, is naturally renormalized at some low scale, 
whereas the short-distance objects contain physics at higher scales. The 
problem of large logarithms arising from the presence of disparate mass 
scales can be dealt with using renormalization-group equations. 
Proceeding in three steps, our strategy will be as follows: 

i) At a high scale $\mu_h\sim m_b$ we match QCD onto SCET and extract 
matching conditions for the hard functions $H_{ij}$. The corresponding 
one-loop expressions have been given in (\ref{Hres}). At that scale, they 
are free of large logarithms and so can be reliably computed using 
perturbation theory. We then evolve the hard functions down to an 
intermediate hard-collinear scale $\mu_i\sim\sqrt{m_b\Lambda_{\rm QCD}}$ 
by solving the renormalization-group equation
\begin{equation}\label{Hevol}
   \frac{d}{d\ln\mu}\,H_{ij}(\bar n\cdot p,\mu)
   = 2\gamma_J(\bar n\cdot p,\mu)\,H_{ij}(\bar n\cdot p,\mu) \,,
\end{equation}
where $\gamma_J$ is the anomalous dimension of the semileptonic current 
in SCET.

ii) Next, we start from a model for the shape function $S(\omega,\mu_0)$ 
at some low scale $\mu_0=\mbox{few}\times\Lambda_{\rm QCD}$ large enough 
to trust perturbation theory. Such a model could be provided by a 
QCD-inspired approach such as QCD sum rules or lattice QCD, or it could 
be tuned to experimental data. We then solve the integro-differential 
evolution equation
\begin{equation}\label{Sevol}
   \frac{d}{d\ln\mu}\,S(\omega,\mu)
   = - \int d\omega'\,\gamma_S(\omega,\omega',\mu)\,S(\omega',\mu) 
\end{equation}
to obtain the shape function at the intermediate scale $\mu_i$.

iii) Finally, at the scale $\mu_i$ we combine the results for the hard 
functions and for the shape function with the jet function $J$ in 
(\ref{Jres}), which at that scale is free of large logarithms and so has 
a reliable perturbative expansion. The dependence on the matching scales 
$\mu_h$ and $\mu_i$ cancels in the final result (to the order at which we 
are working). 

We now discuss these three steps in detail.

\subsection{Evolution of the hard functions}

At one-loop order, the anomalous dimension $\gamma_J$ for the SCET 
current is twice the coefficient of the $1/\epsilon$ pole in the 
renormalization factor $Z_J$ in (\ref{ZJres}). More generally
\cite{Bauer:2000yr,Bosch:2003fc},
\begin{equation}
   \gamma_J(\bar n\cdot p,\mu)
   = - \Gamma_{\rm cusp}(\alpha_s)\,\ln\frac{\mu}{\bar n\cdot p}
    + \gamma'(\alpha_s)
   = \frac{C_F\alpha_s}{\pi}
    \left( - \ln\frac{\mu}{\bar n\cdot p} - \frac54 \right)
    + \dots \,,
\end{equation}
where $\Gamma_{\rm cusp}=C_F\alpha_s/\pi+\dots$ is the universal cusp 
anomalous dimension governing the ultra-violet singularities of Wilson 
lines with light-like segments \cite{Korchemsky:wg}. The exact solution 
to the evolution equation (\ref{Hevol}) can be written as
\begin{equation}
   H_{ij}(\bar n\cdot p,\mu_i) =  H_{ij}(\bar n\cdot p,\mu_h)\,
   \exp{U_H(\bar n\cdot p,\mu_h,\mu_i)} \,,
\end{equation}
where
\begin{equation}
   U_H(\bar n\cdot p,\mu_h,\mu_i)
   = 2 \int\limits_{\alpha_s(\mu_h)}^{\alpha_s(\mu_i)}\!\!
   \frac{d\alpha}{\beta(\alpha)}\,\Bigg[ \Gamma_{\rm cusp}(\alpha)\,
   \Bigg( \ln\frac{\bar n\cdot p}{\mu_h}
   - \int\limits_{\alpha_s(\mu_h)}^{\alpha}\!
   \frac{d\alpha'}{\beta(\alpha')} \Bigg) + \gamma'(\alpha)
   \Bigg] \,,
\end{equation}
and $\beta(\alpha_s)=d\alpha_s/d\ln\mu$ is the QCD $\beta$-function. 
Defining as usual 
\begin{equation}
   \Gamma_{\rm cusp}(\alpha_s) = \sum_{n=0}^\infty \Gamma_n
    \left( \frac{\alpha_s}{4\pi} \right)^{n+1} , \qquad
   \beta(\alpha_s) = -2\alpha_s \sum_{n=0}^\infty \beta_n
    \left( \frac{\alpha_s}{4\pi} \right)^{n+1} ,
\end{equation}
and similarly for all other anomalous dimensions, setting 
$r_1=\alpha_s(\mu_i)/\alpha_s(\mu_h)>1$, and expanding the evolution
function to $O(\alpha_s)$, we obtain
\begin{equation}
   e^{U_H(\bar n\cdot p,\mu_h,\mu_i)}
   = e^{V_H(\mu_h,\mu_i)} \left( \frac{\bar n\cdot p}{\mu_h}
   \right)^{-\frac{\Gamma_0}{\beta_0} \ln r_1}
   \left[ 1 - \frac{\alpha_s(\mu_h)}{4\pi}\,\frac{\Gamma_0}{\beta_0}
   \left( \frac{\Gamma_1}{\Gamma_0} - \frac{\beta_1}{\beta_0} \right)
   (r_1-1)\,\ln\frac{\bar n\cdot p}{\mu_h} \right] ,
\end{equation}
where
\begin{eqnarray}\label{VHres}
   V_H(\mu_h,\mu_i) 
   &=& \frac{\Gamma_0}{2\beta_0^2} \left[ \frac{4\pi}{\alpha_s(\mu_h)}
    \left( 1 - \frac{1}{r_1} - \ln r_1 \right)
    + \frac{\beta_1}{2\beta_0}\,\ln^2 r_1
    - \left( \frac{\Gamma_1}{\Gamma_0} - \frac{\beta_1}{\beta_0} \right) 
    (r_1-1-\ln r_1) \right] \nonumber\\
   &&\mbox{}- \frac{\gamma_0'}{\beta_0}\,\ln r_1
    + O\Big[(r_1-1)\,\alpha_s(\mu_h)\Big] \,.
\end{eqnarray}
Let us briefly explain the structure of resummed perturbation theory in
applications with Sudakov double logarithms. In renormalization-group 
improved perturbation theory the parameter $r_1$ is treated as a 
quantity of $O(1)$. The terms proportional to $1/\alpha_s(\mu_h)$ in 
$V_H$ resum the leading, double logarithmic terms to all orders in 
perturbation theory. The remaining $O(1)$ terms in $V_H$ contribute at 
leading, single-logarithmic order. Note that these effects are not 
suppressed by any small parameter. We therefore refer to the combination 
of these two terms as the ``leading order''. At next-to-leading order, 
the corrections proportional to the coupling $\alpha_s(\mu_h)$ are 
included. In our case, the only piece missing for a complete resummation 
at next-to-leading order is the $O(\alpha_s)$ contribution to $V_H$, 
which is independent of the kinematic variable $\bar n\cdot p$ and 
vanishes for $\mu_i\to\mu_h$. To compute these terms would require to 
calculate the cusp anomalous dimension to three loops and the anomalous 
dimension $\gamma'$ to two loops. The fact that the corresponding 
expansion coefficients are unknown implies a universal, 
process-independent small uncertainty in the normalization of inclusive 
$B$-decay spectra in the shape-function region. We stress, however, that 
this uncertainty cancels in all ratios of decay distributions, even 
between $\bar B\to X_u\,l^-\bar\nu$ and $\bar B\to X_s\gamma$ spectra.

In phenomenological applications of our results the intermediate scale 
$\mu_i$ will be of order $m_c$, because a restriction to hadronic 
invariant masses below the charm threshold is used to separate 
$\bar B\to X_u\,l^-\bar\nu$ from $\bar B\to X_c\,l^-\bar\nu$ decays. It 
is then appropriate to perform the running between $\mu_h$ and $\mu_i$ in 
a theory with $n_f=4$ light quark flavors. The relevant expansion 
coefficients are $\Gamma_0=\frac{16}{3}$, 
$\Gamma_1=\frac{2576}{27}-\frac{16}{3}\pi^2$, 
$\gamma_0'=-\frac{20}{3}$, and $\beta_0=\frac{25}{3}$, 
$\beta_1=\frac{154}{3}$.

\subsection{Evolution of the shape function}

At one-loop order, the anomalous dimension for the shape function is 
twice the coefficient of the $1/\epsilon$ pole in the renormalization 
factor $Z_S$. From (\ref{zres}), we obtain
\begin{equation}\label{ourgamma}
   \gamma_S(\omega,\omega',\mu)
   = \frac{C_F\alpha_s}{\pi} \left[
   \left( 2\ln\frac{\mu}{\bar\Lambda-\omega} - 1 \right)
   \delta(\omega-\omega')
   - 2 \left( \frac{\theta(\omega'-\omega)}{\omega'-\omega}
   \right)_{\!+} \right] .
\end{equation}
Let us briefly compare this result with previous calculations of this
anomalous dimension published in the literature. In 
\cite{Balzereit:1998yf}, Balzereit et al.\ obtained a result that almost 
agrees with ours, except for the factors of 2 in front of the logarithmic 
term and of the plus distribution. (Their expression is, however, given 
in a rather different form.) Aglietti and Ricciardi tried to extract the 
renormalized shape function by matching QCD directly onto HQET, without 
including a jet function \cite{Aglietti:1999ur}. They obtained results 
for the renormalization factor $Z_S$ (and hence for the anomalous 
dimension) in two different regularization schemes. Their expression 
obtained in dimensional regularization disagrees with our findings. Bauer 
et al.\ computed the anomalous dimension of the shape function using SCET
\cite{Bauer:2000ew}. They interpreted the renormalization factor in 
(\ref{zres}) as a distribution in $\omega$ rather than $\omega'$ and 
studied the renormalization-group equation for the short-distance 
coefficients $H_{ij}\cdot J$ instead of that for the shape function. If 
this is done, the logarithm in the anomalous dimension in (\ref{ourgamma})
must be replaced by $\ln[\mu/(\Lambda_{\rm had}+\omega')]$, where 
$\Lambda_{\rm had}$ is the lower cutoff on the integral over $\omega$ 
(see the discussion in Section~\ref{sec:SFrenorm}). The value of this 
cutoff is process dependent and set by kinematics or by virtue of some 
experimental cut. In the shape-function region, 
$\Lambda_{\rm had}=O(\Lambda_{\rm QCD})$. Instead, in 
\cite{Bauer:2000ew} the combination $(\Lambda_{\rm had}+\omega')$ is 
identified with the $b$-quark mass, which is incorrect.

The evolution equation (\ref{Sevol}) can be solved analytically using a
general method developed in \cite{Lange:2003ff}. It is convenient to 
change variables from $\omega$ to 
$\hat\omega=\bar\Lambda-\omega\in[0,\infty[$ and denote
$\hat S(\hat\omega)\equiv S(\bar\Lambda-\hat\omega)$. The 
renormalization-group equation then reads
\begin{equation}\label{oureqn}
   \frac{d}{d\ln\mu}\,\hat S(\hat\omega,\mu)
   = - \int_0^\infty\!d\hat\omega'\,
   \hat\gamma_S(\hat\omega,\hat\omega',\mu)\,
   \hat S(\hat\omega',\mu) \,,
\end{equation}
where the anomalous dimension can be written in the general form
\begin{equation}\label{hatGs}
   \hat\gamma_S(\hat\omega,\hat\omega',\mu)
   = 2 \left[ \,\Gamma_{\rm cusp}(\alpha_s)\,\ln\frac{\mu}{\hat\omega}
   + \gamma(\alpha_s) \right] \delta(\hat\omega-\hat\omega')
   + 2 {\cal G}(\hat\omega,\hat\omega',\alpha_s) \,.
\end{equation}
The logarithmic term containing the cusp anomalous dimension has a 
geometric origin. Since the heavy-quark field $h(x)$ in HQET can be 
represented as a Wilson line along the $v$ direction, the field $\H(x)$ 
entering the SCET formalism contains the product of a light-like Wilson 
line (along $n$) and a time-like Wilson line (along $v$), which form a 
cusp at point $x$. The shape function contains two such cusps. According 
to the renormalization theory of Wilson lines with light-like segments, 
each cusp produces a contribution to the anomalous dimension proportional 
to $\Gamma_{\rm cusp}\,\ln\mu$ \cite{Korchemsky:wg}. The one-loop 
coefficients of the remaining terms in (\ref{hatGs}) are
\begin{equation}\label{G1l}
   \gamma_0 = - 2C_F \,, \qquad
   {\cal G}_0(\hat\omega,\hat\omega') = - \Gamma_0 \left(
    \frac{\theta(\hat\omega-\hat\omega')}
         {\hat\omega-\hat\omega'} \right)_{\!+} .
\end{equation}
The general solution of (\ref{oureqn}) can be obtained using the fact 
that on dimensional grounds
\begin{equation}\label{Fdef}
   \int_0^\infty\!d\hat\omega'\,
   {\cal G}(\hat\omega,\hat\omega',\alpha_s)\,(\hat\omega')^a
   \equiv \hat\omega^a\,\F(a,\alpha_s) \,,
\end{equation}
where the function $\F$ only depends on the exponent $a$ and the coupling 
constant. We set $\F(0,\alpha_s)=0$ by definition, thereby determining 
the split between the terms with $\gamma$ and ${\cal G}$ in 
(\ref{hatGs}). The integral on the left-hand side is convergent as long 
as $\mbox{Re}\,a>-1$. At one-loop order we find from (\ref{G1l})
\begin{equation}\label{F1loop}
   \F(a,\alpha_s) = \Gamma_0\,\frac{\alpha_s}{4\pi}\,
   \Big[ \psi(1+a) + \gamma_E \Big] + \dots \,,
\end{equation}
where $\psi(z)$ is the logarithmic derivative of the Euler $\Gamma$ 
function. Relation (\ref{Fdef}) implies that the ansatz 
\cite{Lange:2003ff}
\begin{equation}
   f(\hat\omega,\mu,\mu_0,\tau)
   = \left( \frac{\hat\omega}{\mu_0} \right)^{\tau+2g(\mu,\mu_0)}
   \exp U_S(\tau,\mu,\mu_0)
\end{equation}
with
\begin{equation}\label{myeqs}
\begin{aligned}
   g(\mu,\mu_0) &= \int\limits_{\alpha_s(\mu_0)}^{\alpha_s(\mu)}\!
    d\alpha\,\frac{\Gamma_{\rm cusp}(\alpha)}{\beta(\alpha)} \,, \\
   U_S(\tau,\mu,\mu_0)
   &= - 2\!\int\limits_{\alpha_s(\mu_0)}^{\alpha_s(\mu)}\!
    \frac{d\alpha}{\beta(\alpha)}\,\Big[
    g(\mu,\mu_\alpha) + \gamma(\alpha) + 
    \F\big(\tau + 2g(\mu_\alpha,\mu_0),\alpha \big) \Big] \,,
\end{aligned}
\end{equation}
provides a solution to the evolution equation (\ref{oureqn}) with initial 
condition $f(\hat\omega,\mu_0,\mu_0,\tau)=(\hat\omega/\mu_0)^\tau$ at 
some scale $\mu_0$. Here $\mu_\alpha$ is defined such that 
$\alpha_s(\mu_\alpha)=\alpha$, and $\tau$ can be an arbitrary complex 
parameter. Note that $g(\mu,\mu_0)>0$ if $\mu>\mu_0$. We now assume that 
the shape function $\hat S(\hat\omega,\mu_0)$ is given at the low scale 
$\mu_0$ and define its Fourier transform with respect to 
$\ln(\hat\omega/\mu_0)$ through
\begin{equation}\label{S0def}
   \hat S(\hat\omega,\mu_0)
   = \frac{1}{2\pi} \int_{-\infty}^\infty\!dt\,{\cal S}_0(t)
   \left( \frac{\hat\omega}{\mu_0} \right)^{it} \,.
\end{equation}
The exact result for the shape function at a different scale $\mu$ is 
then given by
\begin{equation}\label{Sexact}
   \hat S(\hat\omega,\mu)
   = \frac{1}{2\pi} \int_{-\infty}^\infty\!dt\,{\cal S}_0(t)\,
   f(\hat\omega,\mu,\mu_0,it) \,. 
\end{equation}

With the help of this formula, it is straightforward to derive explicit 
expressions for the evolution of the shape function from the hadronic 
scale $\mu_0$ up to the intermediate scale $\mu_i$ at any order in 
renormalization-group improved perturbation theory. Setting 
$r_2=\alpha_s(\mu_0)/\alpha_s(\mu_i)>1$, we obtain for the evolution 
function at leading order
\begin{equation}\label{NLOapprox}
   f(\hat\omega,\mu_i,\mu_0,it)
   = e^{V_S(\mu_i,\mu_0)} \left( \frac{\hat\omega}{\mu_0}
   \right)^{it+\frac{\Gamma_0}{\beta_0}\ln r_2}
   \frac{\Gamma(1+it)}{\Gamma(1+it+\frac{\Gamma_0}{\beta_0}\,\ln r_2)} \,,
\end{equation}
where
\begin{eqnarray}\label{VS}
   V_S(\mu_i,\mu_0) 
   &=& \frac{\Gamma_0}{2\beta_0^2} \left[
    - \frac{4\pi}{\alpha_s(\mu_0)}\,(r_2 -1 - \ln r_2)
    + \frac{\beta_1}{2\beta_0}\,\ln^2 r_2
    + \left( \frac{\Gamma_1}{\Gamma_0} - \frac{\beta_1}{\beta_0} \right) 
    \left( 1 - \frac{1}{r_2} - \ln r_2 \right) \right] \nonumber\\
   &&\mbox{}- \frac{\Gamma_0}{\beta_0}\,\gamma_E\,\ln r_2
    - \frac{\gamma_0}{\beta_0}\,\ln r_2
    + O\Big[(r_2-1)\,\alpha_s(\mu)\Big] \,.
\end{eqnarray}
This result is valid as long as $(\Gamma_0/\beta_0)\,\ln r_2<1$, which is 
the case for all reasonable parameter values. Missing for a resummation 
at next-to-leading order are the $O(\alpha_s)$ contributions to $V_S$, 
which vanish for $\mu\to\mu_0$. Since these corrections have an unknown 
dependence on $t$ via the two-loop contribution to the function 
$\F(\alpha_s,a)$, they will affect the $\hat\omega$ dependence of the 
final result. There are also some known $O(\alpha_s)$ corrections to 
(\ref{NLOapprox}) proportional to $\ln(\hat\omega/\mu_0)$, which we have 
omitted for consistency. For all practical purposes, given the intrinsic 
uncertainties in our knowledge of the shape function, it will be 
sufficient to use the equations given above. As mentioned earlier, we 
typically have $\mu_i\sim m_c$, and so the running between $\mu_i$ and 
$\mu_0$ should be performed in a theory with $n_f=3$ light quark flavors. 
The relevant expansion coefficients are then $\Gamma_0=\frac{16}{3}$, 
$\Gamma_1=\frac{304}{3}-\frac{16}{3}\pi^2$, $\gamma_0=-\frac{8}{3}$, and 
$\beta_0=9$, $\beta_1=64$.

The leading-order result presented above can be simplified further. When 
(\ref{NLOapprox}) is inserted into (\ref{Sexact}), the integration over 
$t$ can be performed analytically. Setting 
$\eta=(\Gamma_0/\beta_0)\,\ln r_2>0$, the relevant integral is
\begin{equation}
   I = \frac{1}{2\pi} \int_{-\infty}^\infty\!dt\,{\cal S}_0(t)
    \left( \frac{\hat\omega}{\mu_0} \right)^{it}
    \frac{\Gamma(1+it)}{\Gamma(1+it+\eta)} \,,
\end{equation}
where
\begin{equation}
   {\cal S}_0(t) = \int_0^\infty\!\frac{d\hat\omega'}{\hat\omega'}\,
    \hat S(\hat\omega',\mu_0)
    \left( \frac{\hat\omega'}{\mu_0} \right)^{-it}
\end{equation}
is the Fourier transform of the shape function as defined in 
(\ref{S0def}). The integrand of the $t$-integral has poles on the 
positive imaginary axis located at $t=in$ with $n\ge 1$ an integer. For 
$\hat\omega<\hat\omega'$ the integration contour can be closed in the 
lower half-plane avoiding all poles, hence yielding zero. For
$\hat\omega>\hat\omega'$ we use the theorem of residues to obtain
\begin{equation}
   I = \int_0^{\hat\omega}\!d\hat\omega'\,R(\hat\omega,\hat\omega')\,
   \hat S(\hat\omega',\mu_0) \,,
\end{equation}
where
\begin{equation}
   R(\hat\omega,\hat\omega')
   = \frac{1}{\hat\omega} \sum_{j=0}^\infty
    \left( - \frac{\hat\omega'}{\hat\omega} \right)^j
    \frac{1}{\Gamma(j+1)\,\Gamma(\eta-j)}
   = \frac{1}{\Gamma(\eta)}\,
    \frac{1}{\hat\omega^{\eta}\,(\hat\omega-\hat\omega')^{1-\eta}} \,.
\end{equation}
Note that $R(\hat\omega,\hat\omega')\to\delta(\hat\omega-\hat\omega')$ in 
the limit $\eta\to 0$, corresponding to $\mu_i\to\mu_0$, as it should be. 
Our final result for the shape function at the intermediate 
hard-collinear scale, valid at leading order in renormalization-group 
improved perturbation theory, can now be written in the simple form 
(valid for $\mu_i>\mu_0$, so that $\eta>0$)
\begin{equation}\label{wow}
   \hat S(\hat\omega,\mu_i) = e^{V_S(\mu_i,\mu_0)}\,
   \frac{1}{\Gamma(\eta)} \int_0^{\hat\omega}\!d\hat\omega'\,
   \frac{\hat S(\hat\omega',\mu_0)}
        {\mu_0^{\eta}\,(\hat\omega-\hat\omega')^{1-\eta}} \,,
\end{equation}
with $V_S$ as given in (\ref{VS}). A similar analytic result for the 
renormalized shape function was obtained in \cite{Balzereit:1998yf} using 
a different strategy to solve the evolution equation for the shape 
function. However, these authors miss a factor 2 in the expression for 
$\eta$, and we disagree with their expression for the function $V_S$.

From the above equation one can derive scaling relations for the 
asymptotic behavior of the shape function for $\hat\omega\to 0$ and
$\hat\omega\to\infty$ (corresponding to $\omega\to\bar\Lambda$ and 
$\omega\to-\infty$). If the function $\hat S(\hat\omega,\mu_0)$ at the 
low scale $\mu_0$ vanishes proportional to $\hat\omega^\zeta$ near the 
endpoint, the shape function at a higher scale $\mu_i>\mu$ vanishes 
faster, proportional to $\hat\omega^{\zeta+\eta}$. Similarly, if 
$\hat S(\hat\omega,\mu_0)$ falls off like $\hat\omega^{-\xi}$ for 
$\hat\omega\to\infty$, the shape function renormalized at a higher scale 
vanishes like $\hat\omega^{-\min(1,\xi)+\eta}$. Irrespective of the 
initial behavior of the shape function, evolution effects generate a 
radiative tail that falls off slower than $1/\hat\omega$. This fact 
implies that the normalization integral of $\hat S(\hat\omega,\mu)$ as 
well as all positive moments are ultra-violet divergent. The 
field-theoretic reason is that the bilocal shape-function operator in 
(\ref{SFops}) contains ultra-violet singularities as $z_-\to 0$, which 
are not subtracted in the renormalization of the shape function. The 
situation is analogous to the case of the $B$-meson light-cone 
distribution amplitude discussed in \cite{Lange:2003ff,Grozin:1996pq}. 
These divergences are never an obstacle in practice. Convolution 
integrals with the shape function are always cut off at some finite value 
of $\hat\omega$ by virtue of phase-space or some experimental cut.

\section{Properties of the shape function}
\label{sec:properties}

In this section we discuss how moments of the shape function are related
with HQET parameters. This will lead us to propose a new, physical 
scheme for defining a running heavy-quark mass, which is most appropriate 
for the study of inclusive spectra in the shape-function region. We will
also present a model-independent result for the asymptotic behavior of 
the renormalized shape function (defined in the $\overline{\rm MS}$ 
scheme), finding that it is {\em not\/} positive definite. 

Most of our discussion in this section is phrased in terms of the 
original (unhatted) shape function $S(\omega,\mu)$. At the end, we 
formulate the resulting constraints on the function 
$\hat S(\hat\omega,\mu)$.

\subsection{Shape-function moments in the pole scheme}

Naively, ignoring renormalization effects, the moments 
$M_N=\int_{-\infty}^{\bar\Lambda} d\omega\,\omega^N S(\omega)$ are given 
by hadronic parameters defined in terms of $B$-meson matrix elements of 
local HQET operators \cite{Neubert:1993ch}. In particular, $M_0=1$ fixes 
the normalization of the shape function, $M_1=0$ vanishes by the HQET 
equation of motion, and $M_2=-\lambda_1/3$ is determined by the matrix 
element of the kinetic-energy operator. The vanishing of the first moment 
is connected with the implicit definition of the heavy-quark pole mass 
built into the HQET Lagrangian via the equation of motion 
$iv\cdot D\,h=0$. These moment constraints have been implemented in 
various model parameterizations for the shape function suggested in the 
literature \cite{Mannel:1994pm,Kagan:1998ym,Bigi:2002qq}. Typically, one 
makes an ansatz for the shape function depending on a few HQET 
parameters such as $\bar\Lambda$ and $\lambda_1$, and determines the 
values of these parameters from a fit to experimental data.

Beyond tree level, all moments $M_N$ with $N\ge 0$ receive ultra-violet 
divergences from the region $\omega\to-\infty$ (or 
$\hat\omega\to\infty$). However, as we have mentioned above, the values 
of $\omega$ needed for the description of physical decay rates are always 
restricted to a finite interval. It is thus sufficient for all purposes 
to define the moments of the renormalized shape function as
\begin{equation}\label{MNdef}
   M_N(\Lambda_{\rm UV},\mu) = \int_{-\Lambda_{\rm UV}}^{\bar\Lambda}\!
   d\omega\,\omega^N S(\omega,\mu) \,.
\end{equation}
The dependence of these moments on the renormalization scale $\mu$ is
controlled by the evolution equation (\ref{Sevol}). In addition, the 
moments depend on the lower cutoff on the $\omega$ integral. The choice 
of $\Lambda_{\rm UV}$ is a matter of convenience, and so we are free to 
pick a value that is numerically (if not parametrically) large compared 
with $\Lambda_{\rm QCD}$. In this case, as we will now show, the 
dependence on $\Lambda_{\rm UV}$ can also be controlled using 
short-distance methods. 

For sufficiently large values of $\Lambda_{\rm UV}$ it is possible to 
expand the moments $M_N(\Lambda_{\rm UV},\mu)$ in a series of $B$-meson 
matrix elements of local HQET operators. If for simplicity we set 
$\Gamma=1$ in the shape-function operator (which is legitimate, since the 
Dirac structure is unaltered in HQET), the operators in question are 
Lorentz-scalar, ``leading-twist'' operators containing $\bar h\dots h$ 
\cite{Neubert:1993ch,Bigi:1993ex}. These are the operators that mix with 
$\bar h\,(in\cdot D)^N h$ under renormalization. It is straightforward to 
find the corresponding operators of a given dimension. The unique 
dimension-3 operator is $\bar h h$. The two operators of dimension~4 are 
$\bar h\,in\cdot D\,h$ (class-1) and $\bar h\,iv\cdot D\,h$ (class-2). 
The class-2 operator vanishes by the HQET equation of motion. The 
possible dimension-5 operators are
\begin{equation}
\begin{aligned}
   \mbox{class-1:} \qquad
   &\bar h\,(in\cdot D)^2 h \,, \quad
    \bar h\,(iD_\perp)^2 h \,, \\
   \mbox{class-2:} \qquad
   &\bar h\,(iv\cdot D)^2 h \,, \quad
    \bar h\,in\cdot D\,iv\cdot D\,h \,, \quad
    \bar h\,iv\cdot D\,in\cdot D\,h \,,
\end{aligned}
\end{equation}
where again the class-2 operators vanish by the equation of motion. 
Moreover, it follows from the Feynman rules of HQET that the two class-1 
operators do not mix under renormalization, so the operator 
$\bar h\,(iD_\perp)^2 h$ can be ignored. From dimension~6 on the 
situation is more complicated, because several class-1 operators exist 
that can mix with $\bar h\,(in\cdot D)^N h$. For $N=3$ these are of the 
form $\bar h\,iD\,G\,h$ or $\sum_q \bar h\dots q\,\bar q\dots h$, where 
we omit Lorentz and color indices. We will restrict our discussion to 
operators of dimension less than 6.

For the operator product expansion of the moments in (\ref{MNdef}) we 
need the forward matrix elements 
\begin{equation}
   \langle O\rangle
   = \frac{\langle\bar B(v)|\,O\,|\bar B(v)\rangle}{2M_B}
\end{equation}
of the leading-twist operators between $B$-meson states in HQET. Using 
the equation of motion, it can be shown that $\langle\bar hh\rangle=1$, 
$\langle\bar h\,in\cdot D\,h\rangle=0$, and 
$\langle\bar h\,(in\cdot D)^2 h\rangle=-\lambda_1/3$ 
\cite{Neubert:1993ch}. We can thus write an expansion of the form
\begin{equation}\label{beauty}
   M_N(\Lambda_{\rm UV},\mu) = \Lambda_{\rm UV}^N
   \left\{ K_0^{(N)}(\Lambda_{\rm UV},\mu)
   + K_2^{(N)}(\Lambda_{\rm UV},\mu)\,\cdot
    \frac{(-\lambda_1)}{3\Lambda_{\rm UV}^2}
   + O\bigg[ \left( \frac{\Lambda_{\rm QCD}}{\Lambda_{\rm UV}} \right)^3
   \bigg] \right\} .
\end{equation}
This expansion is useful as long as the cutoff $\Lambda_{\rm UV}$ is 
chosen much larger than the typical hadronic scale characterizing the 
matrix elements of the local operators. The matching coefficients 
$K_n^{(N)}$ in this relation can be calculated using on-shell external 
$b$-quark states with residual momentum $k$. For operators of dimension 
up to 5 it suffices to calculate two-point functions. (Three and 
four-point functions would have to be considered at dimension~6.) We 
first evaluate the moments of the renormalized shape function in 
(\ref{Sonshell}), finding at one-loop order
\begin{eqnarray}\label{mom1}
   M_N^{\rm parton}(\Lambda_{\rm UV},\mu)
   &=& (n\cdot k)^N\,\Bigg\{ 1 - \frac{C_F\alpha_s}{\pi} \left( 
    \ln^2\frac{\Lambda_{\rm UV}+n\cdot k}{\mu}
    + \ln\frac{\Lambda_{\rm UV}+n\cdot k}{\mu} + \frac{\pi^2}{24}
    \right) \nonumber\\
   &&\mbox{}- \frac{C_F\alpha_s}{\pi}
    \sum_{j=1}^N \frac{1}{j} \left( 1
    + 2\ln\frac{\Lambda_{\rm UV}+n\cdot k}{\mu}
    - \sum_{l=j}^N \frac{2}{l} \right)
    \left[ \left( - \frac{\Lambda_{\rm UV}}{n\cdot k} \right)^j - 1
    \right] \Bigg\} \,. \qquad
\end{eqnarray}
We then expand this result in powers of $n\cdot k/\Lambda_{\rm UV}$. 
Keeping the first three terms in the expansion, we obtain
\begin{equation}\label{mom2}
\begin{aligned}
   M_0^{\rm parton}(\Lambda_{\rm UV},\mu) &= 1 - \frac{C_F\alpha_s}{\pi}
    \left( \ln^2\frac{\Lambda_{\rm UV}}{\mu}
    + \ln\frac{\Lambda_{\rm UV}}{\mu} + \frac{\pi^2}{24} \right) \\
   &\quad\mbox{}- \frac{C_F\alpha_s}{\pi} \left[
    \frac{n\cdot k}{\Lambda_{\rm UV}}
    \left( 2\ln\frac{\Lambda_{\rm UV}}{\mu} + 1 \right)
    + \frac{(n\cdot k)^2}{\Lambda_{\rm UV}^2}
    \left( - \ln\frac{\Lambda_{\rm UV}}{\mu} + \frac12 \right) + \dots
    \right] , \\
   M_1^{\rm parton}(\Lambda_{\rm UV},\mu) &= n\cdot k \left[ 1
    - \frac{C_F\alpha_s}{\pi}
    \left( \ln^2\frac{\Lambda_{\rm UV}}{\mu}
    - \ln\frac{\Lambda_{\rm UV}}{\mu} + \frac{\pi^2}{24} - 1 \right) 
    \right] \\
   &\quad\mbox{}- \frac{C_F\alpha_s}{\pi} \left[
    \Lambda_{\rm UV}
    \left( - 2\ln\frac{\Lambda_{\rm UV}}{\mu} + 1 \right)
    + \frac{(n\cdot k)^2}{\Lambda_{\rm UV}}\,
    2\ln\frac{\Lambda_{\rm UV}}{\mu} + \dots \right] , \\
   M_2^{\rm parton}(\Lambda_{\rm UV},\mu) &= (n\cdot k)^2 \left[ 1
    - \frac{C_F\alpha_s}{\pi}
    \left( \ln^2\frac{\Lambda_{\rm UV}}{\mu}
    - 2\ln\frac{\Lambda_{\rm UV}}{\mu} + \frac{\pi^2}{24} - \frac12
    \right) \right] \\
   &\quad\mbox{}- \frac{C_F\alpha_s}{\pi} \left[
    \Lambda_{\rm UV}^2\,\ln\frac{\Lambda_{\rm UV}}{\mu}
    + n\cdot k\,\Lambda_{\rm UV}
    \left( - 2\ln\frac{\Lambda_{\rm UV}}{\mu} + 3\right) + \dots
    \right] .
\end{aligned}
\end{equation}
In the next step, we calculate the one-loop matrix elements of the local
operators $\bar h\,(in\cdot D)^N h$ between heavy-quark states with 
residual momentum $k$. The relevant diagrams are the same as in 
Figure~\ref{fig:hqet}, where now the black square represents the local 
operators. Keeping $v\cdot k$ non-zero to regularize infra-red 
singularities, we obtain for the bare matrix elements
\begin{eqnarray}
   \langle\bar h\,(in\cdot D)^N h\rangle
   &=& (n\cdot k)^N \Bigg\{ 1 - \frac{4C_F g_s^2}{(4\pi)^{2-\epsilon}}\,
    (-2v\cdot k)^{-2\epsilon} \sum_{j=1}^N
    \bigg( \begin{array}{c} N \\ j \end{array} \bigg)
    \left( \frac{2v\cdot k}{n\cdot k} \right)^j \nonumber\\
   &&\hspace{2.5cm} \times 
    (j-1-\epsilon)\,\Gamma(j-\epsilon)\,\Gamma(2\epsilon-j) \Bigg\} \,.
\end{eqnarray}
While the individual diagrams are infra-red divergent, taking the limit 
$v\cdot k\to 0$ in the sum of all contributions is possible without
encountering singularities. Then the one-loop contributions vanish, and 
the matrix elements simply reduce to their tree-level values. In other
words, the one-loop contributions correspond to a mixing with class-2
operators, whose hadronic matrix elements vanish by the equations of 
motions. It follows that in (\ref{mom2}) we must identify 
$(n\cdot k)^n\to\langle\bar h\,(in\cdot D)^n h\rangle$. Substituting the 
results for the HQET matrix elements given earlier, we obtain for the 
Wilson coefficients of the first three moments
\begin{equation}\label{Dresults}
\begin{aligned}
   K_0^{(0)} &= 1 - \frac{C_F\alpha_s}{\pi}
    \left( \ln^2\frac{\Lambda_{\rm UV}}{\mu}
    + \ln\frac{\Lambda_{\rm UV}}{\mu} + \frac{\pi^2}{24} \right) ,
    \qquad
   K_2^{(0)} = \frac{C_F\alpha_s}{\pi} 
    \left( \ln\frac{\Lambda_{\rm UV}}{\mu} - \frac12 \right) , \\
   K_0^{(1)} &= \frac{C_F\alpha_s}{\pi} 
    \left( 2\ln\frac{\Lambda_{\rm UV}}{\mu} - 1 \right) , \hspace{3.5cm}
   K_2^{(1)} = -2\,\frac{C_F\alpha_s}{\pi}\, 
    \ln\frac{\Lambda_{\rm UV}}{\mu} \,, \\
   K_0^{(2)} &= - \frac{C_F\alpha_s}{\pi}\,
    \ln\frac{\Lambda_{\rm UV}}{\mu} \,, \hspace{1.9cm}
   K_2^{(2)} = 1 - \frac{C_F\alpha_s}{\pi}
    \left( \ln^2\frac{\Lambda_{\rm UV}}{\mu}
    - 2\ln\frac{\Lambda_{\rm UV}}{\mu} + \frac{\pi^2}{24} - \frac12 
    \right) .
\end{aligned}
\end{equation}
At tree level, this reproduces the naive moment relations mentioned at 
the beginning of this section. Beyond tree level, the moments get 
corrected by calculable short-distance effects, which can be controlled 
using fixed-order perturbation theory as long as the ratio 
$\Lambda_{\rm UV}/\mu$ is of $O(1)$. In particular, the renormalized 
first moment no longer vanishes, but is proportional to the cutoff 
$\Lambda_{\rm UV}$ up to small power corrections. 

As mentioned earlier, the value of the first moment is connected with the 
definition of the heavy-quark mass (see also 
\cite{Mannel:1999gs,Mannel:2000aj}). The first moment of the renormalized 
shape function can be made to vanish to all orders in perturbation theory 
by choosing an appropriate scheme for the definition of $m_b$. So far our 
calculations have assumed the definition of the heavy-quark mass as a 
pole mass, $m_b^{\rm pole}$, which is implied by the HQET equation of 
motion $iv\cdot D\,h=0$. Results such as (\ref{Dresults}) are valid in 
this particular scheme. A more general choice is to allow for a residual 
mass term $\delta m$ in HQET, such that $iv\cdot D\,h=\delta m\,h$ with 
$\delta m=O(\Lambda_{\rm QCD})$ \cite{Falk:1992fm}. It is well known that 
the pole mass is an ill-defined concept, which suffers from infra-red 
renormalon ambiguities \cite{Bigi:1994em,Beneke:1994sw}. The parameter 
$\bar\Lambda_{\rm pole}=M_B-m_b^{\rm pole}$, which determines the support 
of the shape function in the pole-mass scheme, inherits the same 
ambiguities. It is therefore advantageous to eliminate the pole mass in 
favor of some short-distance mass. For the analysis of inclusive 
$B$-meson decays, a proper choice is to use a so-called low-scale 
subtracted heavy-quark mass $m_b(\mu_f)$ \cite{Bigi:1996si}, which is 
obtained from the pole mass by removing a long-distance contribution 
proportional to a subtraction scale 
$\mu_f=\mbox{few}\times\Lambda_{\rm QCD}$,
\begin{equation}\label{mdef}
   m_b^{\rm pole} = m_b(\mu_f)
   + \mu_f\,\,g\Big( \alpha_s(\mu), \frac{\mu_f}{\mu} \Big)
   \equiv m_b(\mu_f) + \delta m \,.
\end{equation}
As long as $m_b(\mu_f)$ is defined in a physical way, the resulting
perturbative expressions after elimination of the pole mass are 
well-behaved and not plagued by renormalon ambiguities. Replacing the 
pole mass by the physical mass shifts the values of $n\cdot k$ and 
$\omega$ by an amount $\delta m$, since 
$n\cdot(m_b^{\rm pole} v+k)=m_b(\mu_f)+(n\cdot k+\delta m)$, and because 
the covariant derivative in the definition of the shape function in 
(\ref{Sdef}) must be replaced by $in\cdot D-\delta m$ \cite{Falk:1992fm}. 
At the same time, $\bar\Lambda_{\rm pole}=\bar\Lambda(\mu_f)-\delta m$,
where $\bar\Lambda(\mu_f)=M_B-m_b(\mu_f)$ is a physical parameter. Note 
that this leaves the parameter $\hat\omega=\bar\Lambda-\omega$ and hence 
the shape function $\hat S(\hat\omega,\mu)$ invariant. This follows 
since $\bar\Lambda_{\rm pole}-\omega_{\rm pole}=\bar\Lambda(\mu_f)%
-(\omega_{\rm pole}+\delta m)$, where $\omega_{\rm pole}$ denotes the 
value in the pole-mass scheme used so far.

\subsection{Shape-function moments in a physical scheme}

From now on we will adopt a mass scheme defined by some specific choice 
of $\delta m$. Let us denote by $\omega=\omega_{\rm pole}+\delta m$ the 
value of the light-cone momentum variable in that scheme and define 
``physical'' moments $M_N^{\rm phys}$ as in (\ref{MNdef}), but with all 
parameters replaced by their values in the new scheme, in particular 
$\bar\Lambda=\bar\Lambda(\mu_f)$. Then the expressions for the moments 
in (\ref{mom1}) and (\ref{mom2}) change according to the replacements 
$n\cdot k\to n\cdot k+\delta m$ and 
$\Lambda_{\rm UV}\to\Lambda_{\rm UV}-\delta m$ everywhere. We now 
{\em choose\/} $\delta m$ such that the first moment vanishes, thereby 
defining a low-scale subtracted heavy-quark mass (with 
$\mu_f=\Lambda_{\rm UV}$) to all orders in perturbation theory. We will 
refer to this mass as the ``shape-function mass'' $m_b^{\rm SF}$. This is 
a ``physical'', short-distance mass in the sense that it is free of 
renormalon ambiguities. (However, the definition of the shape-function 
mass depends on the renormalization scheme used to define the shape 
function.) From (\ref{mom2}) and (\ref{mdef}), it follows that at 
one-loop order
\begin{equation}\label{mSFmpole}
   m_b^{\rm pole} = m_b^{\rm SF}(\mu_f,\mu) 
   + \mu_f\,\frac{C_F\alpha_s(\mu)}{\pi} \left[
   \left( 1 - 2\ln\frac{\mu_f}{\mu} \right) 
   + \frac23\,\frac{(-\lambda_1)}{\mu_f^2}\,\ln\frac{\mu_f}{\mu}
   + \dots \right] .
\end{equation}
Note that after introduction of the shape-function mass the coefficients
$K_n^{(1)}$ in the operator-product expansion for the moments in 
(\ref{beauty}) vanish by definition. However, to first order in 
$\alpha_s$ the values for the coefficients $K_n^{(0)}$ and $K_n^{(2)}$ of 
the zeroth and second moments given in (\ref{Dresults}) remain unchanged, 
since $\delta m=O(\alpha_s)$. This would no longer be true for the 
coefficients of higher moments.

The shape-function mass can be related to any other short-distance mass 
using perturbation theory. For instance, at one-loop order its relations 
to the potential-subtracted mass introduced in \cite{Beneke:1998rk} and 
to the kinetic mass defined in \cite{Bigi:1997fj,Benson:2003kp} read
\begin{equation}\label{mSFmPS}
   m_b^{\rm SF}(\mu_f,\mu_f) = m_b^{\rm PS}(\mu_f)
   = m_b^{\rm kin}(\mu_f) + \mu_f\,\frac{C_F\alpha_s(\mu_f)}{3\pi} \,.
\end{equation}
Note that, in addition to the dependence on the subtraction scale 
$\mu_f$, the shape-function mass depends on the scale $\mu$ at which the 
shape function is renormalized. While it is natural to set $\mu=\mu_f$, 
as we did here, this is not necessary. Given a value for the 
shape-function mass for some choice of scales, we can solve 
(\ref{mSFmpole}) to obtain its value for any other choice, using the fact 
that the pole mass is scale independent.

Proceeding in an analogous way, we can use the second moment to define a
physical kinetic-energy parameter, commonly called $\mu_\pi^2$. This 
quantity can be used to replace the HQET parameter $\lambda_1$, which 
like the pole mass suffers from infra-red renormalon ambiguities 
\cite{Martinelli:1995zw}. At one-loop order, we obtain
\begin{eqnarray}\label{mupi2pole}
   \frac{\mu_\pi^2(\Lambda_{\rm UV},\mu)}{3}
   &\equiv& \frac{M_2^{\rm phys}(\Lambda_{\rm UV},\mu)}
                 {M_0^{\rm phys}(\Lambda_{\rm UV},\mu)} \\
   &=& - \frac{C_F\alpha_s(\mu)}{\pi}\,\Lambda_{\rm UV}^2\,
    \ln\frac{\Lambda_{\rm UV}}{\mu} 
    + \frac{(-\lambda_1)}{3} \left[ 1 + \frac{C_F\alpha_s(\mu)}{\pi}
    \left( 3\ln\frac{\Lambda_{\rm UV}}{\mu} + \frac12 \right) \right]
    + \dots \,. \nonumber
\end{eqnarray}
Taking the ratio of $M_2^{\rm phys}$ and $M_0^{\rm phys}$ has the 
advantage of eliminating the double logarithmic radiative corrections 
from this expression. Our definition is similar to the running parameter 
$\mu_\pi^2$ defined in the kinetic scheme 
\cite{Bigi:1997fj,Benson:2003kp}. At one-loop order, the two parameters 
are related by
\begin{equation}\label{mupi2kin}
   \mu_{\pi}^2(\mu_f,\mu_f) = - \mu_f^2\,\frac{C_F\alpha_s(\mu_f)}{\pi}
   + [\mu_{\pi}^2(\mu_f)]_{\rm kin}
   \left[ 1 + \frac{C_F\alpha_s(\mu_f)}{2\pi} \right] .
\end{equation}
Given a value for the kinetic energy in the shape-function scheme for 
some choice of scales, we can solve (\ref{mupi2pole}) to obtain its value 
for any other choice, using that $\lambda_1$ is scale independent.

Similarly, each new moment of the renormalized shape function can be used 
to define a new physical, scale-dependent parameter 
$A_N(\Lambda_{\rm UV},\mu)\equiv M_N^{\rm phys}(\Lambda_{\rm UV},\mu)/%
M_0^{\rm phys}(\Lambda_{\rm UV},\mu)$, which coincides with the 
corresponding HQET parameter $A_N=\langle\bar h\,(in\cdot D)^N h\rangle$ 
at tree level, and which beyond tree level is related to HQET parameters 
through well-controlled perturbative expressions. Obviously, the presence 
of power divergences implies that higher moments are progressively less 
sensitive to HQET parameters, since they are dominated by the 
perturbative terms of order $\alpha_s\Lambda_{\rm UV}^N$. 

\subsection{Moments of the scheme-independent function 
\boldmath$\hat S(\hat\omega,\mu)$\unboldmath}

It will be useful to rewrite the moment relations derived above in terms 
of the variable $\hat\omega=\bar\Lambda-\omega$, which is invariant under
redefinitions of the heavy-quark mass. Defining a new set of 
scheme-independent moments
\begin{equation}\label{hatMNdef}
   \hat M_N(\mu_f,\mu) = \int\limits_0^{\mu_f+\bar\Lambda(\mu_f,\mu)}\!\!
   d\hat\omega\,\hat\omega^N\,\hat S(\hat\omega,\mu) \,,
\end{equation}
we obtain
\begin{eqnarray}\label{M0toM2}
   \hat M_0(\mu_f,\mu)
   &=& 1 - \frac{C_F\alpha_s(\mu)}{\pi}
    \left( \ln^2\frac{\mu_f}{\mu} + \ln\frac{\mu_f}{\mu} 
    + \frac{\pi^2}{24} \right) + \frac{C_F\alpha_s(\mu)}{\pi} 
    \left( \ln\frac{\mu_f}{\mu} - \frac12 \right) 
    \frac{\mu_\pi^2(\mu_f,\mu)}{3\mu_f^2} + \dots , \nonumber\\
   \frac{\hat M_1(\mu_f,\mu)}{\hat M_0(\mu_f,\mu)}
   &=& \bar\Lambda(\mu_f,\mu) \,, \qquad\qquad
   \frac{\hat M_2(\mu_f,\mu)}{\hat M_0(\mu_f,\mu)}
    = \frac{\mu_\pi^2(\mu_f,\mu)}{3} + \bar\Lambda(\mu_f,\mu)^2 \,,
\end{eqnarray}
where the parameters $\bar\Lambda(\mu_f,\mu)=M_B-m_b^{\rm SF}(\mu_f,\mu)$ 
and $\mu_\pi^2(\mu_f,\mu)$ should be considered as known physical 
quantities. Using the relations in the previous subsection, we have
\begin{eqnarray}\label{mustar}
   m_b^{\rm SF}(\mu_f,\mu)
   &=& m_b^{\rm SF}(\mu_*,\mu_*) + \mu_*\,\frac{C_F\alpha_s(\mu_*)}{\pi}
    \nonumber\\
   &&\mbox{}- \mu_f\,\frac{C_F\alpha_s(\mu)}{\pi} \left[ 
    \left( 1 - 2\ln\frac{\mu_f}{\mu} \right) 
    + \frac23\,\frac{\mu_\pi^2(\mu_f,\mu)}{\mu_f^2}\,
    \ln\frac{\mu_f}{\mu} \right] , \\
   \mu_\pi^2(\mu_f,\mu)
   &=& \mu_\pi^2(\mu_*,\mu_*) \left[ 1 - \frac{C_F\alpha_s(\mu_*)}{2\pi}
    + \frac{C_F\alpha_s(\mu)}{\pi}
    \left( 3\ln\frac{\mu_f}{\mu} + \frac12 \right) \right]
    - 3\mu_f^2\,\frac{C_F\alpha_s(\mu)}{\pi}\,\ln\frac{\mu_f}{\mu} \,,
    \nonumber
\end{eqnarray}
where $\mu_*$ denotes the scale at which initial values for the two
parameters are obtained, for instance using relations such as 
(\ref{mSFmPS}) and (\ref{mupi2kin}). These relations are particularly 
simple if one chooses $\mu_f=\mu$.

In the relations above we have eliminated the unphysical HQET parameter 
$\lambda_1$ in favor of the physical parameter $\mu_\pi^2$ defined in the 
shape-function scheme. At first sight, this seems to threaten the 
convergence of the operator product expansion. For instance, the term 
proportional to $\mu_\pi^2$ in the expression for the zeroth moment 
$\hat M_0$ in (\ref{M0toM2}) contains a {\em leading-power\/} 
perturbative contribution of order $\alpha_s^2(\mu)$, and similar 
contributions would arise from all other terms in the expansion. These 
contributions would have to be subtracted from the Wilson coefficient of 
the first term, if this coefficient were computed to two-loop order. The 
overall convergence of the operator product expansion is unaffected by 
this reorganization of perturbative corrections.

\subsection{Asymptotic behavior of the shape function}

The fact that for sufficiently large values of the cutoff the moments of 
the shape function can be calculated using an operator-product expansion 
implies that a similar expansion can be used to obtain a 
model-independent description of the asymptotic behavior of the shape 
function. Taking the derivative of the zeroth moment $\hat M_0$ in 
(\ref{hatMNdef}) with respect to $\mu_f$, one obtains
\begin{equation}
   \hat S(\hat\omega,\mu) \Big|_{\hat\omega=\mu_f+\bar\Lambda(\mu_f,\mu)}
   = \left( 1 - \frac{dm_b^{\rm SF}(\mu_f,\mu)}{d\mu_f} \right)^{-1}
   \frac{d}{d\mu_f}\,\hat M_0(\mu_f,\mu) \,.
\end{equation}
This relation can be trusted as long as $\mu_f\gg\Lambda_{\rm QCD}$. It 
allows us to determine the behavior of the shape function for large 
values of $\hat\omega$. From (\ref{M0toM2}) we find at one-loop order
\begin{equation}\label{Sasymp}
   \hat S(\hat\omega,\mu) = - \frac{C_F\alpha_s(\mu)}{\pi}\,
   \frac{1}{\hat\omega-\bar\Lambda}
   \left[ \left( 2\ln\frac{\hat\omega-\bar\Lambda}{\mu} + 1 \right)
   + \frac23\,\frac{\mu_\pi^2}{(\hat\omega-\bar\Lambda)^2}
   \left( \ln\frac{\hat\omega-\bar\Lambda}{\mu} - 1 \right)
   + \dots \right] .
\end{equation}
The precise definitions of $\bar\Lambda$ and $\mu_\pi^2$ are not 
specified at this order. (Note that the shape function cannot depend on
the value of the cutoff $\mu_f$.) We have checked that this asymptotic
behavior of the shape function is consistent with the evolution equation 
(\ref{wow}) when expanded to first order in $\alpha_s$.

Relation (\ref{Sasymp}) is a model-independent result as long as 
$\hat\omega\gg\Lambda_{\rm QCD}$. We stress the remarkable fact that this 
radiative tail of the shape function is {\em negative}, in contrast with 
the naive expectation based on a probabilistic interpretation of the 
shape function as a momentum distribution function. The point is that the 
definition of the renormalized shape function requires scheme-dependent 
ultra-violet subtractions. From (\ref{Sasymp}) it follows that the shape 
function must have a zero, which for sufficiently large $\mu$ is located 
at a value $\hat\omega_0\approx\bar\Lambda+\mu/\sqrt{e}$.

\section{Recapitulation}
\label{sec:recap}

We have now completed the conceptual part of this paper. Before turning 
to phenomenological applications, let us briefly summarize the discussion 
so far. The hadronic physics governing the inclusive semileptonic decay
$\bar B\to X_u\,l^-\bar\nu$ is encoded in the structure functions $W_i$
appearing in the Lorentz decomposition of the hadronic tensor 
$W^{\mu\nu}$ in (\ref{Wdecomp}). In the shape-function region, only two
combinations of these functions are required at leading order in 
$\Lambda_{\rm QCD}/m_b$. They follow from the factorization formula 
(\ref{Wres}) using the explicit results for the hard functions $H_{ij}$
and the jet function $J$ derived in this paper. Explicitly, we obtain at 
next-to-leading order in renormalization-group improved perturbation 
theory
\begin{eqnarray}\label{master}
   \frac{W_1}{2} 
   &=& \left\{ 1 + \frac{C_F\alpha_s(m_b)}{4\pi}
    \left[ -4\ln^2 y + (6-c)\ln y - \frac{2\ln y}{1-y} - 4 L_2(1-y)
    - \frac{\pi^2}{6} - 12 \right] \right\} \nonumber\\
   &&\times y^{-1-a}\,e^{V_H(m_b,\mu_i)} \int_0^{n\cdot P_H}\!
    d\hat\omega\,\hat J(\hat p_\omega^2,y,\mu_i)\,
    \hat S(\hat\omega,\mu_i) + \dots \,, \\
   \frac{W_4}{2} + \frac{m_b W_5}{4}
   &=& \frac{C_F\alpha_s(m_b)}{4\pi}\,\frac{2\ln y}{1-y}\,y^{-1-a}\,
    e^{V_H(m_b,\mu_i)} \int_0^{n\cdot P_H}\!d\hat\omega\,
    \hat J(\hat p_\omega^2,y,\mu_i)\,\hat S(\hat\omega,\mu_i) + \dots
    \,, \nonumber
\end{eqnarray}
where the dots represent power corrections in $\Lambda_{\rm QCD}/m_b$. In 
these expressions $y=\bar n\cdot p/m_b$ is a partonic scaling variable, 
while $\hat p_\omega^2=m_b(n\cdot P_H-\hat\omega)$ depends on the 
hadronic variable $n\cdot P_H=E_H-|\vec P_H|$. In all our results, $m_b$ 
denotes the heavy-quark mass defined in the ``shape-function scheme'' 
introduced in Section~\ref{sec:properties} (see also 
Section~\ref{sec:appls} below). The jet function 
$\hat J(\hat p_\omega^2,y,\mu_i)$ at an intermediate hard-collinear scale 
$\mu_i\sim\sqrt{m_b\Lambda_{\rm QCD}}$ can be calculated in fixed-order 
perturbation theory. The relevant expression valid at one-loop order is 
given in (\ref{Jrescale}). The above results contain a variety of 
renormalization-group functions, which arise in the solution of evolution 
equations discussed in Section~\ref{sec:RGevol}. The explicit form of the 
Sudakov exponent $V_H$ can be found in (\ref{VHres}). This function is 
independent of the kinematic variables $y$ and $\hat p_\omega^2$. In 
addition, we need
\begin{equation}\label{acdef}
\begin{aligned}
   a &= \frac{\Gamma_0}{\beta_0}\,\ln r_1
    = \frac{16}{25}\,\ln\frac{\alpha_s(\mu_i)}{\alpha_s(m_b)} \,, \\ 
   c &= \frac{4}{\beta_0} \left( \frac{\Gamma_1}{\Gamma_0}
    - \frac{\beta_1}{\beta_0} \right) (r_1-1)
    = \left( \frac{10556}{1875} - \frac{12\pi^2}{25} \right)
    \left( \frac{\alpha_s(\mu_i)}{\alpha_s(m_b)} - 1 \right) .
\end{aligned}
\end{equation}
For simplicity, we have identified the high-energy matching scale $\mu_h$ 
introduced in Section~\ref{sec:RGevol} with the heavy-quark mass $m_b$. 
Our results are formally independent of the precise choice of 
$\mu_h\sim m_b$. The numerical effect of the residual $\mu_h$ dependence 
remaining after truncation of the perturbative expansion has been studied 
in \cite{Bosch:2003fc} and was found to be small. 

The function $\hat S(\hat\omega,\mu_i)$ in (\ref{master}) is the shape 
function after the transformation of variables from $\omega$ to 
$\hat\omega=\bar\Lambda-\omega$. The limits of integration for the 
variable $\hat\omega$ (i.e., $0\le\hat\omega\le n\cdot P_H$) are set by 
hadronic kinematics and are independent of the definition of the 
heavy-quark mass. The shape function is a non-perturbative object, which 
at present cannot be predicted from first principles. It enters our 
results (\ref{master}) renormalized at the intermediate hard-collinear 
scale $\mu_i$. In (\ref{wow}), we have presented an analytic formula 
(valid at leading order in renormalization-group improved perturbation 
theory) that relates the shape function at a high scale to the shape 
function renormalized at a low hadronic scale. Many properties of the 
shape function that were so far unknown have been derived in 
Section~\ref{sec:properties}. In particular, we have given explicit 
formulae relating the moments of the shape function to HQET parameters, 
and we have proved that the shape function has a negative tail for large 
values of $\hat\omega$, whose explicit form can be calculated using an 
operator product expansion. These new insights about the shape function 
will be very helpful in constructing a realistic model for the function 
$\hat S(\hat\omega,\mu_i)$, which can then be refined by tuning it to 
experimental data such as the photon energy spectrum in inclusive 
$\bar B\to X_s\gamma$ decays.

\section{Differential decay rates and spectra}
\label{sec:rates}

In the shape-function region considered in this work, the hadronic tensor 
is most naturally expressed in terms of the variables $n\cdot P_H$ and
$\bar n\cdot p$, where $p=m_b v-q=P_H-\bar\Lambda v$ is the momentum of
the final-state hadronic jet in the parton picture. This would remain 
true if we worked to higher order in the collinear expansion. It is thus 
useful to derive expressions for the decay rates in terms of these 
variables. Our theoretical results are valid as long as $n\cdot P_H$ can 
be considered as being of order a hadronic scale (say, a
$\mbox{few}\times\Lambda_{\rm QCD}$), whereas $\bar n\cdot p$ is 
integrated over a domain of order $m_b\gg\Lambda_{\rm QCD}$. It is this 
integration which provides a sampling over sufficiently many hadronic 
final states needed to ensure quark--hadron duality. The duality 
hypothesis underlies any description of inclusive decay rates using 
short-distance methods \cite{Bigi:2001ys}. 

Under these conditions, it is appropriate to describe the distribution in 
$\bar n\cdot p$ in terms of a {\em partonic\/} scaling variable 
$y=\bar n\cdot p/m_b$, while the distribution in the orthogonal 
light-cone component is described in terms of the dimensionful 
{\em hadronic\/} variable $P_+\equiv n\cdot P_H=E_H-|\vec{P}_H|$. At 
leading order in $\Lambda_{\rm QCD}/m_b$, we obtain from 
\cite{DeFazio:1999sv} the triple differential decay rate
\begin{equation}\label{triple}
   \frac{d^3\Gamma}{d\bar x\,dy\,dP_+}
   = 12m_b\,\Gamma_{\rm tree}\,y(y-\bar x)
   \left[ (1+\bar x-y)\,\frac{W_1}{2}
   + \bar x \left( \frac{W_4}{2} + \frac{m_b W_5}{4} \right) \right]
   + \dots \,,
\end{equation}
where $\bar x=1-x$, and $x=2E_l/m_b$ is a scaling variable proportional 
to the energy of the charged lepton measured in the $B$-meson rest frame. 
$\Gamma_{\rm tree}=G_F^2|V_{ub}|^2(m_b^{\rm pole})^5/(192\pi^3)$ denotes 
the tree-level expression for the total $\bar B\to X_u\,l^-\bar\nu$ decay 
rate obtained at leading order in the heavy-quark expansion. Phase space 
is such that
\begin{equation}\label{ps1}
   0\le P_+\le M_B-2E_l = m_b\,\bar x + \bar\Lambda \,, \qquad
   \frac{P_+-\bar\Lambda}{m_b}\le\bar x\le y\le 1 \,.
\end{equation}
The structure functions $W_i$ have support only for $y>0$, see 
(\ref{Jrescale}). If the lepton scaling variable $\bar x$ is integrated 
over a domain of order unity (meaning that $E_l$ is integrated over a 
domain of order $m_b\gg\Lambda_{\rm QCD}$), one can replace the second 
condition by $0\le\bar x\le y\le 1$ at leading power in 
$\Lambda_{\rm QCD}/m_b$. If, on the other hand, the lepton energy is 
restricted to be close to its kinematic limit, $E_l\approx M_B/2$, then 
$\bar x=O(\Lambda_{\rm QCD}/m_b)$, and at leading order the rate 
(\ref{triple}) can be simplified to
\begin{equation}\label{triple2}
   \frac{d^3\Gamma}{dE_l\,dy\,dP_+}
   = 24\Gamma_{\rm tree}\,y^2(1-y)\,\frac{W_1}{2} + \dots \,,
\end{equation}
with $0\le y\le 1$ and $0\le P_+\le M_B-2E_l$.

\begin{figure}[t]
\begin{center}
\epsfig{file=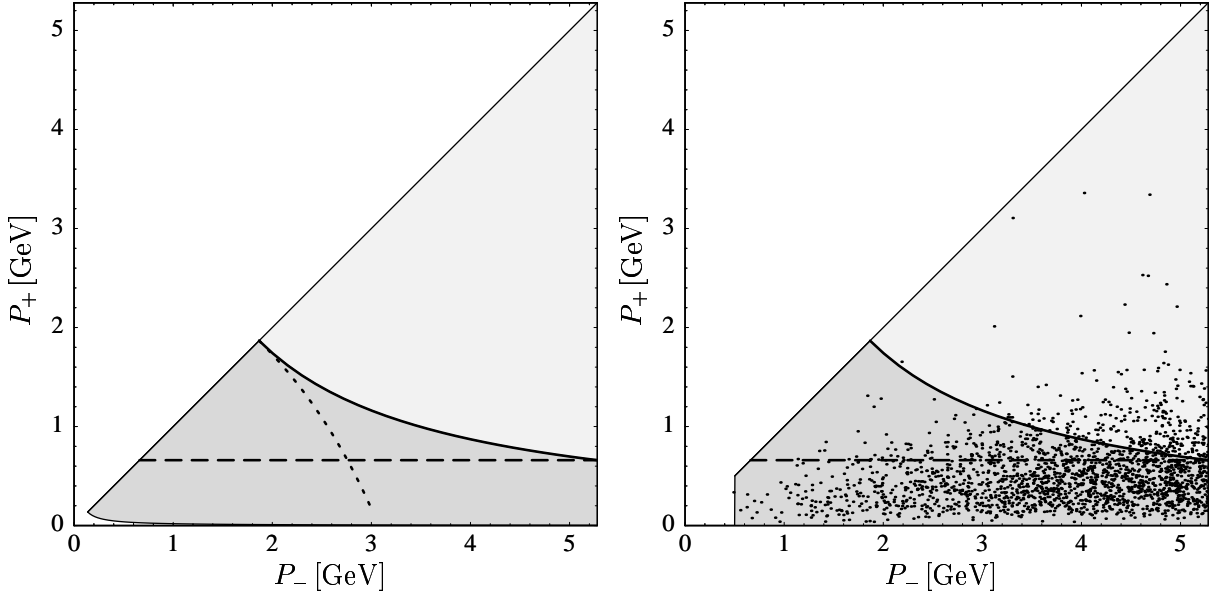,width=16cm}
\end{center}
\centerline{\parbox{14cm}{\caption{\label{fig:phase}
Hadronic phase space for the light-cone variables $P_-$ and $P_+$ (left), 
and theory phase space for $m_b=4.8$\,GeV (right). The scatter points 
indicate the distribution of events as predicted by the model of 
\cite{DeFazio:1999sv}. In each plot the solid line separates the regions 
where $s_H<M_D^2$ (dark gray) and $s_H>M_D^2$ (light gray), whereas the 
dashed line corresponds to $P_+=M_D^2/M_B$. The dotted line in the first 
plot shows the contour where $q^2=(M_B-M_D)^2$.}}}
\end{figure}

It will be useful to develop some intuition for the light-cone momentum 
variables. In general, the hadronic tensor can be described in terms of
the quantities
\begin{equation}\label{PpPmdef}
   P_+ = E_H - |\vec{P}_H| \,, \qquad
   P_- = E_H + |\vec{P}_H| \,,
\end{equation}
whose true phase-space is $M_\pi^2/P_-\le P_+\le P_-\le M_B$, 
corresponding to a triangular region in the $(P_-,P_+)$ plane with a tiny 
portion near the $P_-$ axis left unpopulated. The variable $P_-$ is 
related to our parton variables by 
$P_-=\bar n\cdot p+\bar\Lambda=m_b\,y+\bar\Lambda$. In our theoretical 
description based on quark--hadron duality $P_+$ starts from 0, while the 
small region with $P_-<\bar\Lambda$ is left unpopulated. This is 
illustrated in Figure~\ref{fig:phase}. Contours of constant hadronic or 
leptonic invariant mass in the $(P_-,P_+)$ plane are easy to visualize, 
since
\begin{equation}
   s_H = P_H^2 = P_+ P_- \,, \qquad
   q^2 = (M_B-P_+)(M_B-P_-)
\end{equation}
are given by simple expressions. The solid and dotted lines in the 
left-hand plot in Figure~\ref{fig:phase} show the contours where 
$s_H=M_D^2$ and $q^2=(M_B-M_D)^2$, respectively, which can be used to 
separate $\bar B\to X_u\,l^-\bar\nu$ events from semileptonic decays with 
charm hadrons in the final state. The dashed horizontal line shows the 
maximum allowed value of $P_+$ when a cut $E_l\ge(M_B^2-M_D^2)/(2M_B)$ is 
applied to the charged-lepton energy, which implies $P_+\le M_D^2/M_B$. 
This cut is another way of eliminating the charm background. In the 
right-hand plot, we indicate the density of events in theory phase space 
obtained using the model of \cite{DeFazio:1999sv}.\footnote{While not 
rigorously implementing shape-function effects beyond tree level, the 
model of \cite{DeFazio:1999sv} has the advantage that it interpolates 
between the shape-function region and the remainder of phase space, where 
a local operator product expansion can be employed. On the contrary, our 
more rigorous discussion here is limited to the region of hard-collinear 
jet momenta. We believe that the scatter plot shown in the figure 
provides a reasonably realistic impression of the population in phase 
space.}
The vast majority of events is located in the shape-function region of 
small $P_+$ and large $P_-$.

In the remainder of this section, we present exact analytic results, 
valid at leading order in $\Lambda_{\rm QCD}/m_b$ and at next-to-leading 
order in renormalization-group improved perturbation theory, for a 
variety of spectra in $\bar B\to X_u\,l^-\bar\nu$ decays. They are 
obtained by using our results (\ref{master}) in conjunction with the 
expressions for the differential rates in (\ref{triple}) or 
(\ref{triple2}), as appropriate. Our strategy will always be to integrate 
over the scaling variable $y$ before integrating over the hadronic 
variable $P_+$, changing variables from $P_+$ to 
$\hat p_\omega^2=m_b(P_+-\hat\omega)$. In this step, one must carefully 
evaluate the effect of the star distributions contained in the jet 
function $\hat J(\hat p_\omega^2,y,\mu_i)$, using the definitions in
(\ref{star}). The integral over the shape-function variable $\hat\omega$ 
is left until the end, so that our formulae can be evaluated once an 
explicit form for the shape function is assumed. We will always present
fractional decay rates normalized to the total inclusive rate
\begin{equation}
   \Gamma(\bar B\to X_u\,l^-\bar\nu)\equiv \Gamma_{\rm tot}
   = \Gamma_{\rm tree} \left[ 1 + \frac{C_F\alpha_s(m_b)}{4\pi}
   \left( \frac{25}{2} - 2\pi^2 \right) \right] + \dots \,,
\end{equation}
where the dots represent higher-order perturbative corrections as well as 
power corrections of order $(\Lambda_{\rm QCD}/m_b)^2$ and higher. This 
procedure offers the advantage of eliminating the strong sensitivity to 
the heavy-quark (pole) mass. Replacing $\Gamma_{\rm tree}$ by 
$\Gamma_{\rm tot}$ adds a contribution $(2\pi^2-\frac{25}{2})$ to the 
coefficient of the hard correction to the function $W_1$ in 
(\ref{master}). Our predictions for normalized rate fractions can be 
turned into predictions for absolute rates with the help of an 
independent theoretical prediction for the total 
$\bar B\to X_u\,l^-\bar\nu$ rate obtained using a local operator product 
expansion.

The integrals over the parton variable $y$ encountered in our analysis 
can be reduced to a set of master integrals defined as
\begin{equation}\label{Idef}
\begin{aligned}
   I_1(b,z) &= \int_0^z\!dy\,y^b
    = \frac{z^{1+b}}{1+b} \,, \\
   I_2(b,z) &= \int_0^z\!dy\,y^b\ln y
    = \frac{z^{1+b}}{1+b} \left( \ln z - \frac{1}{1+b} \right) , \\
   I_3(b,z) &= \int_0^z\!dy\,y^b\ln^2 y
    = \frac{z^{1+b}}{1+b} \left( \ln^2 z - \frac{2\ln z}{1+b}
    + \frac{2}{(1+b)^2} \right) , \\
   I_4(b,z) &= \int_0^z\!dy\,y^b \frac{\ln y}{1-y}
    = \sum_{j=0}^\infty \frac{z^{1+b+j}}{1+b+j}
    \left( \ln z - \frac{1}{1+b+j} \right) , \\
   I_5(b,z) &= \int_0^z\!dy\,y^b L_2(1-y)
    = \frac{z^{1+b}}{1+b}\,L_2(1-z) - \frac{I_4(1+b,z)}{1+b} \,,
\end{aligned}
\end{equation}
where $b>-1$  and $z\le 1$ are a arbitrary real numbers. The integral 
$I_4$ can be expressed in terms of the incomplete $\beta$ function 
$B(z,a,b)$ and the Lerch transcendent $\Phi(z,a,b)$ as
\begin{equation}
   I_4(b,z) = \ln z\,B(z,1+b,0) - z^{1+b}\,\Phi(z,2,1+b) \,.
\end{equation}
We note the useful relations
\begin{equation}
   I_4(1+b,z) = I_4(b,z) - I_2(b,z) \,, \qquad
   I_4(b,1) = - \psi'(1+b) \,,
\end{equation}
where $\psi'(z)$ is the derivative of the Euler $\psi$ function.

\subsection{Charged-lepton energy spectrum}

As a first application, we study the distribution of the charged-lepton 
energy near the kinematic endpoint. Specifically, we assume that 
$M_B-2E_l$ is of order a hadronic scale. Starting from the triple 
differential rate in (\ref{triple2}), we obtain for the normalized energy 
spectrum
\begin{eqnarray}\label{Elspectrum}
   \frac{1}{\Gamma_{\rm tot}}\,\frac{d\Gamma}{dE_l}
   &=& \frac{4T(a)}{m_b}\,e^{V_H(m_b,\mu_i)}
    \int_0^{M_B-2E_l}\!d\hat\omega\,\hat S(\hat\omega,\mu_i)\,
    \Bigg\{ 1 + \frac{C_F\alpha_s(m_b)}{4\pi}\,H(a) \nonumber\\
   &&\mbox{}+ \frac{C_F\alpha_s(\mu_i)}{4\pi} \Bigg[
    2\ln^2\frac{m_b(M_B-2E_l-\hat\omega)}{\mu_i^2}
    + \Big( 4f_2(a) - 3 \Big)
    \ln\frac{m_b(M_B-2E_l-\hat\omega)}{\mu_i^2} \nonumber\\
   &&\hspace{2.3cm}\mbox{}
    + \Big( 7 - \pi^2 - 3f_2(a) + 2f_3(a) \Big) \Bigg] \Bigg\} \,.
\end{eqnarray}
Here
\begin{equation}
\begin{aligned}
   T(a) &= 6 \Big[ I_1(1-a,1) - I_1(2-a,1) \Big] \,, \\
   f_n(a) &= \frac{I_n(1-a,1)-I_n(2-a,1)}{I_1(1-a,1)-I_1(2-a,1)} \,, \\
   H(a) &= \frac{11\pi^2}{6} - \frac{49}{2} + (6-c)f_2(a) - 4f_3(a)
    - 2f_4(a) - 4f_5(a) \,.
\end{aligned}
\end{equation}
Using the results for the master integrals in (\ref{Idef}), one readily 
derives the explicit formulae
\begin{equation}
\begin{aligned}
   T(a) &= \frac{6}{(3-a)(2-a)} \,, \\
   H(a) &= \frac{11\pi^2}{6} - \frac{45}{2}
    - \frac{2(146-162a+59a^2-7a^3)}{(3-a)^2(2-a)^2}
    - 4\psi'(2-a) - c\,f_2(a) \,, \\
   f_2(a) &= - \frac{5-2a}{(3-a)(2-a)} \,, \qquad
    f_3(a) = \frac{2(19-15a+3a^2)}{(3-a)^2(2-a)^2} \,.
\end{aligned}
\end{equation}
At leading power in $\Lambda_{\rm QCD}/m_b$ the heavy-quark mass in the 
denominator of the prefactor on the right-hand side of (\ref{Elspectrum}) 
can be replaced by $m_b+\omega=M_B-\hat\omega$, which removes any 
sensitivity to the definition of $m_b$. This replacement can indeed be 
justified by studying power corrections to the shape function
\cite{Bauer:2002yu,Neubert:2002yx}.

All our results for decay rates will have a similar structure, but the 
definitions of the functions $T$, $f_n$, and $H$ will be different in 
each case. The product $T\,e^{V_H}$ resums the leading double and 
single-logarithmic corrections to all orders in perturbation theory. The
tree-level result can be recovered by setting $V_H=0$ and $a=0$, in which
case $T(0)=1$, and the spectrum is simply given in terms of an integral 
over the shape function \cite{Neubert:1993ch}. The next-to-leading order 
terms can be divided into an $\hat\omega$-independent hard function 
$H(a)$, whose structure follows from the form of the hard corrections in 
(\ref{master}), and a sum of hard-collinear radiative corrections, which 
follow from the integration over the jet function.

Using the above result, it is straightforward to calculate the fraction 
$F_E=\Gamma(E_l\ge E_0)/\Gamma_{\rm tot}$ of all 
$\bar B\to X_u\,l^-\bar\nu$ events with charged-lepton energy above a 
threshold $E_0$. Defining $\Delta_E=M_B-2E_0$, we find
\begin{eqnarray}\label{FE}
   F_E(\Delta_E) &=& T(a)\,e^{V_H(m_b,\mu_i)}
    \int_0^{\Delta_E}\!d\hat\omega\,
    \frac{2(\Delta_E-\hat\omega)}{M_B-\hat\omega}\,
    \hat S(\hat\omega,\mu_i)\,
    \Bigg\{ 1 + \frac{C_F\alpha_s(m_b)}{4\pi}\,H(a) \nonumber\\
   &&\mbox{}+ \frac{C_F\alpha_s(\mu_i)}{4\pi} \Bigg[
    2\ln^2\frac{m_b(\Delta_E-\hat\omega)}{\mu_i^2}
    + \Big( 4f_2(a) - 7 \Big)
    \ln\frac{m_b(\Delta_E-\hat\omega)}{\mu_i^2} \nonumber\\
   &&\hspace{2.3cm}\mbox{}
    + \Big( 14 - \pi^2 - 7f_2(a) + 2f_3(a) \Big) \Bigg] \Bigg\} \,.
\end{eqnarray}
The fraction $F_E(\Delta_E)$ is given in terms of a weighted integral 
over the shape function, with a weight factor of order 
$\Lambda_{\rm QCD}/m_b$ that vanishes at the upper end of integration. As 
a result, only a small fraction of events is contained in the lepton 
endpoint region.

\subsection{Hadronic \boldmath$P_+$ spectrum\unboldmath}

A cut on the charged-lepton energy restricts the variable $P_+$ to be 
less than $\Delta_E$. On the contrary, however, a cut on $P_+$ does 
{\em not\/} restrict the lepton energy to be in the endpoint region. In 
fact, according to (\ref{ps1}) the integration over $\bar x$ can be taken 
to run from 0 to 1 at leading power if $P_+$ is small. It follows that 
the fraction of events with $P_+\le\Delta_E$ samples the same hadronic 
phase space as the lepton-endpoint cut, but it contains significantly 
more events. Such a cut therefore offers an excellent opportunity to 
determine the CKM matrix element $|V_{ub}|$.

The calculation of the fraction 
$F_P(\Delta_P)=\Gamma(P_+\le\Delta_P)/\Gamma_{\rm tot}$ of all 
$\bar B\to X_u\,l^-\bar\nu$ events with hadronic light-cone momentum 
$P_+$ below a threshold $\Delta_P$ starts from the expression for the 
triple differential rate in (\ref{triple}). We integrate over $\bar x$
and $y$ in the range $0\le\bar x\le y\le 1$ before integrating over 
$P_+$. The result is
\begin{eqnarray}\label{FP}
   F_P(\Delta_P) &=& T(a)\,e^{V_H(m_b,\mu_i)}
    \int_0^{\Delta_P}\!d\hat\omega\,\hat S(\hat\omega,\mu_i)\,
    \Bigg\{ 1 + \frac{C_F\alpha_s(m_b)}{4\pi}\,H(a) \nonumber\\
   &&\mbox{}+ \frac{C_F\alpha_s(\mu_i)}{4\pi} \Bigg[
    2\ln^2\frac{m_b(\Delta_P-\hat\omega)}{\mu_i^2}
    + \Big( 4f_2(a) - 3 \Big)
    \ln\frac{m_b(\Delta_P-\hat\omega)}{\mu_i^2} \nonumber\\
   &&\hspace{2.3cm}\mbox{}
    + \Big( 7 - \pi^2 - 3f_2(a) + 2f_3(a) \Big) \Bigg] \Bigg\} \,,
\end{eqnarray}
where now
\begin{equation}\label{TforFP}
\begin{aligned}
   T(a) &= 6 I_1(2-a,1) - 4 I_1(3-a,1) \,, \\
   H(a) &= \frac{11\pi^2}{6} - \frac{49}{2} + (6-c)f_2(a) - 4f_3(a)
    - 2 \Big[ f_4(a) - \Delta f_4(a) \Big] - 4f_5(a) \,,
\end{aligned}
\end{equation}
and
\begin{equation}\label{FforFP}
\begin{aligned}
   f_n(a) &= \frac{3I_n(2-a,1)-2I_n(3-a,1)}{3I_1(2-a,1)-2I_1(3-a,1)} \,,
    \\
   \Delta f_4(a) &= \frac{I_4(3-a,1)}{3I_1(2-a,1)-2I_1(3-a,1)} \,.
\end{aligned}
\end{equation}
The contribution $\Delta f_4$ arises from the terms contained in the 
structure functions $W_4$ and $W_5$ in (\ref{master}). Using the analytic 
results for the master integrals yields
\begin{equation}
\begin{aligned}
   T(a) &= \frac{2(6-a)}{(4-a)(3-a)} \,, \\
   H(a) &= \frac{11\pi^2}{6} - \frac{49}{2}
    - \frac{4(486-389a+103a^2-9a^3)}{(6-a)(4-a)^2(3-a)^2}
    - 4\psi'(3-a) - c\,f_2(a) \,, \\
   f_2(a) &=  - \frac{30-12a+a^2}{(6-a)(4-a)(3-a)} \,, \qquad
    f_3(a) = \frac{2(138-90a+18a^2-a^3)}{(6-a)(4-a)^2(3-a)^2} \,.
\end{aligned}
\end{equation}
Comparing the result for $F_P(\Delta_P)$ in (\ref{FP}) with the 
expression for $F_E(\Delta_E)$ in (\ref{FE}) we observe that, as 
anticipated, the cut on hadronic $P_+$ contains a much larger fraction of 
all $\bar B\to X_u\,l^-\bar\nu$ events. In fact, $F_P(\Delta_P)$ is 
directly given in terms of an integral over the shape function, without a 
weight function of order $\Lambda_{\rm QCD}/m_b$. If we neglect radiative
corrections for a moment, we simply have 
$F_P(\Delta_P)=\int_0^{\Delta_P}\!d\hat\omega\,\hat S(\hat\omega)$, where 
the shape function is expected to peak at a position 
$\hat\omega\approx\bar\Lambda\approx 0.5$\,GeV. For the ``optimal cut'' 
$P_+\le M_D^2/M_B\simeq 0.66$\,GeV, which eliminates the charm background 
entirely, we thus expect that more than half of all events are contained 
in the event fraction $F_P(\Delta_P)$. The hadronic $P_+$ spectrum 
therefore offers a very promising new avenue for a high-precision 
measurement of $|V_{ub}|$.

In a realistic measurement, it is often necessary to implement a loose 
cut on the lepton energy in order to eliminate soft leptons, which can be 
difficult to measure in the detector. Let us discuss how our results 
would change in the presence of such a cut. Assume that we require 
$E_l\ge E_0$ with $E_0$ of order 1\,GeV or so, corresponding to 
$\bar x\le\bar x_0=1-2E_0/m_b$ with $\bar x_0=O(1)$. Our expression for
the event fraction $F_P(\Delta_P)$ in (\ref{FP}) remains valid, except
that the master functions now become functions of the parameter
$\bar x_0$. We find that $T(a)$, $f_n(a)$ and $\Delta f_4(a)$ in 
(\ref{TforFP}) and (\ref{FforFP}) get replaced by
\begin{eqnarray}
   T(a,\bar x_0)
   &=& 2\bar x_0^2(3+2\bar x_0) \Big[ I_1(-a,\bar x_0) -\! I_1(-a,1)
    \Big]
    - 12\bar x_0(1+\bar x_0) \Big[ I_1(1-a,\bar x_0) -\! I_1(1-a,1)
    \Big] \nonumber\\
   &&\mbox{}+ 12\bar x_0 \Big[ I_1(2-a,\bar x_0) - I_1(2-a,1) \Big]
    + 6 I_1(2-a,\bar x_0) - 4 I_1(3-a,\bar x_0) \,, \nonumber\\[0.1cm]
   f_n(a,\bar x_0)
   &=& \frac{\mbox{[expression for $T(a,\bar x_0)$ with 
                  $I_1\to I_n$]}}{T(a,\bar x_0)} \,, \\
   \Delta f_4(a,\bar x_0)
   &=& \frac{4\bar x_0^3 \Big[ I_4(-a,\bar x_0) - I_4(-a,1) \Big]\!
            - 6\bar x_0^2 \Big[ I_4(1-a,\bar x_0) - I_4(1-a,1) \Big]
            + 2 I_4(3-a,\bar x_0)}{T(a,\bar x_0)} \,. \nonumber
\end{eqnarray}
For $\bar x_0=1$ these complicated results reduce to the simpler 
expressions given above. Numerically, the cut on the lepton energy has a 
minor effect provided that $\bar x_0$ is larger than about 0.6, 
corresponding to a lower cutoff $E_0$ less than about 1\,GeV.

\subsection{Hadronic invariant mass spectrum}

A cut $\sqrt{s_H}\le M_D$ on the hadronic invariant mass in the final 
state constitutes the ideal separator between $\bar B\to X_u\,l^-\bar\nu$
and $\bar B\to X_c\,l^-\bar\nu$ events, since any final state containing 
a charm hadron has invariant mass above $M_D$. The left-hand plot in 
Figure~\ref{fig:phase} shows that such a cut fully contains the region 
with $P_+\le M_D^2/M_B$ considered earlier. In addition, the hadronic 
invariant mass cut contains a triangle-shaped region of larger $P_+$, 
which culminates in a cusp where $P_+=P_-=M_D$. Near the cusp, 
both light-cone momentum components are of the same order, and hence this
portion of phase space lies outside the shape-function region. In other 
words, the region near the cusp is a dangerous one (a ``Bermuda 
triangle''), where the theoretical description based on the collinear 
expansion breaks down. A priori, then, it is not evident that we can 
compute the fractional rate 
$F_M(s_0)=\Gamma(s_H\le s_0)/\Gamma_{\rm tot}$ in a controlled 
heavy-quark expansion.

To see what happens, it is instructive to first ignore radiative 
corrections. At tree level, it is straightforward to obtain
\begin{equation}\label{FM}
   F_M(s_0) = \int\limits_0^{\Delta_s}\!\!d\hat\omega\,\hat S(\hat\omega)
   + \int\limits_{\Delta_s}^{\sqrt{s_0}}\!\!d\hat\omega\,
   \hat S(\hat\omega) \left( \frac{\Delta_s}{\hat\omega} \right)^3
   \left( 2 - \frac{\Delta_s}{\hat\omega} \right) ,
\end{equation}
where $\Delta_s=s_0/M_B$. The calculation of this event fraction requires 
knowledge of the shape function over a wider range in $\hat\omega$ than 
in the case of the event fraction with a cut on $P_+$. It is therefore 
more difficult to extract the relevant hadronic information from the 
$\bar B\to X_s\gamma$ photon spectrum. The first 
integral is the same as in (\ref{FP}) and corresponds to the region in 
phase space where $P_+\le s_0/M_B$. Note that the ratio 
$\Delta_s=s_0/M_B$ plays the same role as the cutoff $\Delta_P$ on the 
$P_+$ spectrum. The second integral corresponds to the phase space above 
the dashed line in Figure~\ref{fig:phase}. The region near the cusp 
corresponds to the upper integration region in the second integral. Note 
that, due to the rapid fall-off of the integrand, the ``Bermuda 
triangle'' only gives a power-suppressed contribution to the decay rate 
and so can be ignored at leading order in the heavy-quark expansion. We 
will address this point in more detail below. 

When radiative corrections are included, the result for the integrated 
hadronic invariant mass spectrum becomes rather complicated. It is 
convenient to split up the integration region in phase space into a 
box-shaped region with $P_+\le s_0/M_B$ and a triangular shaped region
with $s_0/M_B<P_+\le\sqrt{s_0}$. Writing 
\begin{equation}
   F_M(s_0) = F_M^{\rm box}(s_0) + F_M^{\rm triangle}(s_0) \,, \qquad
   \mbox{with} \quad F_M^{\rm box}(s_0) = F_P(\Delta_s) \,,
\end{equation}
we find that the box contribution is given by the expression for the rate 
fraction $F_P(\Delta_P)$ in (\ref{FP}) evaluated with 
$\Delta_P=\Delta_s=s_0/M_B$. For the remaining contribution from the 
triangular region, we obtain
\begin{eqnarray}\label{Ftrian}
   F_M^{\rm triangle}(s_0)
   &=& e^{V_H(m_b,\mu_i)} \int_{\Delta_s}^{\sqrt{s_0}}\!d\hat\omega\,
    \hat S(\hat\omega,\mu_i) \left[ G_1(\Delta_s/\hat\omega)
    + \frac{C_F\alpha_s(\mu_i)}{4\pi}\,G_2(\Delta_s,\hat\omega) \right]
    \nonumber\\
   &&\mbox{}+ e^{V_H(m_b,\mu_i)} \int_0^{\Delta_s}\!d\hat\omega\,
    \hat S(\hat\omega,\mu_i)\,\frac{C_F\alpha_s(\mu_i)}{4\pi}\,
    G_3(\Delta_s,\hat\omega) \,,
\end{eqnarray}
where
\begin{equation}
   G_1(z) = T(a,z) \left\{ 1 + \frac{C_F\alpha_s(m_b)}{4\pi}\,H(a,z)
   + \frac{C_F\alpha_s(\mu_i)}{4\pi} \left[ 7 - \pi^2 - 3f_2(a,z)
   + 2f_3(a,z) \right] \right\}
\end{equation}
contains the same functions $T$, $H$ and $f_n$ as defined in 
(\ref{TforFP}) and (\ref{FforFP}), but with all master integrals replaced 
by $I_n(b,1)\to I_n(b,z)$. In addition, we need
\begin{eqnarray}\label{G2G3}
   G_2(\Delta_s,\hat\omega)
   &=& \int_0^{\mu_i^2/m_b}\!\frac{dP}{P} \left\{
    \ln\frac{m_b P}{\mu_i^2} \left[
    k_1\!\left( \frac{\Delta_s}{P+\hat\omega} \right)
    - k_1\!\left( \frac{\Delta_s}{\hat\omega} \right) \right]
    + \left[ k_2\!\left( \frac{\Delta_s}{P+\hat\omega} \right)
    - k_2\!\left( \frac{\Delta_s}{\hat\omega} \right) \right] \right\}
    \nonumber\\
   &&\mbox{}+ \int_{\mu_i^2/m_b}^{\sqrt{s_0}-\hat\omega}\!\frac{dP}{P}
    \left[ \ln\frac{m_b P}{\mu_i^2}\,
    k_1\!\left( \frac{\Delta_s}{P+\hat\omega} \right)
    + k_2\!\left( \frac{\Delta_s}{P+\hat\omega} \right) \right] , \\
   G_3(\Delta_s,\hat\omega)
   &=& \int_{\Delta_s}^{\sqrt{s_0}}\!\frac{dP}{P-\hat\omega}
    \left[ \ln\frac{m_b(P-\hat\omega)}{\mu_i^2}\,
    k_1\!\left( \frac{\Delta_s}{P} \right)
    + k_2\!\left( \frac{\Delta_s}{P} \right) \right] , \nonumber
\end{eqnarray}
where
\begin{equation}
\begin{aligned}
   k_1(z) &= 4 \Big[ 6 I_1(2-a,z) - 4 I_1(3-a,z) \Big] = 4 T(a,z) \,, \\
   k_2(z) &= 4 \Big[ 6 I_2(2-a,z) - 4 I_2(3-a,z) \Big] 
    - 3 \Big[ 6 I_1(2-a,z) - 4 I_1(3-a,z) \Big] \,.
\end{aligned}
\end{equation}

As mentioned above, the phase-space region near the cusp where 
$\hat\omega\sim\sqrt{s_0}$ or $P\sim\sqrt{s_0}$ gives a power-suppressed 
contribution to the decay rate. Using the explicit results for the 
functions $G_i$ together with the asymptotic form of the shape function 
in (\ref{Sasymp}), we find that the corresponding term is
\begin{equation}\label{Bermuda}
   F_M(s_0)\owns e^{V_H(m_b,\mu_i)}\,
   \frac{C_F\alpha_s(\mu_i)}{\pi}\,\frac{6}{(3-a)^2}
   \left( \frac{\Delta_s}{\sqrt{s_0}} \right)^{3-a}
   \left( \frac74 + \frac{3}{3-a} \right) + \dots \,,
\end{equation}
where the dots represent higher-order power corrections. Whereas the 
fraction $F_M(s_0)$ is of $O(1)$ in power counting, the result 
(\ref{Bermuda}) scales like $(\Lambda_{\rm QCD}/m_b)^{(3-a)/2}$ for
$s_0\sim m_b\Lambda_{\rm QCD}$. In the derivation of the formula for the 
event fraction we have neglected other power-suppressed terms with the 
same scaling. For consistency, we should therefore omit the term in 
(\ref{Bermuda}), which can be done by replacing all occurrences of 
$\sqrt{s_0}$ in upper integration limits in (\ref{Ftrian}) and 
(\ref{G2G3}) with $\infty$. Only in that way we ensure that our 
calculations provide the unique leading-power contribution in the 
heavy-quark limit. We will use this prescription in our numerical 
analysis in Section~\ref{sec:appls}.

Because of the presence of power corrections from a region in phase space 
where the collinear expansion breaks down, it is not clear to us how one 
would construct a systematic heavy-quark expansion for the fraction 
$F_M(s_0)$ beyond the leading order. Clarification of this point deserves 
further study.

\subsection{Combined cuts on hadronic and leptonic invariant mass}
\label{sec:comb}

\begin{figure}
\begin{center}
\epsfig{file=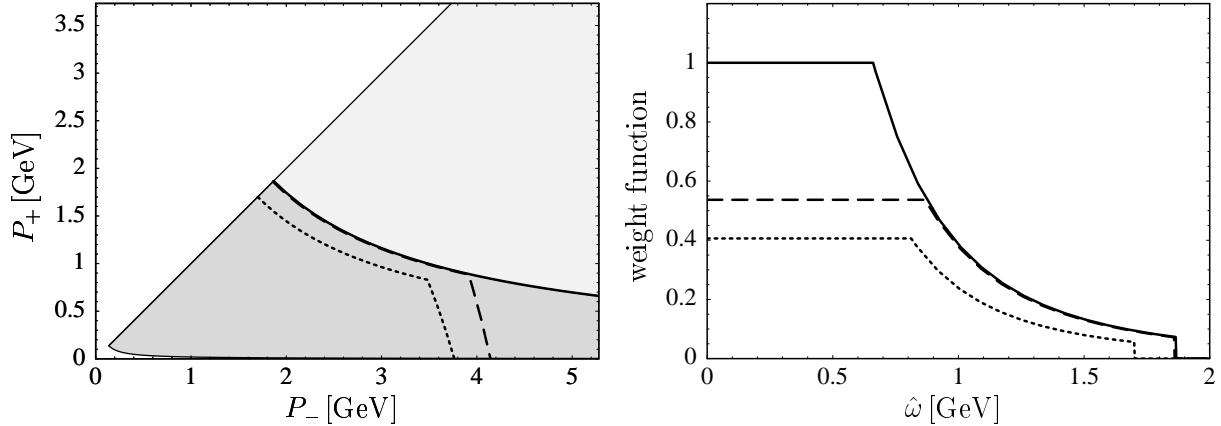,width=16cm}
\end{center}
\centerline{\parbox{14cm}{\caption{\label{fig:combined}
Phase-space constraints (left) and weight functions (right) for combined 
cuts on the hadronic and leptonic invariant mass: 
$(s_0,\,q_0^2)=(M_D^2,\,0)$ (solid), $(M_D^2,\,6\,\mbox{GeV}^2)$ 
(dashed), and $((1.7\,\mbox{GeV})^2,\,8\,\mbox{GeV}^2)$ (dotted).}}}
\end{figure}

Bauer et al.\ have proposed to reduce the sensitivity to shape-function 
effects in the extraction of $|V_{ub}|$ by combining a cut on hadronic
invariant mass with a cut $q^2\ge q_0^2$ on the invariant mass squared of 
the lepton pair \cite{Bauer:2001rc}. The first plot in 
Figure~\ref{fig:combined} shows that this eliminates a large portion of
the events with large $P_-$. It is straightforward to study the effects 
of such a combined cut in the approximation where radiative corrections 
are neglected. For the corresponding event fraction at tree level, we 
obtain
\begin{equation}
   F_{\rm comb}(s_H\le s_0, q^2\ge q_0^2)
   = y_0^3\,(2-y_0)\! \int\limits_0^{\Delta_s/y_0}\!d\hat\omega\,
   \hat S(\hat\omega)
   + \int\limits_{\Delta_s/y_0}^{\sqrt{s_0}}\!\!d\hat\omega\,
   \hat S(\hat\omega) \left( \frac{\Delta_s}{\hat\omega} \right)^3
   \left( 2 - \frac{\Delta_s}{\hat\omega} \right) ,
\end{equation}
where $y_0=1-q_0^2/(m_b M_B)$, and $\Delta_s=s_0/M_B$ as above. For a 
fixed hadronic-mass cut $s_0$, the effect of the cut on $q^2$ is to 
broaden the support of the first integral, while at the same time 
reducing its weight due to the prefactor. To illustrate this point, we 
show in the second plot in Figure~\ref{fig:combined} the weight functions 
under the integral with the shape function for three different choices of 
$(s_0,\,q_0^2)$. The sensitivity to the precise form of the shape 
function is reduced because the weight functions become progressively 
more shallow as the value of $q_0^2$ is raised. However, this reduction 
comes at the price of a significant reduction of the rate, raising 
questions about the validity of the assumption of quark--hadron duality. 
We will see in Section~\ref{sec:appls} that the {\em relative\/} 
uncertainty due to shape-function effects is not strongly reduced when 
imposing an additional cut on $q^2$. We are therefore not convinced that 
it is worth paying this price.

\subsection{Comparison with the literature}

Before concluding this section, let us comment on results for inclusive 
$B$-decay spectra in the shape-function region previously published in 
the literature. Detailed predictions for $\bar B\to X_u\,l^-\bar\nu$ 
decay distributions including shape-function effects were presented in 
\cite{Dikeman:1997es} and \cite{DeFazio:1999sv}. In these papers, 
$O(\alpha_s)$ corrections were included at the level of the underlying 
parton spectra. In the last reference fully differential distributions 
are presented, which can be used to calculate arbitrary spectra and 
implement experimental cuts. Dedicated studies of the hadronic invariant 
mass spectrum can also be found in 
\cite{Bigi:1997dn,Falk:1997gj,Jezabek:2001pg}. Similar analyses for the 
photon energy spectrum in $\bar B\to X_s\gamma$ decays were presented in 
\cite{Kagan:1998ym,Dikeman:1995ad,Leibovich:1999xf}. In all these works, 
shape-function effects are implemented by convolving parton-model 
distributions with a primordial structure function, typically by 
replacing the $b$-quark mass in expressions for the parton spectra by a 
new variable $m_b^*=m_b+\omega$ \cite{Mannel:1994pm}. This procedure is 
correct at tree level. However, it is no longer fully consistent when 
radiative corrections are included, because part of the $O(\alpha_s)$ 
corrections in the expressions for parton-level spectra are absorbed into 
the renormalization of the shape function. In particular, this changes 
the sign of the leading Sudakov logarithm, as can be seen by comparing 
the coefficients of the $\ln p^2/p^2$ terms in (\ref{Fulvia}) and 
(\ref{Jres}). This point has also been emphasized in \cite{Bauer:2003pi}. 
An attempt to include radiative corrections to the spectra in a 
systematic way was made in \cite{Aglietti:2001br}, where in particular 
the evolution of the shape function has been addressed. Since our 
evolution equation does not agree with the one found by this author 
\cite{Aglietti:1999ur}, our results for decay spectra are also in
disagreement.

In \cite{Bauer:2003pi}, an expression has been presented for the double 
differential rate in the variables $E_l$ and $E_H$, which (apart from a 
typo) is consistent with our findings. However, in this paper no 
distinction between $\alpha_s(m_b)$ and $\alpha_s(\mu_i)$  has been made 
in the next-to-leading order corrections, which entails significant 
perturbative uncertainties. Also, the important question of a physical 
definition of the $b$-quark mass has not been addressed. Whereas in our 
case all quantities are defined in a physical subtraction scheme and no 
reference to the $b$-quark mass is left in the final expressions 
for decay distributions (except as arguments of running couplings), the 
results of \cite{Bauer:2003pi} contain explicit reference to the 
$b$-quark pole mass.

\section{Model-independent relations between spectra}
\label{sec:relations}

The decay spectra and event fractions discussed in the previous section
are given in terms of weighted integrals of perturbative expressions with 
a non-perturbative shape function, which encodes hadronic physics related 
to the bound-state properties of the $B$ meson. To use these formulae, 
one must take recourse to a model for the shape function, or (better) 
extract the shape function from a fit to experimental data. 
Alternatively, it is possible to derive model-independent relations 
between different decay distributions in which the shape function has 
been eliminated \cite{Neubert:1993ch}. This makes use of the fact that at 
leading power in $\Lambda_{\rm QCD}/m_b$ shape-function effects in 
inclusive decays to light hadronic final states are described by a single 
universal (i.e., process independent) function. The most promising 
strategy is to relate event fractions in semileptonic 
$\bar B\to X_u\,l^-\bar\nu$ decays to a weighted integral over the 
$\bar B\to X_s\gamma$ photon spectrum, which at present provides the most 
direct access to the shape function. 

While it is straightforward to derive such relations at tree level, 
radiative corrections introduce non-trivial complications 
\cite{Leibovich:1999xfx,Neubert:2001sk,Leibovich:2001ra}. In the 
following, we illustrate with a concrete example how such shape-function 
independent relations can be derived systematically within our framework. 
Since our formalism has yet to be applied to the $\bar B\to X_s\gamma$ 
photon spectrum, we will instead derive a relation between the 
charged-lepton energy spectrum and a weighted integral over the $P_+$ 
spectrum in $\bar B\to X_u\,l^-\bar\nu$ decays. This relation serves as a 
prototype for all other shape-function independent relations between 
partially integrated decay rates. Note that the properties of the $P_+$ 
distribution are very similar to those of the $\bar B\to X_s\gamma$ 
photon spectrum in the sense that, at tree level, both spectra are 
directly given in terms of the shape function, e.g.\
$(1/\Gamma_{\rm tot})\,d\Gamma/dP_+=\hat S(P_+)+\dots$. This connection,
and its potential usefulness for an extraction of $|V_{ub}|$, has been
noted earlier in \cite{Mannel:1999gs}. 

Specifically, we wish to construct a {\em perturbative\/} weight function 
$w(\Delta,P_+)$ such that at leading power in $\Lambda_{\rm QCD}/m_b$
\begin{equation}\label{raterel}
   \int_{E_0}^{M_B/2}\!dE_l\,\frac{d\Gamma}{dE_l}
   = \int_0^{\Delta}\!dP_+\,w(\Delta,P_+)\,\frac{d\Gamma}{dP_+} \,,
   \qquad \Delta = M_B - 2E_0 \,.
\end{equation}
This relation is independent of the shape function and hence insensitive 
to hadronic physics. The construction of the weight function is 
straightforward order by order in perturbation theory. Using the results 
of the previous section, we find
\begin{eqnarray}\label{wfun}
   w(\Delta,P_+)
   &=& \frac{2(\Delta-P_+)}{M_B-P_+}\,\frac{3(4-a)}{(6-a)(2-a)}\,
    \Bigg\{ 1 + \frac{C_F\alpha_s(m_b)}{4\pi}\,h_1(a) \nonumber\\
   &&\mbox{}+ \frac{C_F\alpha_s(\mu_i)}{4\pi} \left[ h_2(a)\,
    \ln\frac{m_b(\Delta-P_+)}{\mu_i^2} + h_3(a) \right] \Bigg\} \,,
\end{eqnarray}
where
\begin{eqnarray}
   h_1(a) &=& 2 - 2\,\frac{3952-5416a+2988a^2-838a^3+120a^4-7a^5}
                      {(6-a)(4-a)^2(3-a)(2-a)^2}
    + c\,\frac{20-8a+a^2}{(6-a)(4-a)(2-a)} \,, \nonumber\\
   h_2(a) &=& - 4\,\frac{20-8a+a^2}{(6-a)(4-a)(2-a)} \,, \\
   h_3(a) &=& \frac{5056-6744a+3556a^2-942a^3+127a^4-7a^5}
                   {(6-a)(4-a)^2(3-a)(2-a)^2} \,. \nonumber
\end{eqnarray}
Large logarithms of the form $[\alpha_s\ln(m_b/\mu_i)]^n$ and 
$\alpha_s[\alpha_s\ln(m_b/\mu_i)]^n$ are resummed exactly at 
next-to-leading order in renormalization-group improved perturbation 
theory. They enter the coefficient functions through the parameters $a$ 
and $c$ defined in (\ref{acdef}). 

In the equations for decay rates and spectra presented in 
Section~\ref{sec:rates}, the dependence of the perturbative coefficients 
on the intermediate matching scale $\mu_i$ cancels against the scale 
dependence of the renormalized shape function $\hat S(\hat\omega,\mu_i)$. 
In practice, it is difficult to trace this cancellation if a model for 
the shape function at a fixed scale is employed. On the contrary, in the 
present case the weight function is formally independent of the scale 
$\mu_i$, because there is nothing to cancel a potential $\mu_i$ 
dependence in (\ref{raterel}). This fact can also be shown explicitly 
using the formulae given above. Expanding the resummed result for the 
weight function to first order in $\alpha_s$, we obtain the simple 
expression
\begin{equation}\label{w1loop}
   w(\Delta,P_+)\big|_{\rm 1-loop}
   = \frac{2(\Delta-P_+)}{M_B-P_+} \left[ 1 + \frac{C_F\alpha_s}{4\pi}
   \left( - \frac53\,\ln\frac{\Delta-P_+}{m_b} - \frac{17}{36} \right)
   \right] ,
\end{equation}
in which the dependence on $\mu_i$ has canceled. However, since this 
formula contains a large logarithm and the scale to be used in 
$\alpha_s$ is undetermined, it should not be used for phenomenological 
applications.

\begin{figure}
\begin{center}
\epsfig{file=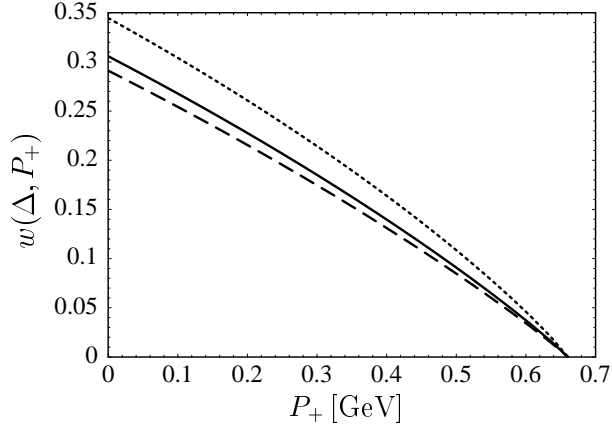,width=8cm}
\end{center}
\centerline{\parbox{14cm}{\caption{\label{fig:wfun}
Weight function $w(\Delta,P_+)$ entering the rate relation 
(\ref{raterel}) for $\Delta=M_D^2/M_B$ and three different choices of the
intermediate scale, namely $\mu_i=1.5$\,GeV (solid), 2.0\,GeV (dashed), 
and 1.0\,GeV (dotted). The weight function is formally independent of 
$\mu_i$.}}}
\end{figure}

Figure~\ref{fig:wfun} shows results for the resummed weight function in 
the case where $\Delta=M_D^2/M_B\simeq 0.66$\,GeV. The three curves refer 
to different values of the scale $\mu_i$. The stability with respect to 
variations of the intermediate scale is very good except for the case of 
a very low scale ($\mu_i=1$\,GeV), for which the convergence of 
perturbation theory is expected to be poor.

The result (\ref{wfun}) settles an old argument about the form of the 
weight function in relations such as (\ref{raterel}). Leibovich et al.\ 
have presented a form for the weight function in a relation between the 
resummed $\bar B\to X_s\gamma$ and $\bar B\to X_u\,l^-\bar\nu$ decay 
rates, in which the kinematic variables corresponding to $P_+$ and 
$\Delta$ above enter in a most complicated form \cite{Leibovich:1999xfx}. 
Their result has the unattractive feature that the integral over the 
weight function contains a Landau-pole singularity, which must be avoided 
by introducing a cutoff on the $P_+$ integral on the right-hand side of 
(\ref{raterel}). A simpler form of the weight function, which is 
equivalent to (\ref{w1loop}) and contains only a single logarithm of the 
kinematic variables, has been promoted in \cite{Neubert:2001sk}. Our 
exact result (\ref{wfun}) obtained after renormalization-group 
improvement retains a simple form with only a single logarithm. Note, in 
particular, that our result does not exhibit any unphysical Landau 
singularities. We believe the resolution of the discrepancy has to do 
with the choice of the intermediate matching scale $\mu_i$. Since the 
weight function is formally independent of $\mu_i$, we are free to choose 
any value $\mu_i\sim\sqrt{m_b\Lambda_{\rm QCD}}$. Taking the particular 
choice $\mu_i^2=m_b(\Delta-P_+)$ would eliminate the logarithmic term in 
(\ref{wfun}), however at the price of introducing a very complicated 
dependence on the kinematic variables $\Delta$ and $P_+$ via the 
dependence of the weight function on the quantity 
$a=\frac{16}{25}\ln[\alpha_s(\mu_i)/\alpha_s(m_b)]$. Also, with this
choice the weight function would develop a Landau pole singularity at the 
point where the coupling $\alpha_s(\sqrt{m_b(\Delta-P_+)})$ gets strong, 
which happens for $(\Delta-P_+)\sim\Lambda_{\rm QCD}^2/m_b$. All these 
unwanted features are avoided by using a fixed value for the intermediate 
scale, as we did in (\ref{wfun}).

\section{Numerical results}
\label{sec:appls}

We are now ready to study the implications of our analysis for 
phenomenology. We start by deriving the numerical values for the 
shape-function mass and kinetic energy including errors. We then present 
a model for the shape function which satisfies all theoretical 
constraints, and study its behavior under renormalization-group 
evolution. Finally, we present numerical results for the various decay 
rates and spectra investigated in Section~\ref{sec:rates}.

Throughout this paper we use the two-loop running coupling constant in 
the $\overline{\rm MS}$ scheme, normalized such that 
$\alpha_s(M_Z)=0.119$. We take $m_b=4.65$\,GeV as the default value for
the $b$-quark mass in the shape-function scheme (see below), and
$\mu_i=1.5$\,GeV as the standard choice of the intermediate matching 
scale. This corresponds to setting $\mu_i^2=m_b\Lambda_{\rm had}$ with a 
typical hadronic scale $\Lambda_{\rm had}\approx 0.5$\,GeV. Note that 
this choice eliminates the appearance of the $b$-quark mass from the 
arguments of the logarithmic terms in the expressions for the decay 
rates. The values of the strong coupling evaluated at these scales are
$\alpha_s(m_b)\simeq 0.222$ and $\alpha_s(\mu_i)\simeq 0.375$. The 
corresponding values of the perturbative parameters $a$ and $c$ defined 
in (\ref{acdef}) are $a\simeq 0.335$ and $c\simeq 0.614$. Finally, the 
leading-order Sudakov factor in (\ref{master}) takes the value 
$e^{V_H(m_b,\mu_i)}\simeq 1.21$.

\subsection{Shape-function mass and kinetic energy}

A value for the shape-function mass can be obtained by 
combining the relations (\ref{mSFmpole}) or (\ref{mSFmPS}) with existing
predictions for the $b$-quark mass in the relevant renormalization 
schemes. The potential-subtracted mass at the scale $\mu_f=2$\,GeV has 
been determined from moments of the $b\bar b$ cross section and the mass 
of the $\Upsilon(1S)$ state \cite{Beneke:1999fe}. Using the first 
relation in (\ref{mSFmPS}), the result of this paper implies
$m_b^{\rm SF}(2\,\mbox{GeV},2\,\mbox{GeV})=m_b^{\rm PS}(2\,\mbox{GeV})%
=(4.59\pm 0.08)$\,GeV. From a similar analysis the kinetic mass has 
been determined at the scale $\mu_f=2$\,GeV to be 
$m_b^{\rm kin}(1\,\mbox{GeV})=(4.57\pm 0.06)$\,GeV \cite{Benson:2003kp}. 
From the second relation in (\ref{mSFmPS}) it then follows that 
$m_b^{\rm SF}(1\,\mbox{GeV},1\,\mbox{GeV})=(4.65\pm 0.06)$\,GeV. Using
relation (\ref{mustar}) to compute the scale dependence of the 
shape-function mass, we obtain at the intermediate scale the values 
$m_b^{\rm SF}(\mu_i,\mu_i)=(4.61\pm 0.08)$\,GeV and
$m_b^{\rm SF}(\mu_i,\mu_i)=(4.65\pm 0.06)$\,GeV, respectively. 
Alternatively, we may use relation (\ref{mSFmpole}) in conjunction with 
an experimental determination of the $b$-quark pole mass from moments of 
inclusive $\bar B\to X_c\,l^-\bar\nu$ and $\bar B\to X_s\gamma$ decay 
spectra. Using the average value 
$\bar\Lambda_{\rm pole}=(0.375\pm 0.065)$\,GeV obtained from
\cite{Chen:2001fj,Cronin-Hennessy:2001fk,Mahmood:2002tt,Aubert:2003dr}, 
we find $m_b^{\rm SF}(\mu_i,\mu_i)=(4.67\pm 0.07)$\,GeV. It is quite
remarkable that these different determinations of the shape-function mass,
which use rather different physics input, give highly consistent results.
Combining them, we quote our default value for the shape-function mass
at the intermediate scale $\mu_i=1.5$\,GeV as
\begin{equation}\label{mbdefault}
   m_b^{\rm SF}(\mu_i,\mu_i) = (4.65\pm 0.07)\,\mbox{GeV} \,.
\end{equation}
The corresponding $\bar\Lambda$ parameter is 
$\bar\Lambda(\mu_i,\mu_i)=(0.63\pm 0.07)$\,GeV.

A value of the kinetic-energy parameter in the shape-function scheme can 
be obtained from (\ref{mupi2pole}) or (\ref{mupi2kin}). Using the first 
relation combined with the experimental value 
$-\lambda_1=(0.25\pm 0.06)$\,GeV$^2$ 
\cite{Cronin-Hennessy:2001fk,Mahmood:2002tt,Aubert:2003dr} yields
$\mu_\pi^2(\mu_i,\mu_i)=(0.271\pm 0.064)$\,GeV$^2$. Alternatively, we may 
use the result for the kinetic-energy parameter obtained in the kinetic 
scheme, $[\mu_\pi^2(1\,\mbox{GeV})]_{\rm kin}=(0.45\pm 0.10)$\,GeV$^2$ 
\cite{Benson:2003kp}, to get from (\ref{mupi2kin}) the value
$\mu_\pi^2(\mu_i,\mu_i)=(0.254\pm 0.107)$\,GeV$^2$. Again, the two
determinations are in very good agreement with each other. Combining 
them, we obtain
\begin{equation}\label{mupi2default}
   \mu_\pi^2(\mu_i,\mu_i) = (0.27\pm 0.07)\,\mbox{GeV}^2 \,.
\end{equation}

\subsection{Model shape functions}

In our analysis of decay rates below, we will adopt a model for the shape 
function $\hat S(\hat\omega,\mu_i)$ at the intermediate scale. We stress,
however, that ultimately the shape function could be extracted from a 
fit to the photon spectrum in $\bar B\to X_s\gamma$ decays. This would
largely reduce the theoretical uncertainties in our predictions.

For the purpose of illustration, we use a two-component ansatz for the 
shape function that is a generalization of the model employed in 
\cite{DeFazio:1999sv,Kagan:1998ym}. The form we propose is
\begin{equation}\label{SFmodel}
   \hat S(\hat\omega,\mu)
   = \frac{N}{\Lambda} \left( \frac{\hat\omega}{\Lambda} \right)^{b-1}
   \exp\left( - b\,\frac{\hat\omega}{\Lambda} \right)
   - \frac{C_F\alpha_s(\mu)}{\pi}\,
   \frac{\theta(\hat\omega-\Lambda-\mu/\sqrt{e})}{\hat\omega-\Lambda}
   \left( 2\ln\frac{\hat\omega-\Lambda}{\mu} + 1 \right) ,
\end{equation}
where $\Lambda$ and $b$ are model parameters, and $\Lambda$ differs from 
the pole-scheme parameter $\bar\Lambda_{\rm pole}$ by an amount of 
$O(\alpha_s(\mu))$. In the limit $\alpha_s(\mu)\to 0$ this function 
reduces to the one used in \cite{DeFazio:1999sv,Kagan:1998ym}. The 
radiative tail ensures the correct leading asymptotic 
behavior of the shape function as displayed in (\ref{Sasymp}). This in 
turn gives the correct power-like dependence of shape-function moments on 
the integration cutoff. In our model, this tail is glued onto a 
``primordial'', exponential function such that the combined result is 
continuous. 

There are several non-trivial constraints on the parameters of the model.
The normalization factor $N$ is given by
\begin{equation}
   N = \left[ 1 - \frac{C_F\alpha_s(\mu)}{\pi}
   \left( \frac{\pi^2}{24} - \frac14 \right) \right]
   \frac{b^b}{\Gamma(b)} ,
\end{equation}
which is determined such that the integral over the shape function from
$\hat\omega=0$ to $\mu_f+\bar\Lambda(\mu_f,\mu)$ coincides with the 
first expression in (\ref{M0toM2}) up to second-order power corrections
and exponentially small terms of order $e^{-\mu_f/\Lambda}$, which are 
negligible whenever $\mu_f$ is sufficiently large to trust our moment 
relations. By evaluating the first moment of the model shape function, we 
find that
\begin{equation}
   \Lambda = \bar\Lambda_{\rm pole}
   + \frac{C_F\alpha_s(\mu)}{\pi}\,\frac{2\mu}{\sqrt e}
\end{equation}
to first order in $\alpha_s$. From (\ref{mSFmpole}) it then follows that
\begin{equation}\label{Lambdavalues}
   \Lambda = \bar\Lambda(\mu_i,\mu_i)
   + \mu_i \left( \frac{2}{\sqrt e} - 1 \right)
   \frac{C_F\alpha_s(\mu_i)}{\pi}
   \simeq \bar\Lambda(\mu_i,\mu_i) + 51\,\mbox{MeV} \,.
\end{equation}
Finally, the model parameter $b$ can be adjusted to reproduce a given 
value for the second moment of the shape function. 

\begin{table}
\centerline{\parbox{14cm}{\caption{\label{tab:SFproperties}
Parameters and moments of the model shape functions at the intermediate 
scale $\mu_i$. The running quantities $m_b^{\rm SF}$, $\bar\Lambda$, and 
$\mu_\pi^2$ are defined in the shape-function scheme and evaluated at 
$\mu_f=\mu=\mu_i=1.5$\,GeV.}}}
\vspace{0.1cm}
\begin{center}
{\tabcolsep=0.3cm
\begin{tabular}{|cc|ccc|cc|}
\hline\hline
Model & Lines & $m_b^{\rm SF}$\,[GeV] & $\bar\Lambda$\,[GeV]
 & $\mu_\pi^2$\,[GeV$^2$] & $\Lambda$\,[GeV] & $b$ \\
\hline\hline
S1 & Dotted & 4.72 & 0.56 & 0.20 & 0.611 & 2.84 \\
S2 & & & & 0.27 & 0.617 & 2.32 \\
S3 & & & & 0.34 & 0.626 & 1.92 \\
\hline
S4 & Solid & 4.65 & 0.63 & 0.20 & 0.680 & 3.57 \\
S5 & & & & 0.27 & 0.685 & 2.93 \\
S6 & & & & 0.34 & 0.692 & 2.45 \\
\hline
S7 & Dashed & 4.58 & 0.70 & 0.20 & 0.751 & 4.40 \\
S8 & & & & 0.27 & 0.753 & 3.61 \\
S9 & & & & 0.34 & 0.759 & 3.03 \\
\hline\hline
\end{tabular}}
\end{center}
\end{table}

\begin{figure}[!ht]
\begin{center}
\epsfig{file=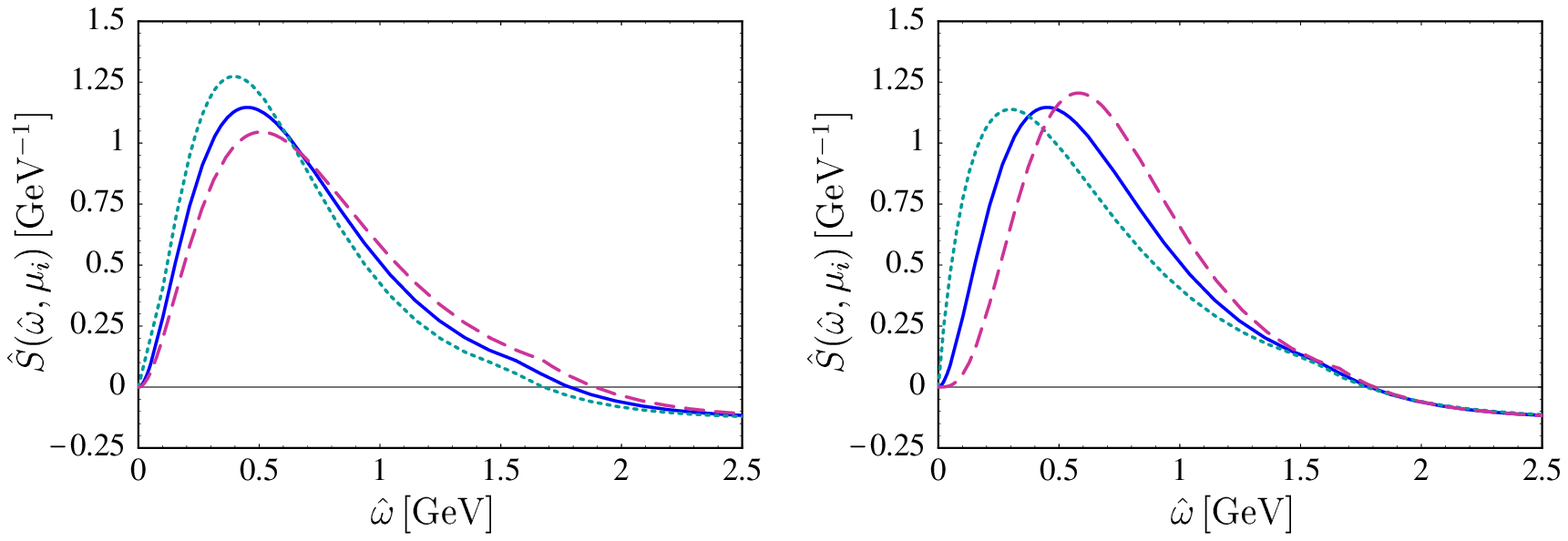,width=16cm}
\end{center}
\centerline{\parbox{14cm}{\caption{\label{fig:SFatmui}
Various models for the shape function at the intermediate scale 
$\mu_i=1.5$\,GeV, corresponding to different parameter settings in 
Table~\ref{tab:SFproperties}. Left: Functions S1, S5, S9 with 
``correlated'' parameter variations. Right: Functions S3, S5, S7 with 
``anti-correlated'' parameter variations.}}}
\vspace{0.3cm}
\begin{center}
\epsfig{file=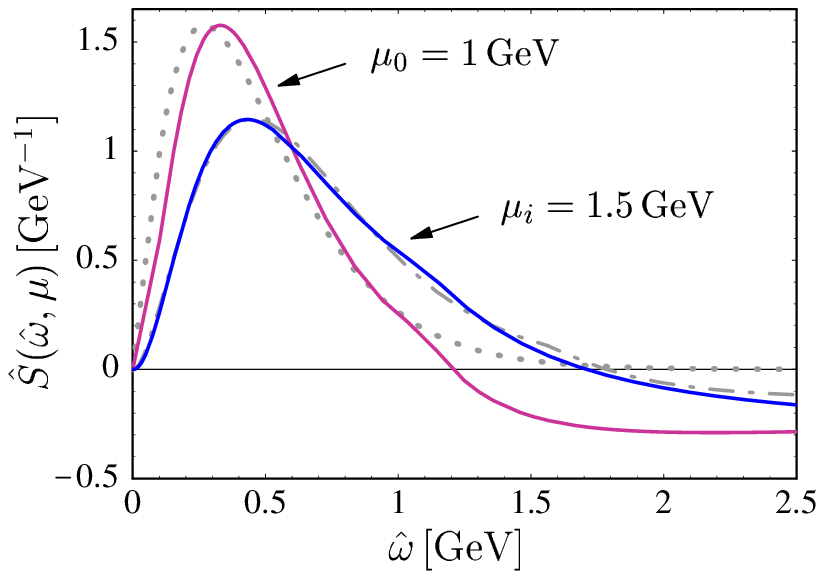,width=8cm}
\end{center}
\centerline{\parbox{14cm}{\caption{\label{fig:SFrge}
Renormalization-group evolution of a model shape function from a low 
scale $\mu_0$ (sharply peaked solid curve) to the intermediate scale 
$\mu_i$ (broad solid curve). See the text for an explanation of the other 
curves.}}}
\end{figure}

Table~\ref{tab:SFproperties} collects the parameters of the model shape 
functions  at the intermediate scale $\mu_i=1.5$\,GeV corresponding to 
different values of $\bar\Lambda(\mu_i,\mu_i)$ and 
$\mu_\pi^2(\mu_i,\mu_i)$,  as computed from the moment relations 
(\ref{M0toM2}). These quantities are varied within their respective error 
ranges given in (\ref{mbdefault}) and (\ref{mupi2default}). Note that the 
$\Lambda$ values in the table are in very good agreement with those 
obtained from the relation (\ref{Lambdavalues}). The left-hand 
(right-hand) plot in Figure~\ref{fig:SFatmui} shows three models for the 
shape function obtained by varying the parameters $\bar\Lambda$ and 
$\mu_\pi^2$ in a correlated (anti-correlated) way. In both cases, the 
solid, dashed, and dotted curves refer to different values of 
$\bar\Lambda$, as indicated in the table.

In Figure~\ref{fig:SFrge} we illustrate how the shape function behaves 
under renormalization-group evolution. The sharply peaked solid line 
shows our model function evaluated with $\Lambda=0.495$\,GeV and $b=3.0$, 
which we use as an ansatz for the function $\hat S(\hat\omega,\mu_0)$ at 
the low scale $\mu_0=1$\,GeV. For comparison, the dotted gray curve shows 
the default choice for the shape function adopted in 
\cite{DeFazio:1999sv,Kagan:1998ym}, which exhibits a very similar shape
except for the missing radiative tail. The broad solid curve gives the
shape function at the intermediate scale $\mu_i=1.5$\,GeV as obtained 
from the evolution equation (\ref{wow}). The barely visible dashed-dotted 
curve shows our default model for the function 
$\hat S(\hat\omega,\mu_i)$, which coincides with the solid line in the
left-hand plot. The beautiful agreement of the two curves gives us
confidence in the consistency of our models adopted for the shape 
function at the intermediate scale.

\subsection{Predictions for decay spectra and event fractions}

We are now ready to present our results for the decay spectra and 
partially integrated event fractions in $\bar B\to X_u\,l^-\bar\nu$ 
decays. In order to illustrate the sensitivity to shape-function effects 
we use the nine shape functions S1 through S9 in 
Table~\ref{tab:SFproperties}, thereby taking into account the full range 
of allowed values for the parameters $\bar\Lambda(\mu_i,\mu_i)$ and 
$\mu_\pi^2(\mu_i,\mu_i)$. For each physical quantity we draw three bands
corresponding to the three different values of $\bar\Lambda$. The width
of each band reflects the sensitivity to the variation of $\mu_\pi^2$. 
While one might in principle consider other functional forms for the 
shape function (which, however, must be consistent with the asymptotic 
behavior as predicted by the operator product expansion), we believe that 
the variation of our results corresponding to the various models presents 
a realistic estimate of the shape-function sensitivity (see below).

The following predictions for spectra and rate fractions refer to the 
leading term in the heavy-quark expansion. Using our formalism, power 
corrections can be computed systematically to any order in 
$\Lambda_{\rm QCD}/m_b$, by extending the two-step matching 
QCD\,$\to$\,SCET\,$\to$\,HQET to the appropriate order. These power
corrections fall into several categories, including corrections from 
phase space, kinematic factors, and subleading shape functions 
\cite{Bauer:2001mh} (see 
\cite{Bauer:2002yu,Leibovich:2002ys,Burrell:2003cf} for tree-level 
discussions of these effects on various $\bar B\to X_u\,l^-\bar\nu$ 
spectra). Estimating the corrections from phase space alone, we find 
that they typically change the leading-order predictions for partially 
integrated event fractions by about 10\%. A more careful investigation of 
power corrections is left for future work. Finally, we note that our 
calculations would break down if the cuts on kinematic variables were 
taken to be too strict, because then the spectra would become dominated 
by hadronic resonance effects. Parametrically, this happens when the 
quantities $\Delta_P$, $\Delta_s$, or $\Delta_E$ become of order 
$\Lambda_{\rm QCD}^2/M_B\sim 50$\,MeV.

\begin{figure}[t]
\begin{center}
\epsfig{file=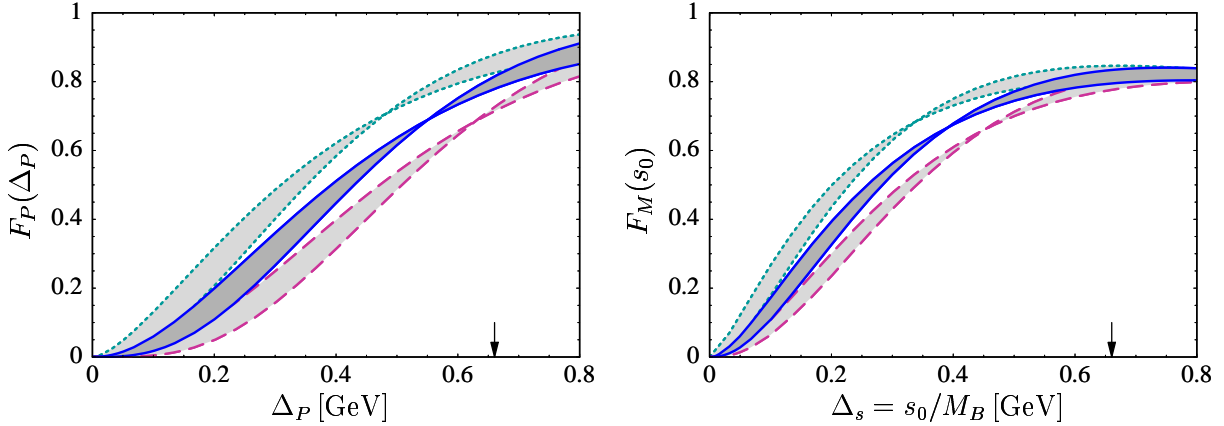,width=16cm}
\end{center}
\centerline{\parbox{14cm}{\caption{\label{fig:FPandFM}
Fraction of $\bar B\to X_u\,l^-\bar\nu$ events with hadronic light-cone 
momentum $P_+\le\Delta_P$ (left), and fraction of events with hadronic 
invariant mass $s_H\le s_0$ (right). In each plot, the three bands 
correspond to the values $\bar\Lambda=0.63$\,GeV (solid curves), 
0.70\,GeV (dashed curves), and 0.56\,GeV (dotted curves). Their width 
reflects the sensitivity to the value of $\mu_\pi^2$ varied in the range 
between 0.20 and 0.34\,GeV$^2$. The arrow indicates the point at which 
the charm background starts.}}}
\end{figure}

In Figure~\ref{fig:FPandFM} we show predictions for the fractions of all
$\bar B\to X_u\,l^-\bar\nu$ events with hadronic light-cone momentum 
$P_+\le\Delta_P$, and with hadronic invariant mass squared $s_H\le s_0$. 
Recall that, for $\Delta_P=\Delta_s=s_0/M_B$, the hadronic invariant mass 
fraction $F_M$ differs from the fraction $F_P$ by the contribution of the 
events in the triangular region above the dashed line in 
Figure~\ref{fig:phase}. Comparing the two plots, we observe that this 
additional contribution is predicted to be very small. (Note that for 
large values of $\Delta_s$ we even find a negative contribution to the 
rate from the triangle region for some choices of the shape function. 
This feature is unphysical and should be fixed by the inclusion of 
power corrections to our leading-order predictions.) The arrows on the 
horizontal axes indicate the points $\Delta_{P,s}=M_D^2/M_B$, beyond 
which final states containing charm hadrons are kinematically allowed. 
With this choice of the cut, both rate fractions capture about 80\% of 
all events. While it is well known that a hadronic invariant mass cut 
$\sqrt{s_H}\le M_D$ provides a very efficient discrimination against 
charm background \cite{Bigi:1997dn,Falk:1997gj,Jezabek:2001pg}, here we 
observe that the same is true for a cut on the $P_+$ variable. Cutting on 
$P_+$ offers the additional advantage of a ``buffer zone'' against charm 
background. Whereas the region in which charm final states are 
kinematically allowed borders the region with $\sqrt{s_H}\le M_D$, it 
touches the phase-space region with $P_+\le M_D^2/M_B$ at only a single 
point (see Figure~\ref{fig:phase}). 

\begin{figure}
\begin{center}
\epsfig{file=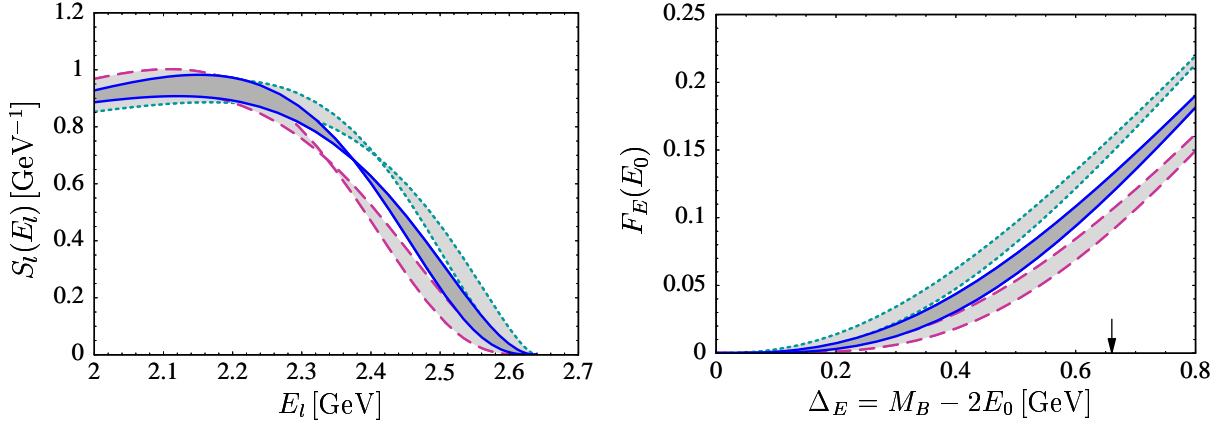,width=16cm}
\end{center}
\centerline{\parbox{14cm}{\caption{\label{fig:FE}
Charged-lepton energy spectrum in the region near the kinematic endpoint 
(left), and fraction of events with charged-lepton energy $E_l\ge E_0$ 
(right). The meaning of the bands and the arrow is the same as in 
Figure~\ref{fig:FPandFM}.}}}
\end{figure}

Our results for the charged-lepton energy spectrum 
$S_l(E_l)=(1/\Gamma_{\rm tot})\,(d\Gamma/dE_l)$, and for the event 
fraction with a cut $E_l\ge E_0$, are displayed in Figure~\ref{fig:FE}. 
The right-hand plot shows that with $\Delta_E=M_D^2/M_B$ only about 
10--15\% of all events are retained, and the theoretical calculation is 
very sensitive to shape-function effects. Such a cut is therefore much 
less efficient than the cuts on hadronic invariant mass or $P_+$. As a 
result, an extraction of $|V_{ub}|$ from the charged-lepton endpoint 
region is theoretically disfavored.

Let us now comment in more detail on the quantitative features of the
results for the various event fractions. The most significant observation 
drawn from Figure~\ref{fig:FPandFM} is that the shape-function 
sensitivity is rather small for values of $\Delta_P$ and $\Delta_s$ near
the charm threshold. This is to some extent a consequence of our improved 
knowledge of the shape-function parameters. For instance, whereas in 
previous analyses the heavy-quark mass was varied in the range 
$m_b=(4.8\pm 0.2)$\,GeV \cite{DeFazio:1999sv}, the mass defined in the 
shape-function scheme is known with much better accuracy, see 
(\ref{mbdefault}). A similar statement applies to the width of the shape 
function and, more importantly, to its asymptotic behavior. Yet, the 
good convergence of the three bands in each plot seems puzzling at first 
sight, given that the model shape functions in Figure~\ref{fig:SFatmui} 
are still rather different at $\hat\omega\approx 0.7$\,GeV. The event 
fraction $F_P(\Delta_P)$, in particular, is {\em at tree level\/} given
by the area under the shape-function curves between 0 and $\Delta_P$. 
Remarkably, an interesting ``focusing mechanism'' arises beyond the tree 
approximation, which has its origin in a subtle interplay between the 
shape function and the jet function under the convolution integral in 
(\ref{FP}). The point is that the next-to-leading order corrections to 
the jet function (the terms proportional to $\alpha_s(\mu_i)$) contain a 
double-logarithmic singularity at the endpoint of the integration domain 
(at $\hat\omega=\Delta_P$), which comes with a {\em positive\/} 
coefficient. Consider now the integrals over the model shape functions 
from 0 to $\Delta_P$, assuming that $\Delta_P$ is well beyond the maximum 
of the curves. The function with the smallest (largest) area takes the 
largest (smallest) value at the endpoint. When the shape functions are 
weighted with the jet function, the logarithmic spike at 
$\hat\omega=\Delta_P$ gives a contribution to the integral that is 
proportional to $\hat S(\Delta_P)$ and so is largest (smallest) for the 
function with the smallest (largest) area. The net result is to balance 
the differences in the areas and make the integrals converge more quickly 
toward a single value. The workings of this mechanism are nicely 
illustrated by comparing the results for the event fractions $F_P$ and 
$F_E$ in Figures~\ref{fig:FPandFM} and \ref{fig:FE}. Whereas focusing 
takes place in the former case, the weight function under the integral 
for $F_E$ in (\ref{FE}) vanishes at $\hat\omega=\Delta_E$, thereby 
suppressing the contribution from the logarithms in the jet function. 
This explains why no focusing is observed for the event fraction with 
a cut on charged-lepton energy.

Another way of thinking about this mechanism is to notice that the 
broadening of the shape function under renormalization-group evolution 
from a low scale up to the intermediate scale (right-hand plot in 
Figure~\ref{fig:SFatmui}) is a perturbative effect, which should not 
lead to an increased shape-function sensitivity. Because the 
convolution of the shape function with the jet and hard functions is
independent of the scale $\mu_i$, the broadening of the shape function
must be compensated by perturbative logarithms in the jet function. 

Note that it is crucial for this mechanism that the logarithmic terms 
in the jet function give a {\em positive\/} contribution near 
$\hat\omega=\Delta_P$. This focusing effect did not take place in 
earlier studies such as 
\cite{DeFazio:1999sv,Dikeman:1997es,Bigi:1997dn,Falk:1997gj,Jezabek:2001pg}, 
where parton-model spectra were convoluted with a primordial shape 
function. As mentioned earlier, in the parton model the leading Sudakov 
logarithm comes with the opposite (negative) sign, hence causing an 
anti-focusing effect of the radiative corrections. This also explains 
why our prediction for the hadronic invariant mass fraction $F_M$ 
exhibits a smaller shape-function sensitivity than what has been found 
in most previous analyses.

\begin{figure}
\begin{center}
\epsfig{file=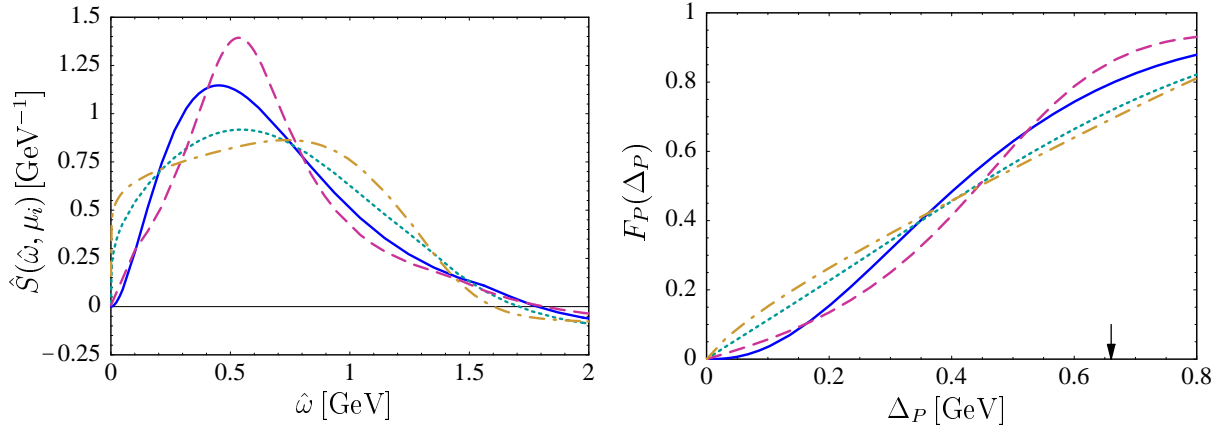,width=16cm}
\end{center}
\centerline{\parbox{14cm}{\caption{\label{fig:new}
Left: Four examples of shape functions with identical normalization and 
first two moments, corresponding to $\bar\Lambda(\mu_i,\mu_i)=0.63$\,GeV 
and $\mu_\pi^2(\mu_i,\mu_i)=0.27$\,GeV$^2$, but different functional 
form. Right: Corresponding results for the event fraction 
$F_P(\Delta_P)$.}}}
\end{figure}

In addition to their dependence on the moment parameters $\bar\Lambda$ 
and $\mu_\pi^2$, our predictions are also sensitive to the functional 
form adopted for the shape function. In order to study this sensitivity 
we have constructed four shape functions with identical zeroth, first, 
and second moments, but rather different functional forms. They are shown 
in the left-hand plot in Figure~\ref{fig:new}. Some of these choices are 
intentionally rather extreme, given what is known about the shape 
function from the measurement of the $\bar B\to X_s\gamma$ photon 
spectrum \cite{Chen:2001fj}. The point is to illustrate that even drastic 
variations of the functional form do not invalidate the predictions 
presented earlier in this section. The right-hand plot in 
Figure~\ref{fig:new} shows, as an example, the results for the event 
fraction $F_P(\Delta_P)$ obtained using the four functions. Comparing 
this plot with the corresponding one in Figure~\ref{fig:FPandFM} shows 
that at present the variation of the parameters $\bar\Lambda$ and 
$\mu_\pi^2$ covers the uncertainty in the functional form. We have 
checked that this is also true for the other decay distributions studied 
here. However, if in the future the errors on $\bar\Lambda$ and 
$\mu_\pi^2$ were reduced significantly, one should also include the 
sensitivity to the functional form in the error estimate.

\begin{table}[t]
\centerline{\parbox{14cm}{\caption{\label{tab:comp}
Comparison of different theoretical methods using inclusive $B$-decay 
rates to extract the CKM matrix element $|V_{ub}|$. The error on the 
efficiency represents the sensitivity to the shape function only. All
results refer to the leading term in the heavy-quark expansion.}}}
\vspace{0.1cm}
\begin{center}
{\tabcolsep=0.3cm\renewcommand{\arraystretch}{1.35} 
\begin{tabular}{|c|c|c|}
\hline\hline
Method & Cut & Efficiency \\
\hline\hline
Hadronic invariant mass & $s_H\le M_D^2$ & $(81.4_{\,-3.7}^{\,+3.2})\%$ \\
 & $s_H\le(1.7\,\mbox{GeV})^2$ & $(78.2_{\,-5.2}^{\,+4.9})\%$ \\
 & $s_H\le(1.55\,\mbox{GeV})^2$ & $(72.7_{\,-6.3}^{\,+6.4})\%$ \\
\hline
Hadronic $P_+$ & $P_+\le\frac{M_D^2}{M_B}=0.66$\,GeV
 & $(79.6_{\,-8.2}^{\,+8.2})\%$ \\
 & $P_+\le 0.55$\,GeV & $(69.0_{\,-12.1}^{\,+\phantom{1}9.7})\%$ \\
\hline
Charged-lepton energy & $E_l\ge\frac{M_B^2-M_D^2}{2M_B}=2.31$\,GeV
 & $(12.5_{\,-3.5}^{\,+3.4})\%$ \\
 & $E_l\ge 2.2$\,GeV & $(22.2_{\,-3.6}^{\,+3.2})\%$ \\
\hline
Combined ($s_H$, $q^2$) cuts & $s_H\le M_D^2$, ~$q^2\ge 0$
 & $(74.6_{\,-5.1}^{\,+5.1})\%$ \\
\mbox{[tree level only]} & $s_H\le M_D^2$, ~$q^2\ge 6\,\mbox{GeV}^2$
 & $(45.7_{\,-2.0}^{\,+1.8})\%$ \\
 & $s_H\le(1.7\,\mbox{GeV})^2$, ~$q^2\ge 8\,\mbox{GeV}^2$
 & $(33.4_{\,-1.8}^{\,+1.6})\%$ \\
\hline\hline
\end{tabular}}
\end{center}
\end{table}

A summary of our phenomenological results can be found in 
Table~\ref{tab:comp}. For a variety of different experimental cuts we 
report the fractions of the contained $\bar B\to X_u\,l^-\bar\nu$ events 
and indicate how these fractions vary under the variation of the 
shape-function models. In all cases the goal is to reject the charm 
background as efficiently as possible, while preserving a sufficiently 
large fraction of the signal events so as to obtain a reliable 
determination of the CKM matrix element $|V_{ub}|$. We emphasize that 
our results refer to the leading term in the heavy-quark expansion, 
and that the rates are expected to be modified somewhat by power 
corrections. The uncertainties quoted on the contained event fractions 
reflect their sensitivity to shape-function effects only. No other 
theoretical uncertainties are included in these estimates.

The first portion of the table contains results for the event fractions 
with a cut on hadronic invariant mass. The first line corresponds to the 
``ideal'' cut $\sqrt{s_H}\le M_D$, which in theory eliminates the charm
background entirely. This cut retains about 80\% of all 
$\bar B\to X_u\,l^-\bar\nu$ events. In practice, spill-over from final
states containing charm hadrons in a realistic measurement requires to 
lower the cut on hadronic invariant mass to values slightly below $M_D$. 
Two typical choices are also covered in the table. Lowering the cut 
reduces the contained event fraction by modest amounts, while the 
sensitivity to shape-function effects increases markedly. Yet, even with 
a cut at 1.55\,GeV the uncertainty on  $|V_{ub}|$ would still amount to 
only 4\%.

The second portion of the table shows two examples of cuts on the 
hadronic light-cone momentum $P_+$. With the ``ideal'' cut 
$P_+\le M_D^2/M_B$ about 80\% of all events are contained, which is as 
good as with the hadronic invariant mass cut. The sensitivity to 
shape-function effects is somewhat more pronounced but still at a very 
acceptable level. The corresponding uncertainty on $|V_{ub}|$ is about 
5\%. Since the $P_+$ spectrum is directly related to the photon 
energy spectrum in $\bar B\to X_s\gamma$ decays, this uncertainty can be 
reduced further by using experimental information on the photon spectrum. 
Due to the fact that there is a ``buffer zone'' separating the signal 
events from the charm background, we expect less spill-over than in the 
case of a cut on hadronic invariant mass. It may therefore be 
experimentally feasible to implement the cutoff near the optimal value. 
However, even if the value of $\Delta_P$ must be lowered by a small 
amount, the efficiency remains high and the shape function sensitivity 
at an acceptable level (see the second entry in the table).

The third portion of the table shows results for the case of a cut on the 
energy of the charged lepton. The efficiency is obviously much reduced in 
this case, even if the cut can be relaxed into the region where some 
charm background is present (second entry). The theoretical calculations 
are therefore more prone to uncertainties from other effects such as weak 
annihilation \cite{Voloshin:2001xi}. 

In the lower portion of the table we give results for some combined cuts 
on hadronic and leptonic invariant mass, which have been briefly 
discussed in Section~\ref{sec:comb}. Contrary to the other cases, these 
numbers refer to the tree-level approximation and so should be taken with 
caution. For reference, we quote again the (tree-level) result for the 
pure hadronic invariant mass cut, which differs significantly from the 
corresponding result including radiative corrections. While the 
additional cut on leptonic $q^2$ reduces the shape-function sensitivity, 
it comes along with a strong reduction of the efficiency. For instance, 
the combined cut $\sqrt{s_H}\le 1.7$\,GeV and $q^2\ge 8$\,GeV$^2$ 
employed in a recent analysis of the Belle Collaboration 
\cite{Kakuno:2003fk} has an efficiency of about 33\% (at tree level and 
leading order in $\Lambda_{\rm QCD}/m_b$), which is much smaller than the 
efficiency of the pure hadronic invariant mass cut 
$\sqrt{s_H}\le 1.7$\,GeV. However, the sensitivity to shape-function 
effects is only slightly better in the case of the combined cut.

As a final remark, let us comment on the applicability of the theoretical 
framework developed in this work. According to Figure~\ref{fig:phase} 
most of the $\bar B\to X_u\,l^-\bar\nu$ events are located in the 
shape-function region of large $P_-$ and small to moderate $P_+$. Our 
approach is based on a systematic heavy-quark expansion valid in that 
region of phase space. It allows us to calculate inclusive decay rates 
integrated over domains $\Delta P_-\sim M_B$ and $\Delta P_+\ll M_B$, 
where typically $\Delta P_+\sim\Lambda_{\rm QCD}$. (In the examples 
above, $\Delta P_+=\Delta_E$, $\Delta_P$, or $\Delta_s$, respectively.)
While the corresponding predictions for decay spectra and event fractions 
are sufficient to analyse experimental data over most of the phase space 
relevant to measurements of the CKM matrix element $|V_{ub}|$, it would
be of interest to extend the validity of the theoretical description 
outside the shape-function region. In the case where 
$\Delta P_-\sim\Delta P_+\sim M_B$ are both large, the decay spectra can 
be computed using a local operator product expansion. An interesting 
question is whether it will be possible to match these two approaches in
some intermediate region of $\Delta P_+$ values that are numerically (but 
not parametrically) large compared with $\Lambda_{\rm QCD}$. If the two 
predictions were to agree in an overlap region, this could be used to 
construct a theoretical description of inclusive 
$\bar B\to X_u\,l^-\bar\nu$ decay distributions that is valid over the 
entire phase space. While this is an exciting prospect, we note that 
performing a systematic operator product expansion in the overlap region 
is far from trivial. Because of the hierarchy of scales 
$\Delta P_+^2\ll\Delta P_+\,M_B\ll M_B^2$, again a two-step procedure is 
in order. An example of such an approach can be found in 
\cite{Neubert:2001ib}.

\section{Conclusions}

We have calculated differential spectra and partially integrated event 
fractions for the inclusive semileptonic decays 
$\bar B\to X_u\,l^-\bar\nu$ at next-to-leading order in 
renormalization-group improved perturbation theory, and at leading 
power in the heavy-quark expansion. The hadronic tensor entering the 
differential decay rates has been factorized into perturbatively 
calculable hard and jet functions, $H_{ij}$ and $J$, which we give to 
one-loop order, and a universal shape function $S$ containing 
non-perturbative physics below an intermediate scale 
$\mu_i\sim\sqrt{m_b\Lambda_{\rm QCD}}$. This factorization has been 
obtained by matching QCD onto soft-collinear effective theory to 
integrate out hard fluctuations of order $m_b$, and by matching the 
result further onto heavy-quark effective theory (HQET) to integrate out 
hard-collinear modes at the intermediate mass scale. Large logarithms 
have been resummed by solving the renormalization-group equation for the 
hard kernels, evolving the functions $H_{ij}$ from the hard scale 
$\mu_h\sim m_b$ down to the intermediate scale $\mu_i$. In order to make 
use of the resulting expression for the hadronic tensor, the shape 
function is needed at this intermediate scale $\mu_i$. We have derived 
the anomalous dimension of the shape function at one-loop order and given 
an exact analytic solution of the resulting renormalization-group 
equation. Our solution can be applied to any model for the shape function 
obtained at any scale. Alternatively, if the shape function is treated as 
a phenomenological quantity, renormalizing it at the scale $\mu_i$ avoids 
doing perturbation theory at a low hadronic scale.

Moments of the shape function are often identified with HQET parameters 
such as $\bar\Lambda$ and $\lambda_1$. This identification was so far 
only understood at tree level. We have discovered the appearance of a 
radiative tail of the shape function that vanishes slower than 
$1/|\omega|$. This feature renders all shape-function moments (including 
the normalization integral) ultra-violet divergent. To define the moments
consistently we have introduced a hard ultra-violet cutoff 
$\Lambda_{\rm UV}$ on the integrals over $\omega$. This is natural, 
since in any physical process the shape function will only be integrated 
over a finite interval. The dependence of the moments on this cutoff can
be controlled using a local operator product expansion, which has enabled 
us to derive reliable perturbative relations between shape-function 
moments and HQET parameters. We have also obtained a formula for the 
normalization integral over the shape function as a function of 
$\Lambda_{\rm UV}$, and from it a model-independent prediction for the 
asymptotic behavior of the shape function for large values of $|\omega|$. 
Surprisingly, this analysis reveals that the shape function has a 
negative tail and so is not positive definite, contrary to common 
expectation. 

Due to the fact that the $b$-quark pole mass, and with it the HQET 
parameter $\bar\Lambda$, suffers from infra-red renormalon ambiguities, 
it is favorable to replace $m_b^{\rm pole}$ by a short-distance mass 
defined in a physical subtraction scheme. We have argued that the most
natural definition for applications to inclusive spectra is to define 
a ``shape-function mass'' $m_b^{\rm SF}(\Lambda_{\rm UV},\mu)$ by 
enforcing that the first moment of the renormalized shape function 
vanish for any given value of the cutoff $\Lambda_{\rm UV}$. The 
dependence of the shape-function mass on the cutoff $\Lambda_{\rm UV}$ 
and on the dimensional regularization scale $\mu$ is controlled by 
evolution equations, which can be trusted as long as both scales are much 
larger than $\Lambda_{\rm QCD}$. Using a similar approach, we have 
defined a running kinetic-energy parameter 
$\mu_\pi^2(\Lambda_{\rm UV},\mu)$ in the shape-function scheme, which can 
be used to replace the (ambiguous) HQET parameter $\lambda_1$. We relate 
our new parameters to some previous, physical definitions of $m_b$ and 
$\mu_\pi^2$. All of these new insights restrict model building of the 
shape function drastically. We give expressions for some model shape 
functions that are fully consistent with all constraints, albeit leaving 
enough freedom to accommodate future experimental constraints from the 
$\bar B\to X_s\gamma$ photon spectrum.

In the second part of the paper we have applied our results to make
predictions for several interesting $\bar B\to X_u\,l^-\bar\nu$ decay 
distributions and spectra in the shape-function region of large hadronic
energy ($E_H\sim M_B$) and small hadronic invariant mass 
($s_H\sim M_B\Lambda_{\rm QCD})$. We have presented results for event 
fractions with cuts on the charged-lepton energy, hadronic invariant 
mass, or hadronic light-cone momentum $P_+=E_H-|\vec{P}_H|$, where in 
each case the cuts are chosen so as to reject background from 
$\bar B\to X_c\,l^-\bar\nu$ decays. These formulae are presented as 
convolution integrals of weight functions with the shape function 
renormalized at the intermediate scale $\mu_i$. Our results are valid at 
next-to-leading order in renormalization-group improved perturbation 
theory and at leading power in the heavy-quark expansion. The discussion 
of these distributions is most transparent in terms of the hadronic phase 
space for the two variables $P_\pm=E_H\mp|\vec{P}_H|$, which we have 
studied in some detail (see Figure~\ref{fig:phase}).

An important outcome of our phenomenological analysis is the finding that 
a cut $P_+\le M_D^2/M_B\simeq 0.66$\,GeV on the hadronic light-cone 
momentum eliminates the charm background while containing the vast 
majority of all $\bar B\to X_u\,l^-\bar\nu$ events. At leading power in 
$\Lambda_{\rm QCD}/m_b$ this cut has an efficiency of about 80\%, which
is almost as efficient as the cut $\sqrt{s_H}\le M_D$ on hadronic 
invariant mass. A cut on $P_+$ offers several advantages over a cut on 
hadronic invariant mass. First, it provides a ``buffer zone'' against 
charm background, which borders the signal region at only a single point 
in phase space. Secondly, the hadronic physics affecting the $P_+$ 
spectrum is directly related to the hadronic physics affecting the 
$\bar B\to X_s\gamma$ photon energy spectrum (the two spectra are 
identical at tree level). This implies that a simple, shape-function 
independent relation between the two distributions can be derived, which 
could be used to eliminate hadronic uncertainties (at least at leading 
power in $\Lambda_{\rm QCD}/m_b$). Finally, as we have pointed out, the 
calculation of the hadronic invariant mass distribution suffers from the 
fact that it includes a region in phase space where the collinear 
expansion breaks down. While this region gives only a power-suppressed 
contribution to the rate, its presence might cause complications if the 
calculation is taken beyond the leading power. The $P_+$ spectrum, on the 
other hand, can be accurately calculated beyond the leading order using 
the methods developed here. It would be a most useful quantity to 
measure.

Using our formalism, shape-function independent relations between 
different decay distributions can be derived in a systematic way, and 
they are free of unphysical Landau singularities. As an example, we have 
derived a relation between the charged-lepton energy spectrum and a 
weighted integral over the $P_+$ spectrum in $\bar B\to X_u\,l^-\bar\nu$ 
decays. This formula serves as a prototype for other relations, including 
the more useful relation between the $P_+$ spectrum and the photon energy 
spectrum in $\bar B\to X_s\gamma$ decays, which can be used for a new, 
high-precision determination of $|V_{ub}|$ once the technology developed 
in this paper has been applied to $\bar B\to X_s\gamma$ decays.

Finally, let us reiterate that the results presented here are valid at 
leading power in $\Lambda_{\rm QCD}/m_b$. While it may be laborious to 
calculate the next term in the expansion, this can be systematically done 
in our formalism (with the possible exception of the hadronic invariant 
mass spectrum), and we believe it must be done in order to achieve a
theoretical accuracy in the prediction of decay spectra at the 10\% 
level. If this level of precision can be achieved, it would for the first 
time lead the way toward a precision measurement of the CKM matrix 
element $|V_{ub}|$ with a theoretical error at the 5\% level. This would 
be worth the effort.

\vspace{0.5cm}\noindent
{\em Note added:\/}
After this work was completed, we became aware of a paper by Grozin and 
Korchemsky \cite{Grozin:1994ni}, where the evolution kernel for the shape
function in (\ref{hatGs}) was calculated at two-loop order. Our one-loop 
results in (\ref{G1l}) agree with their findings.

\vspace{0.5cm}\noindent
{\em Acknowledgments:\/}
This research was supported by the National Science Foundation under 
Grant PHY-0098631.

\end{document}